\documentclass[lettersize,journal,twoside]{IEEEtran}
\usepackage{amsmath,amsfonts}
\usepackage{algorithmic}
\usepackage{algorithm}
\usepackage{array}
\usepackage[caption=false,font=normalsize,labelfont=sf,textfont=sf]{subfig}
\usepackage{colortbl}
\usepackage{textcomp}
\usepackage{stfloats}
\usepackage{url}
\usepackage{verbatim}
\usepackage{graphicx}
\usepackage{cite}
\usepackage{bm}
\usepackage{fix-cm}
\usepackage{amssymb}
\usepackage{color}
\usepackage{booktabs}
\usepackage[colorlinks, linkcolor=red, anchorcolor=blue, citecolor=blue]{hyperref}
\usepackage{balance}
\usepackage{orcidlink}
\usepackage{multirow}
\usepackage{multicol}
\usepackage{bbding}
\usepackage{makecell}
\makeatletter
\newcommand*{\revise}{\@ifnextchar\bgroup{\revise@}{\color{blue}}}
\newcommand*{\revise@}[1]{{\textcolor{blue}{#1}}}
\makeatother

\begin{document}

\title{Advancing Fluid Antenna-Assisted Non-Terrestrial Networks in 6G and Beyond: Fundamentals, State of the Art, and Future Directions}

\author{Tianheng Xu, \IEEEmembership{Member, IEEE,}
Runke Fan, \IEEEmembership{Student Member, IEEE,}
Jie Zhu, \IEEEmembership{Student Member, IEEE,}\\
Pei Peng, \IEEEmembership{Member, IEEE,}
Xianfu Chen, \IEEEmembership{Senior Member, IEEE,}
Qingqing Wu, \IEEEmembership{Senior Member, IEEE,}\\
Ming Jiang, \IEEEmembership{Senior Member, IEEE,} 
Celimuge Wu, \IEEEmembership{Senior Member, IEEE,}\\ 
and Kai-Kit Wong, \IEEEmembership{Fellow, IEEE}
\vspace{-2ex}

\thanks{Tianheng Xu, Runke Fan, Jie Zhu and Xianfu Chen are with the Shanghai Advanced Research Institute, Chinese Academy of Sciences, Shanghai 201210, China (e-mails: xuth@sari.ac.cn; fanrunke25@mails.ucas.ac.cn; zhujie2024@sari.ac.cn; xianfu.chen@ieee.org).}
\thanks{Runke Fan and Jie Zhu are also with the University of Chinese Academy of Sciences, Beijing 100049, China.}
\thanks{Xianfu Chen is also with the Shenzhen CyberAray Network Technology Co., Ltd., Shenzhen 518038, China.}
\thanks{Pei Peng is with the School of Telecommunication and Information Engineering, Nanjing University of Posts and Telecommunications, Nanjing 210003, China (e-mail: pei.peng@njupt.edu.cn).}
\thanks{Qingqing Wu is with the Department of Electronic Engineering, Shanghai Jiao Tong University, Shanghai 200240, China (e-mail: qingqingwu@sjtu.edu.cn).}
\thanks{Ming Jiang is with the School of Electronics and Information Technology (School of Microelectronics), Sun Yat-sen University, Guangzhou 510275, China (e-mail: jiangm7@mail.sysu.edu.cn).}
\thanks{Celimuge Wu is with the Graduate School of Informatics and Engineering, The University of Electro-Communications, Tokyo 1828585, Japan (e-mail: celimuge@uec.ac.jp).}
\thanks{
Kai-Kit Wong is affiliated with the Department of Electronic and Electrical Engineering, University College London, Torrington Place, WC1E 7JE, United Kingdom and he is also affiliated with the Department of Electronic Engineering, Kyung Hee University, Yongin-si, Gyeonggi-do 17104, Korea (e-mail: kai-kit.wong@ucl.ac.uk).}
\thanks{Corresponding author: Xianfu Chen.}
}

\markboth{IEEE Communications Surveys and Tutorials, ~Vol.~14, No.~8, May~2026}%
{Xu \MakeLowercase{\textit{et al.}}: Advancing Fluid Antenna-Assisted Non-Terrestrial Networks in 6G and Beyond}


\maketitle
\pagestyle{headings}
\begin{abstract}
With the surging demand for ultra-reliable, low-latency, and ubiquitous connectivity in Sixth-Generation (6G) networks, Non-Terrestrial Networks (NTNs) emerge as a key complement to terrestrial networks by offering flexible access and global coverage. Despite the significant potential, NTNs still face critical challenges, including dynamic propagation environments, energy constraints, and dense interference. As a key 6G technology, Fluid Antennas (FAs) can reshape wireless channels by reconfiguring radiating elements within a limited space, such as their positions and rotations, to provide higher channel diversity and multiplexing gains. Compared to fixed-position antennas, FAs can present a promising integration path for NTNs to mitigate dynamic channel fading and optimize resource allocation. This paper provides a comprehensive review of FA-assisted NTNs. We begin with a brief overview of the classical structure and limitations of existing NTNs, the fundamentals and advantages of FAs, and the basic principles of FA-assisted NTNs. We then investigate the joint optimization solutions, detailing the adjustments of FA configurations, NTN platform motion modes, and resource allocations. We also discuss the combination with other emerging technologies and explore FA-assisted NTNs as a novel network architecture for intelligent function integrations. Furthermore, we delve into the physical layer security and covert communication in FA-assisted NTNs. Finally, we highlight the potential future directions to empower broader applications of FA-assisted NTNs.
\end{abstract}


\section{Introduction}\label{intro}
\subsection{Background}
The forthcoming Sixth-Generation (6G) wireless networks aim to provide seamless connectivity and personalized services for massive users anywhere and anytime. The International Telecommunication Union Telecommunication Radiocommunication Sector (ITU-R) has identified six 6G usage scenarios for 2030 and beyond in International Mobile Telecommunications (IMT)-2030 \cite{0}: immersive communication, Hyper-Reliable Low-Latency Communication (HRLLC), massive communication, AI-and-communication integration, ubiquitous connectivity, and Integrated Sensing And Communication (ISAC). In the 6G era, mobile networks are expected to evolve beyond incremental enhanced connectivity toward globally covered intelligent networks. This evolution is driven by immersive/holographic services, massive machine collaboration, digital twins, and native AI, demanding higher throughput, capacity, and reliability, as well as deeper convergence of communication, sensing, and computing. However, traditional Terrestrial Networks (TNs) are inadequate for addressing uninterrupted global coverage, particularly in dense areas, remote areas, and high-mobility scenarios \cite{1}. For example, in urban areas, the limited number and fixed locations of terrestrial base stations fail to satisfy the increasing throughput requirements. In remote areas, the scarcity of network infrastructure hinders the provision of real-time communications. In high-mobility scenarios, frequent cell handovers and severe channel fadings lead to poor link quality. Additionally, the planar topology of terrestrial cellular networks poses challenges in facilitating the collaborative interaction of communication, sensing, and computation between users and their surroundings, which constrains the development of digital applications such as immersive/holographic communications.

\begin{figure*}[t]
    \centering
    \includegraphics[width=0.9\textwidth]{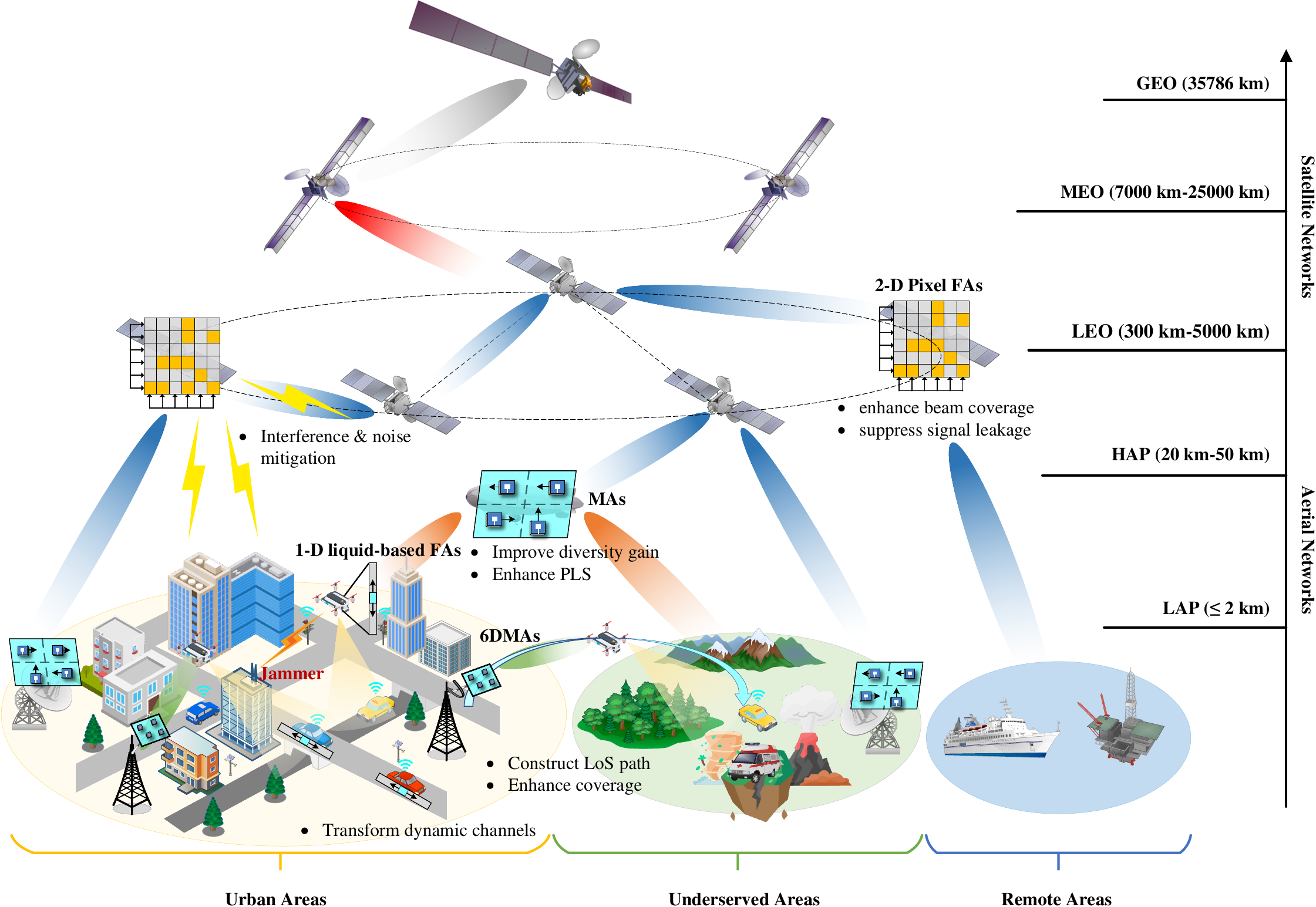}
    \caption{
    A potential FA-assisted NTN architecture. FAs integrated with different NTN platforms can provide global signal coverage and flexible resource allocations.}
    \label{fig:fa-ntn}
    \vspace{-2ex}
\end{figure*}

After years of research, Non-Terrestrial Networks (NTNs) are considered to be a cornerstone technology in 6G, which complement traditional TNs and help overcome the aforementioned implementation constraints \cite{2}. NTNs can establish multi-layer heterogeneous networks that offer ubiquitous connectivity and flexible deployment by leveraging the collaborative operations of Low Altitude Platforms (LAPs), High Altitude Platforms (HAPs), and satellite networks \cite{3}. Each type of NTN platform can function as an aerial base station, relay, or service terminal, forming an adaptive and scalable network framework. As illustrated in Fig. \ref{fig:fa-ntn}, NTNs serving as a complementary extension to TNs, can meet diverse requirements in dense urban areas, underserved areas, and remote areas \cite{4}. Firstly, the flexibility of NTNs offers specific user devices reliable and low-latency connectivity, which is evident in the case of moving platforms (e.g., car, train, airplane, etc.) and mission-critical communications. Secondly, in cooperation with NTNs, TNs can support a range of essential functionalities by equipping adequate communication, sensing, and computing capabilities, including positioning and navigation, environmental sensing, autonomous and real-time control, as well as collaborative computation offloading. Consequently, the integration of NTNs expands the performance boundaries and application scenarios of TNs, where conventional users' experiences can be transferred into interconnected, intelligent, and personalized paradigms. Furthermore, NTNs can guarantee global 6G service coverage, covering underserved areas or difficult to be covered by TNs (e.g., remote areas). Under this context, recognizing the transformative impact of NTNs on existing Fifth-Generation (5G) TNs is essential to unlock new revenue streams of the low-altitude economy within the broader 6G landscape.

\subsection{Limitations of Current NTNs}
Although NTNs exhibit potential in providing massive and ubiquitous connectivity for 6G, they still face several technical challenges \cite{3}, including dynamic propagation environment, the tradeoff between latency and dynamic network topology, dense interference, resource and energy constraints, to mention a few. 

\subsubsection{Dynamic Propagation Environment}
With the heterogeneity and dynamics of communication environments for NTN platforms, it is crucial to ensure the stability and security of communication under complex or extreme conditions. 
The first challenge is the severe path loss and fading due to varying path or orbital distances between TNs and NTN platforms at different altitudes.
The second challenge is the time-varying channels. The high mobility of nodes in NTNs can cause changing Doppler shifts and scattering environments, which aggravate data links and received signal strength  \cite{6}. Notably, the estimated Channel State Information (CSI) may become outdated due to node position changes, reducing algorithm performance \cite{7}.

\subsubsection{Tradeoff between Latency and Dynamic Network}
The tradeoff between low latency and dynamic network topology also complicates NTN sidelinks \cite{4}. 6G communication networks aim to reduce end-to-end latency to the microsecond level to support time-sensitive applications such as autonomous driving. Compared to the traditional static topology of TNs, the communication environments, devices, and tasks across platforms in NTNs vary frequently over time. When the network topology changes, such as satellites relocating to new locations or LAPs altering their cruise routes, the channel conditions may change rapidly. To find an optimal signal propagation scheme, the network must recalculate the route, which inevitably induces additional latency. To address such tradeoffs, researchers focus on exploring advanced air-interface technologies and optimizing deployment strategies in NTNs. Furthermore, precise positioning and optimal navigation are crucial challenges, particularly for LAPs, which are urgent for the real-time adaptability.

\subsubsection{Dense Interference}
NTNs suffer from co-channel and adjacent-channel interference due to the scarcity of frequency spectrum resources \cite{3}. Interference sources mainly include neighboring cellular cells, nearby NTN platforms, and multi-domain communication links, forming a complex three-Dimensional (3-D) interference environment. Specifically, in a cellular network, NTNs can provide stronger Line-of-Sight (LoS) channels to improve the reception of signals from neighboring cells but this results in inter-cell interference. In Unmanned Aerial Vehicle (UAV) clusters, deploying multiple UAVs in the same or adjacent airspace introduces additional interference and aggravates sum-rate fluctuation.
Accordingly, NTNs require advanced physical-layer techniques that enhance diversity gains to optimize throughput and transmission efficiency, as well as effective mitigation methods to cancel co-channel interference, environmental noise, and external attacks, thereby improving the Signal-to-Interference-plus-Noise Ratio (SINR).

\subsubsection{Resource and Energy Constraints}
Spectrum resource and energy efficiency constraints are critical bottlenecks in NTNs as well. Limited spectrum resource directly leads to transmission interference \cite{5}, necessitating advanced intelligent spectrum sensing and multiplexing to enhance utilization efficiency. Concurrently, NTN platforms face computation, sensing, and storage constraints due to physical size and power constraints, where resource orchestration is essential to support multi-task coordination. Existing research has been centered on lightweighting the system, such as by integrating functional modules and optimizing algorithms to reduce hardware demands. Meanwhile, with the rapid rise of Artificial Intelligence (AI) techniques, NTNs are increasingly combined with large models and edge computing to facilitate intelligence \cite{8}. Therefore, how to process model training and inference efficiently on resource-constrained aerial platforms is a promising topic.

\subsection{Motivations of FA-Assisted NTNs}
The Fixed-Position Antennas (FPAs) struggle to address the challenges in precedent subsection due to the limited Degree of Freedom (DoF), while advanced antenna techniques like massive Multiple Input Multiple Output (MIMO) are too complex and costly. \revise{Fluid Antenna (FA), presenting a promising physical-layer technology for NTNs, has attracted significant attention in academia \cite{23, 24, 25}. FAs broadly refer to any fluid, conductive, or dielectric structure that can be controlled by software for dynamic reconfiguration of their position, rotation, size, shape, and radiation characteristics based on real-time channel conditions \cite{9, 52-nn, 53}. The positions where the radiating elements of antennas can move are known as ports. Over the past years, many studies have investigated how to exploit the reconfigurability of FAs in dynamic NTNs \cite{94,17,18, 100, 18-n,19-n,104,16,103,20-n,80,21-n,22-n,102,23-n}. In the following, we provide the key insights of the performance gains brought by FAs in several representative NTN application scenarios.}

\subsubsection{Advantages in Urban Areas}
Although NTN platforms usually provide a direct LoS link, rich multipath remains prevalent in urban areas because surrounding reflectors (e.g., buildings, vegetation, and moving vehicles) induce wavelength-scale fluctuations in the channel response. In this context, FA-assisted NTNs exploit extra spatial DoFs and perform port selection to attain more favorable channel response, harvesting spatial diversity without enlarging the physical aperture. Meanwhile, urban networks feature high user density and pronounced heterogeneity in user locations (indoor/outdoor, varying elevations) and mobility. By reconfiguring positions and rotation angles, FAs can reshape the effective steering vectors to suppress interference while preserving the array and spatial multiplexing gains \cite{94,17}. This is beneficial for cellular-connected UAV links, where LoS propagation amplifies interference from non-associated base stations \cite{18}. For data collection in aerial Internet of Things (IoT) networks, FA-enabled UAVs can further co-design trajectories and reconfiguration to improve beam alignment and channel gains, reducing collection latency or transmit energy under reliable connectivity \cite{100, 18-n}. Such reconfigurability is also attractive for self-interference mitigation in ISAC \cite{19-n} and Full-Duplex (FD) systems \cite{104}, as it can alter coupling characteristics and enhance spatial resolution via beam adaptation. Furthermore, FAs give rise to the concept of Fluid Antenna Multiple Access (FAMA) \cite{16, 103}, which can support simultaneous connections for hundreds, even thousands of users in LOS/scattering environments, thereby improving spectrum utilization efficiency.

\subsubsection{Advantages in Underserved Areas}
In underserved areas (e.g., post-disaster or rural areas), local infrastructure and payload resources are often inadequate to deliver reliable broadband service. In such cases, NTN platforms can form flexible relaying networks, for example, deploying UAVs as aerial relays to establish long-distance links between users and sparsely distributed gateways \cite{20-n}. However, UAV motion, platform jitter, and attitude variations frequently induce rapid channel fluctuations and beam misalignment. FA-assisted UAVs can alleviate these effects by reconfiguring their position and orientation, thereby stabilizing the effective channel and maintaining robust beam alignment \cite{80, 21-n}. Cell-free UAV networking is another promising paradigm for such regions. Here, FA-enabled local port/rotation adjustments support agile beam steering and interference suppression, improving spectrum sharing while reducing fronthaul demands and the processing burden of centralized coordination \cite{22-n}. For direct user links to HAPs or satellites, FAs can further reshape the effective steering vectors and beamforming patterns to enhance alignment with intended users and limit energy leakage toward interference-prone directions \cite{102}. This improves coverage consistency and link budget under constrained spectrum and dense co-channel interference.

\subsubsection{Advantages in Remote Areas}
In remote areas, deploying terrestrial base stations is costly and often infeasible. Satellite networks are expected to provide essential connectivity in coordination with the ground gateways. Satellite companies such as Starlink, Telesat, and Amazon are adopting the FA structures to change the position of the beam quickly and flexibly. As thousands of satellites operate across multiple orbital planes, a single gateway’s LoS region may be simultaneously covered by multiple satellites, which can introduce severe inter-satellite interference and limit the overall network capacity. Deploying FAs at satellites and gateways enables high-gain directional reception/transmission toward the intended link while placing deep nulls toward dominant interferers, thereby improving interference mitigation and increasing network capacity \cite{102, 23-n}. Moreover, severe Doppler dynamics and long propagation delays in satellite networks accelerate CSI aging and increase pilot/feedback requirements. FAs can exploit port selection and rotation to transform the effective channels close to static channels \cite{80}, thus reducing pilot overhead and CSI estimation latency in highly dynamic satellite environments.

{\revise
\subsection{Comparison of FAs with Other Antenna Technologies}
With the rapid emergence of cutting-edge antenna technologies for 6G, this subsection aims to highlight the key similarities and differences between FAs and some related technologies.

\subsubsection{MIMO and Antenna Selection}
Over the past few decades, MIMO has improved the capacity and spectral efficiency of wireless systems through space-time coding and multiuser-MIMO beamforming. Massive MIMO represents the advanced MIMO paradigm in current 5G networks for higher spatial resolution \cite{12-nn, 13-nn}. However, scaling up the number of antennas also brings significant challenges in terms of hardware cost, energy consumption, and signal processing complexity. To balance the performance and the overhead, antenna selection has been widely investigated, which activates only a subset of high-performance antennas within a MIMO array. Nevertheless, existing MIMO and antenna selection schemes rely on conventional FPAs, where radiating elements are separated by the half-wavelength and cannot be repositioned after deployment.

In contrast to MIMO or antenna selection based on FPAs, FAs can utilize the entire 3-D channel variation within a designated space through position and rotation adjustments, even in spaces smaller than half a wavelength. This flexibility offers additional spatial DoFs, higher diversity and channel gains \cite{11, 12}. Additionally, in multiuser scenarios, MIMO or antenna selection cannot adjust the coherence between the steering vectors of users after FPA deployment. FAs can reconfigure the steering vectors to achieve the tradeoff between signal maximization and interference mitigation \cite{14-n}. Due to the spectrum scarcity, more advanced MIMO architectures such as extremely Large-scale (XL)-MIMO have been explored \cite{14-nn}. The core idea is to deploy an extremely large number of antennas in a compact space in which the antenna aperture could be discrete or continuous. From the perspective of system design, FAs can be viewed as an alternative implementation of compact XL-MIMO with antenna selection, which reduces hardware costs and power consumption while maintaining the same efficiency \cite{14}. 

\subsubsection{Movable Antenna}
The concept of Movable Antenna (MA) was introduced and rigorously studied for wireless communication in 2022 \cite{52,52-nn}. The basic idea is to adjust antenna positions within a spatial region to improve channel conditions and communication performance, with motor-driven mechanical movement being one practical implementation. MAs exploit additional spatial DoFs through position reconfiguration for signal enhancement. Furthermore, by incorporating 3-D rotation, the MA can be extended to the six-Dimensional Movable Antenna (6DMA) \cite{53}. Since early studies on FAs and MAs mainly focus on position reconfiguration, the two paradigms often share similar mathematical models. Consequently, the optimization methods and analytical results can be transferable to jointly promote the development of flexible antenna technologies for 6G \cite{10}. More recently, the concepts of FAs and MAs have gradually been adopted within their respective frameworks: the MA can be viewed as a solid-based FA \cite{23}, whereas the FA can be regarded as a liquid-based MA \cite{24}. Accordingly, unless otherwise specified, this paper treats MA as a solid-based form of FAs for detailed analysis.

Although FAs and MAs are becoming similar in a broader concept, their characteristics should still be distinguished according to the specific hardware designs. For example, pixel-based FAs rely on a large number of densely distributed discrete ports for observation and activation, which makes mutual coupling and circuit effects important in FAs even when some pixels are inactive. By contrast, MAs typically assume continuous antenna movement, i.e., equivalently infinite candidate ports, while maintaining the half-wavelength spacing to avoid mutual coupling. Moreover, since MA more emphasizes physical movement, the mechanical motion and rotation control also require in-depth discussion. Notably, when antenna shape, size, or other radiation characteristics become reconfigurable \cite{10-nn, 11-nn}, further distinctions between FA and MA should be analyzed according to the specific implementation.

\subsubsection{Rotatable Antenna}
This is an emerging flexible antenna technology that reconfigures antenna orientation/boresight without changing the physical antenna position. By reshaping the spatial distribution of radiated energy, Rotatable Antennas (RAs) introduce additional DoFs beyond conventional FPA beamforming for more accurate beam focusing, more effective interference avoidance, and more flexible 3-D coverage adaptation \cite{15-nn}. RA implementation mainly includes two types \cite{16-nn}. The first is mechanically driven RA, inspired by MA/6DMA, which directly adjusts antenna orientation via micromotors or gimbals. It offers a wide rotation range and high angular resolution, but suffers from actuation latency, mechanical wear, and high control complexity. The second is electronically driven RA, which is closely related to pixel-based FAs, reconfigurable feeding networks, and reactively controlled directive arrays. By reconfiguring the feed networks for current redistribution without rotating the antenna, this structure enables faster response and higher integration. However, the accuracy and pattern consistency of beamforming are limited.

With the continuous evolution of flexible antenna technologies, RAs and FAs have become conceptually related in several respects. In particular, FAs can achieve RA directional reconfiguration by incorporating rotational capability. Meanwhile, advances in pixel-based FAs also enable electronically driven RAs to achieve more flexible beam steering and higher scanning resolution. In this paper, the FA is treated as a broader, hardware-agnostic system concept that emphasizes the reconfigurability of radiating elements. Accordingly, some rotation-related characteristics of RAs are also discussed in the modeling and optimization of FAs.}

\begin{table*}[htbp!]
\revise{
    \caption{The Summary of Existing Surveys Related to FA-Assisted NTNs}
    \label{tab:my_label}}
    \centering
    \resizebox{\textwidth}{!}{
    \begin{tabular}{p{1cm}<{\centering}|p{15cm}}
        \hline
        \multicolumn{1}{c|}{\textbf{Ref.}} & \multicolumn{1}{c}{\textbf{Main Contributions}}\\
         \hline
         \hline
         \multirow{5}{*}{\cite{3}} & \textbf{Summary:} This work reviewed the evolution trajectory of developments of NTNs from 5G to 6G, providing an analysis of the integration between NTNs and TNs to deliver global coverage, high capacity, and low-latency services.\\ 
         & \textbf{Related Contributions to FA-Assisted NTNs:}\\
         & a) It offered comprehensive insights into the design and deployment of NTNs within 6G networks.\\
         & b) It discussed the limitations of FPA-MIMO in NTNs and underscored the potential of flexible-antenna technologies.\\
         \hline
         \multirow{6}{*}{\cite{19}} & \textbf{Summary:} This work provided a systematic introduction to MIMO satellite communications, summarizing typical MIMO satellite scenarios and reviewing the MIMO beamforming, modulations, and channel estimations to address Doppler shifts, propagation delays, and the overhead of MIMO arrays in satellite networks. \\
         & \textbf{Related Contributions to FA-Assisted NTNs:}\\
         & a) It reviewed the channel modeling, estimation, and resource optimization methods for MIMO-satellite systems.\\
         & b) It outlined the potential research opportunities in FA-assisted satellite networks. \\
         \hline
         \multirow{6}{*}{\cite{20}} & \textbf{Summary:} This work delved into the multi-satellite MIMO systems, covering channel modeling, frequency bands, and antenna designs, reviewing the collaborative paradigms under severe Doppler, long delays, and inter-satellite interference.\\
         & \textbf{Related Contributions to FA-Assisted NTNs:}\\
         & a) It discussed channel modeling of satellite-terrestrial links and inter-satellite links\\
         & b) It presented a summary of satellite antenna designs.\\
         & c) It provided a systematic framework and guidance for FAs incorporating in 6G NTN networks.\\
         \hline
         \multirow{4}{*}{\cite{21}} & \textbf{Summary:} This work leveraged AI to explore optimal channel estimation, port selection, and joint optimization of antenna position and beamforming for FAs, further examining the superiority of AI-assisted FA-assisted ISAC systems.\\
         & \textbf{Related Contributions to FA-Assisted NTNs:}\\
         & a) It explored the potential of AI for optimization in FA-assisted UAV networks.\\
         \hline
         \multirow{5}{*}{\cite{22}} & \textbf{Summary:} This work incorporated RIS, FAs, and UAVs to adjust the dynamic characteristics of propagation channels in ISAC systems, highlighting their potential to achieve less system overhead, predictive resource allocation, and anti-blocking capability.\\
         & \textbf{Related Contributions to FA-Assisted NTNs:}\\
         & a) It presented the cooperative applications of FAs and RIS in NTNs.\\
         & b) It proposed an advanced ISAC collaborative intelligent communication environment framework of FA-RIS-assisted NTNs.\\
         \hline
         \multirow{5}{*}{\cite{23}} & \textbf{Summary:} This work conducted a comprehensive study of FAs, including all current fundamental paradigms of FAs and channel modeling, including advanced information theory, optimization techniques, hardware designs, and future directions.\\
         & \textbf{Related Contributions to FA-Assisted NTNs:}\\
         & a) It significantly contributed to the maturation and evolution of FA technology frameworks.\\
         & b) It emphasized the promising application scenarios of FAs in NTNs.\\
         \hline
         \multirow{5}{*}{\cite{24}} & \textbf{Summary:} This work highlighted substantial benefits of MAs in communication, sensing, and computation applications, developed various system models for MAs that further enrich the breadth and depth of the FA concept.\\
         & \textbf{Related Contributions to FA-Assisted NTNs:}\\
         & a) It introduced a mechanical implementation of FAs and comprehensively reviewed its fundamentals and advancements.\\
         & b) It offered an innovative design paradigm for the future applications of FA-assisted NTNs.\\
         \hline
        \multirow{5}{*}{\cite{25}} &  \textbf{Summary:} This work conducted a systematic overview of the 6DMA technology from the perspective of its evolution in 6G, the promising architectures, and critical application scenarios.\\
        & \textbf{Related Contributions to FA-Assisted NTNs:}\\
        & a) It investigated system models, optimizations, and applications of FAs with jointly reconfigurable position and rotation.\\
        & b) It emphasized the optimization for interference mitigations in 6DMA-assisted UAV networks.\\
         \hline
    \end{tabular}}
\end{table*}

\vspace{-2ex}
\revise{
\subsubsection{Pinching Antenna}
Pinching antenna (PA) system is a waveguide-based flexible antenna technology consisting of dielectric waveguides and separate dielectric PAs along them \cite{17-nn, 18-nn, 19-nn}. Signals first propagate through a low-loss waveguide before being radiated into free space at selected PA locations. Compared to FAs with wavelength-scale reposition, PAs can extend their waveguide length to tens of meters, which enables flexible activation of multiple PAs for transmission and reception over a wider area. This facilitates mitigating not only small-scale fading but also large-scale path loss by reshaping the propagation path. Specifically, PA systems can deploy the PAs closer to users' locations, in which the part of the propagation occurs inside the low-loss waveguide rather than entirely in free space. In this sense, PAs shift the wireless communication paradigm from the “last mile” to the “last meter” and offer a promising solution for establishing near-field channels \cite{18-nn}. In NTNs, PAs are more suitable as ground-side auxiliary infrastructure to enhance access, backhaul, and coverage capabilities \cite{19-nn}.}

\begin{table*}[htbp!]
\caption{
Table with Comparisons between Existing Surveys and Ours (Un-Covered: \textcolor{red}{\textbf{\XSolidBrush}}, Partially Covered:\textcolor{blue}{\textbf{\dag}}, Covered: \textcolor{green}{\Checkmark})}
\label{tab:compare}
\centering
\renewcommand{\arraystretch}{2}
\resizebox{\textwidth}{!}{
\begin{tabular}
{>{\centering\arraybackslash}m{1.5cm}|>{\centering\arraybackslash}m{2cm}|>{\centering\arraybackslash}m{2cm}|>{\centering\arraybackslash}m{2cm}|>{\centering\arraybackslash}m{2cm}|>{\centering\arraybackslash}m{2cm}|>{\centering\arraybackslash}m{2cm}|>{\centering\arraybackslash}m{2cm}|>{\centering\arraybackslash}m{2cm}|>{\centering\arraybackslash}m{2cm}|>{\centering\arraybackslash}m{2cm}}
\hline
\textbf{Ref.} & \textbf{Potentials of FA-assisted NTNs} & \textbf{FA selection in NTNs} & \textbf{FA channel modeling} & \textbf{NTN channel modeling} & \textbf{FA-assisted NTN channel modeling} & \textbf{channel estimation} & \textbf{Joint optimization of FAs \& NTNs} & \textbf{Compatibility with B5G technologies} & \textbf{Intelligent function integration} & \textbf{Security}\\
\hline
\hline
\cite{3} & \textcolor{green}{\Checkmark} & \textcolor{red}{\XSolidBrush} & \textcolor{red}{\XSolidBrush} & \textcolor{green}{\Checkmark} & \textcolor{red}{\XSolidBrush} & \textcolor{red}{\XSolidBrush} & \textcolor{red}{\XSolidBrush} & \textcolor{blue} {\textbf{\dag}} & \textcolor{blue} {\textbf{\dag}} & \textcolor{blue} {\textbf{\dag}} \\ \hline
\cite{19} & \textcolor{green}{\Checkmark} & \textcolor{red}{\XSolidBrush} & \textcolor{red}{\XSolidBrush} & \textcolor{green}{\Checkmark} & \textcolor{red}{\XSolidBrush} &  \textcolor{blue} {\textbf{\dag}} & \textcolor{blue} {\textbf{\dag}} & \textcolor{red}{\XSolidBrush} & \textcolor{red}{\XSolidBrush} & \textcolor{blue} {\textbf{\dag}} \\ \hline
\cite{20} & \textcolor{green}{\Checkmark} & \textcolor{blue} {\textbf{\dag}} & \textcolor{red}{\XSolidBrush} & \textcolor{green}{\Checkmark} & \textcolor{red}{\XSolidBrush} &  \textcolor{blue} {\textbf{\dag}} & \textcolor{blue} {\textbf{\dag}} & \textcolor{red}{\XSolidBrush} & \textcolor{red}{\XSolidBrush} & \textcolor{blue} {\textbf{\dag}} \\ \hline
\cite{21} & \textcolor{green}{\Checkmark} & \textcolor{red}{\XSolidBrush} & \textcolor{green}{\Checkmark} & \textcolor{blue} {\textbf{\dag}} & \textcolor{blue} {\textbf{\dag}} &  \textcolor{red}{\XSolidBrush} & \textcolor{blue} {\textbf{\dag}} & \textcolor{red}{\XSolidBrush} & \textcolor{red}{\XSolidBrush} & \textcolor{blue} {\textbf{\dag}} \\ \hline
\cite{22} & \textcolor{green}{\Checkmark} & \textcolor{red}{\XSolidBrush} & \textcolor{green}{\Checkmark} & \textcolor{blue} {\textbf{\dag}}& \textcolor{blue} {\textbf{\dag}} &  \textcolor{red}{\XSolidBrush} & \textcolor{blue} {\textbf{\dag}} & \textcolor{red}{\XSolidBrush} & \textcolor{green}{\Checkmark} & \textcolor{blue} {\textbf{\dag}} \\ \hline
\cite{23} & \textcolor{green}{\Checkmark} &  \textcolor{blue} {\textbf{\dag}} & \textcolor{green}{\Checkmark} & \textcolor{red}{\XSolidBrush}& \textcolor{red}{\XSolidBrush} &  \textcolor{blue} {\textbf{\dag}} & \textcolor{blue} {\textbf{\dag}} & \textcolor{blue} {\textbf{\dag}} & \textcolor{blue} {\textbf{\dag}} & \textcolor{blue} {\textbf{\dag}} \\ \hline
\cite{24} & \textcolor{green}{\Checkmark} &  \textcolor{blue} {\textbf{\dag}} & \textcolor{green}{\Checkmark} & \textcolor{red}{\XSolidBrush}& \textcolor{red}{\XSolidBrush} &  \textcolor{blue} {\textbf{\dag}} & \textcolor{blue} {\textbf{\dag}} & \textcolor{blue} {\textbf{\dag}} & \textcolor{blue} {\textbf{\dag}} & \textcolor{blue} {\textbf{\dag}} \\ \hline
\cite{25} & \textcolor{green}{\Checkmark} &  \textcolor{blue} {\textbf{\dag}} & \textcolor{green}{\Checkmark} & \textcolor{blue} {\textbf{\dag}}& \textcolor{blue} {\textbf{\dag}} &  \textcolor{blue} {\textbf{\dag}} & \textcolor{blue} {\textbf{\dag}} & \textcolor{blue} {\textbf{\dag}} & \textcolor{blue} {\textbf{\dag}} & \textcolor{blue} {\textbf{\dag}} \\ \hline
Ours & \textcolor{green}{\Checkmark} &  \textcolor{green}{\Checkmark} & \textcolor{green}{\Checkmark} & \textcolor{green}{\Checkmark}& \textcolor{green}{\Checkmark} &  \textcolor{green}{\Checkmark} & \textcolor{green}{\Checkmark} & \textcolor{green}{\Checkmark} & \textcolor{green}{\Checkmark} & \textcolor{green}{\Checkmark} \\ \hline
\end{tabular}}
\end{table*}

\subsection{Comparison with Other Related Surveys}
Recent efforts started to lay the crucial technical foundation for FA-assisted NTNs.
In \cite{3}, Azari et al. systematically investigated the cooperation of diverse NTN platforms (e.g., LAPs, HAPs, and satellites) and TNs from 5G to 6G. 
The work in \cite{3} demonstrated the urgent need for advanced antenna technologies to empower massive communication and ubiquitous connectivity.
In \cite{19}, Heo et al. provided a systematic introduction to MIMO satellite networks, encompassing channel modeling, channel estimation, beamforming designs, and collaboration with other communication technologies. 
In \cite{20}, Bakhsh et al. focused on the multi-satellite MIMO system, covering channel modeling, frequency bands, satellite antenna designs, and cooperative paradigms to address severe Doppler shifts and propagation delays.
These works lay a solid foundation for channel modeling and hardware design of FAs in NTNs. To optimize the performance of FAs in NTNs, Wang et al. in \cite{21} explored the relationships and applications between AI and FAs. They applied the advanced AI techniques (e.g. Deep Learning (DL), Reinforcement Learning (RL), Deep Reinforcement Learning (DRL), and AI large models) to solve optimization problems in FA applications.
In \cite{22}, Meng et al. incorporated smart propagation engineering, such as FAs, RIS, and NTNs, to address challenges in ISAC systems, including overhead control, resource allocation, and channel blocking. 
To introduce FAs from a comprehensive perspective, New et al. in \cite{23} proposed a complete set of methodologies for FAs, including FA channels, performance analysis, hardware designs, and future application scenarios. 
In \cite{24}, Zhu et al. introduced MAs to highlight the movement and rotation characteristics of FAs. They expanded the field-response channel models and optimizations of MAs in narrowband and wideband systems, as well as far-field and near-field propagation scenarios.
Further considering the 3-D positions and 3-D rotations of MAs, Shao et al. in \cite{25} introduced a 6DMA technology to analyze the architectures, antenna position and rotation optimization, channel estimation, and applications. \revise{The authors in \cite{23,24,25} comprehensively discussed the general channel models, optimization methods, and hardware designs of FAs, laying a technical foundation for FA-assisted NTNs.} Although they all indicated the prospects of integrating FAs into NTNs, they did not conduct an in-depth discussion on the core issues and challenges faced by FA-assisted NTNs, as well as the various applications that arise from the combination. With the exponential growth of architectures and algorithms in FA-assisted NTNs, a comprehensive review of the basic theories and key technologies is necessary. In Table \ref{tab:my_label} and Table \ref{tab:compare}, we provide a summary of existing surveys related to FA-assisted NTNs and a comparison of our survey with existing surveys.

\subsection{Contributions and Structure of This Survey}
As previously mentioned, FA-assisted NTNs are leading to a profound revolution in communication networks for 6G and beyond. These breakthroughs underscore the substantial potential of FAs to enhance the effectiveness, reliability, security, and adaptability of current NTN frameworks. As research on FAs progresses, scholars continually push forward the performance boundaries of NTNs and the 6G ecosystem. FA-assisted NTNs offer diverse technological pathways, profoundly influencing various aspects of communication systems, including massive and robust connectivity, elevated throughput, compatible architecture, and intelligent services. This evolution will enable transformative applications in 6G communication.

This survey aims to provide a comprehensive perspective that exceeds the scope of existing review literature, offering a thorough overview of the technological advancements and applications towards FA-assisted NTNs across multiple performance indicators. In addition, we highlight the future directions that deserve further exploration. The contributions of this survey are summarized as follows:

\begin{itemize}

\item We present a broad overview of current NTN and FA technologies. In particular, the structure and functions of core components in NTNs are analyzed first. Then, we systematically introduce the FAs, covering fundamental principles, hardware designs, spatial correlation channel models, and key performance advantages. 

\item We summarize the FA-assisted NTN channel models to gain a deeper understanding of the advantages of FA-assisted NTNs and highlight the necessary introduction of FAs. Additionally, we comprehensively investigate advanced CSI estimation methods from the perspectives of model-based and AI-based methods.

\item We review the joint optimization of FAs and NTNs in communication systems. Specifically, we introduce three FA-assisted NTN scenarios and State-of-the-Art (SoTA) optimization algorithms in detail, summarize the latest literature, and inspire researchers to exploit the full benefits of FAs.

\item We further discuss the compatibility of FA-assisted NTNs with other emerging Beyond-5G (B5G) technologies, including Cell-Free massive MIMO (CF-mMIMO), FD communication, Next Generation Multiple Access (NGMA), and RIS.

\item We systematically summarize the innovative FA-assisted NTN architectures for integration of intelligent functions, such as integrated communication-computation systems and ISAC systems.

\item We analyze the security in FA-assisted NTNs from the perspectives of Physical Layer Security (PLS) and covert communication. The system challenges and solution in PLS and how to utilize FAs for covert communication are discussed. 

\item We outline future directions and new research opportunities in FA-assisted NTNs, including AI applications, flexible antenna technologies, high-frequency communication, near-field communication, as well as Integrated Sensing, Communication, and Computation (ISCC).
\end{itemize}

\begin{figure*}[!ht]
\revise{
    \centering
    \includegraphics[width = 1\textwidth]{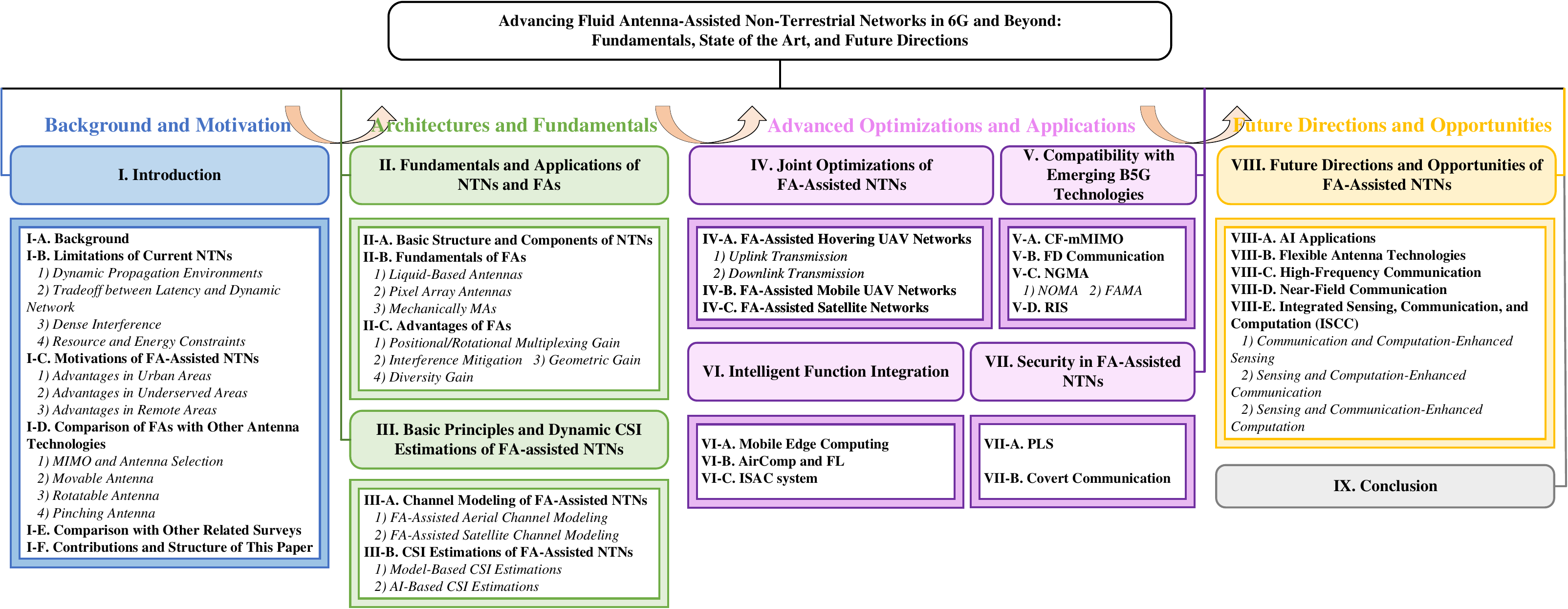}
    \caption{Organization of this survey.}
    \label{fig:content}}
    \vspace{-2ex}
\end{figure*}

The organization of this survey is shown in Fig. \ref{fig:content}. In Section \ref{fund}, we first introduce the basic concepts of NTNs and FAs. Then, Section \ref{csi} summarizes the channel model and CSI estimation methods of FA-assisted NTNs. Section \ref{advance} reviews the joint optimizations of FA-assisted NTNs, detailing the advancing algorithms in the latest literature and the performance indicators. Section \ref{b5g} investigates the effectiveness of combining FA-assisted NTNs with other B5G technologies. Moreover, the works of intelligent-function integration of FA-assisted NTNs are discussed in Section \ref{integration}. In Section \ref{pls}, the security in FA-assisted NTNs is reviewed. We highlight the future directions and new opportunities for FA-assisted NTNs in Section \ref{challenge}. Finally, Section \ref{conclusion} concludes this survey. To facilitate the readers, Table \ref{tab:acronyms} summarizes the list of major acronyms used in this survey.
\begin{table*}[htbp!]
\caption{List of Major Acronyms}
\centering
\renewcommand{\arraystretch}{1.2}
\resizebox{\textwidth}{!}{
\begin{tabular}{p{1.5cm}<{\centering}|p{6.2cm}<{\centering}|p{1.5cm}<{\centering}|p{6.2cm}<{\centering}}
\hline
\textbf{Acronyms} & \textbf{Meaning} & \textbf{Acronyms} & \textbf{Meaning} \\
\hline
\hline
1-D & one-Dimensional & LLM & Large Language Model \\ \hline
2-D & two-Dimensional & LMMSE & Linear Minimum Mean-Square Error \\ \hline
3-D & three-Dimensional & LoS & Line-of-Sight \\ \hline
5G & fifth-Generation & LSTM & Long Short-Term Memory \\ \hline
6DMA & six-Dimensional Movable Antenna & MA & Movable Antenna \\ \hline
6G & sixth-Generation & MEO & Medium Earth Orbit \\ \hline
A2A & Air-to-Air & MIMO & Multiple Input Multiple Output \\ \hline
A2G & Air-to-Ground & MISO & Multiple Input Single Output \\ \hline
AI & Artificial Intelligence & ML & Machine Learning \\ \hline
AirComp & over-the-Air Computation & mmWave & millimeter-Wave \\ \hline
AO & Alternating Optimization & MSE & Mean Squared Error \\ \hline
AP & Access Point & NGMA & Next Generation Multiple Access \\ \hline
B5G & Beyond-5G & NLoS & Non-Line-of-Sight \\ \hline
CF-mMIMO & Cell-Free massive MIMO & NOMA & Non-Orthogonal Multiple-Access \\ \hline
CNN & Convolutional Neural Network & NTN & Non-Terrestrial Network \\ \hline
CPU & Central Processing Unit & OTFS & Orthogonal Time Frequency Space \\ \hline
CSI & Channel State Information & PA & Pinching Antenna \\ \hline
CUMA & Compact Ultra Massive Antenna Array & PLS & Physical Layer Security \\ \hline
DL & Deep Learning & PSO & Particle Swarm Optimization \\ \hline
DoF & Degree of Freedom & RA & Rotatable Antenna \\ \hline
DRL & Deep Reinforcement Learning & RF & Radio Frequency \\ \hline
FA & Fluid Antenna & RIS & Reconfigurable Intelligent Surface \\ \hline
FAMA & Fluid Antenna Multiple Access & RL & Reinforcement Learning \\ \hline
FD & Full-Duplex & RSMA & Rate-Splitting Multiple Access \\ \hline
f-FAMA & fast-FAMA & SCA & Successive Convex Approximation \\ \hline
FL & Federated Learning & s-FAMA & slow-FAMA \\ \hline
FPA & Fixed-position antenna & SIC & Serial Interference Cancellation \\ \hline
FPGA & Field-Programmable Gate Array & SIMO & Single Input Multiple Output \\ \hline
GAN & Generative Adversarial Network & SINR & Signal-to-Interference-plus-Noise Ratio \\ \hline
GEO & Geostationary Orbit & SNR & Signal-to-Noise Ratio \\ \hline
HAP & High Altitude Platform & SISO & Single Input Single Output \\ \hline
ISAC & Integrated Sensing And Communication & THz & Terahertz \\ \hline
ISCC & Integrated Sensing, Communication, and Computation & TN & Terrestrial Network \\ \hline
LAP & Low Altitude Platform & UAV & Unmanned Aerial Vehicle \\ \hline
LEO & Low Earth Orbit & XL-MIMO & extremely Large-scale MIMO \\ \hline
\end{tabular}}
\label{tab:acronyms}
\end{table*}
\section{Fundamentals and Applications of NTNS and FAs}\label{fund}
In this section, we first introduce a brief overview of the basic structure and core components of NTNs. After that, we provide the details of FAs, including their unique hardware structures, spatial-correlation channel models, and performance advantages.

\subsection{Basic Structure and Components of NTNs}
NTNs leverage non-terrestrial platforms to provide various services, including communication, sensing, and computing, targeting global signal coverage and connectivity \cite{26}. As defined by the 3rd Generation Partnership Project (3GPP) \cite{27, 28, 29}, NTN comprises two main segments: aerial and satellite networks, each with diverse platforms operating at varying altitudes.

Aerial networks mainly comprise LAPs and HAPs. LAPs (e.g., UAVs) typically deploy at altitude in the air ranges of 100 m to 2 km, leveraging LoS channels to establish direct connections \cite{30, 31}. With high flexibility and rapid deployment, they can address coverage gaps in traditional TNs. Additionally, LAPs can bridge TNs and satellite networks to enable seamless and reliable communication in underserved areas (e.g., post-disaster or rural areas). Beyond enhancing communication, LAPs can also serve as service terminals for large-scale data collection, environmental sensing, edge computing, and wireless power transfer \cite{32, 33, 34, 35}. 

In contrast, HAPs (e.g., airships), usually operate at altitude from 20 km to 50 km with superior endurance, substantial payload capacity, and broad coverage. These capabilities allow them to be deployed as Ad-hoc networks, providing coverage to specific areas to satisfy various demands. Moreover, HAPs can be an intermediate element between the satellite and the terrestrial receiver. This two step downlink communication will have a first hop between satellite and HAPs, and a second hop between HAPs and terrestrial networks. Furthermore, HAPs hold great potential as distributed data centers to monitor and record the orbital paths of satellites and calculate the probability of a collision. With the powerful payload capacity, HAPs can provide more intelligent service by integrating computing and storage resources to facilitate data processing and storage \cite{36}.

Satellite networks primarily consist of Geostationary Orbit (GEO), Medium Earth Orbit (MEO), and Low Earth Orbit (LEO) satellites, which enhance global coverage and provide massive throughput. GEO satellites are positioned at geostationary orbit, thus maintaining a constant position relative to the Earth's surface. This characteristic enables them to provide persistent coverage over extensive regions, which is beneficial for television broadcasting, telecommunications, and weather monitoring. MEO satellites offer lower orbit altitudes and latency than GEO satellites. They are commonly used for applications such as navigation systems. And LEO satellites orbit between 200 km and 2000 km above the Earth’s surface. This orbital configuration achieves lowest latency, path loss, and launch costs among various types of satellites, supporting various applications such as high-bandwidth internet access, global positioning, and navigation \cite{37, 38}. Therefore, compared to MEO and GEO, LEO satellites are more commonly used for interactions with NTNs and TNs. 

\begin{figure}[t]
    \centering
    \includegraphics[width=0.45\textwidth]{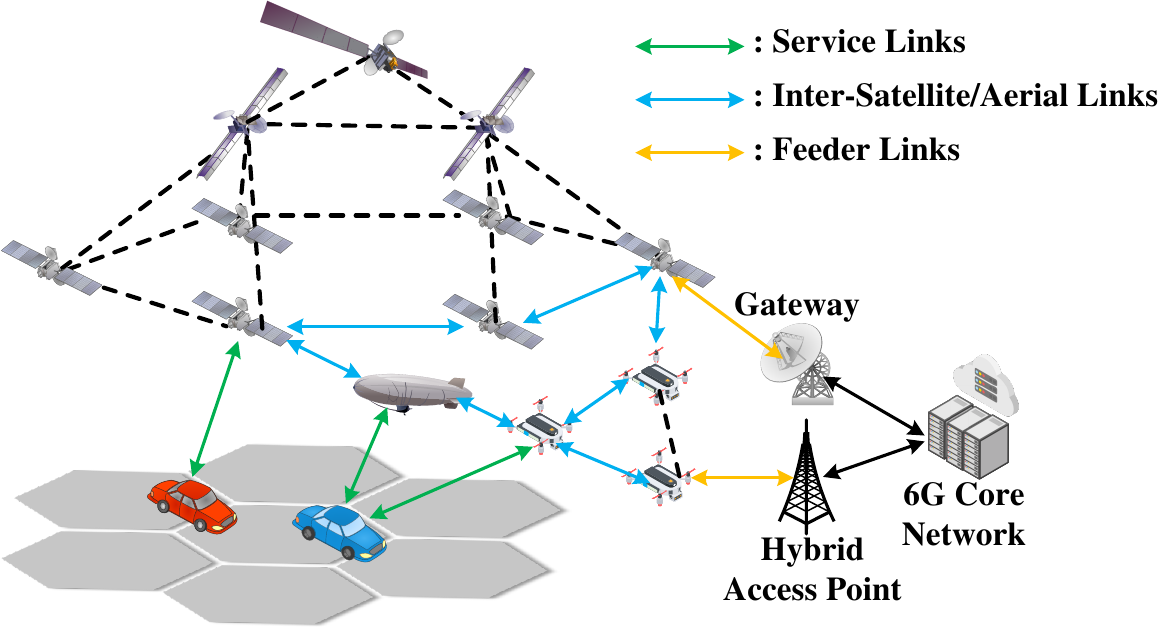}
    \caption{NTN accessing TN architecture. Mobile terminals transmit signals through service links to NTN platforms, which can relay signals via inter-satellite/inter-aerial links until they pass through feeder links to a terrestrial AP for TN core network integration.}
    \label{fig:NTN-TN}
    \vspace{-2ex}
\end{figure}

Fig. \ref{fig:NTN-TN} presents an NTN accessing TN architecture. Mobile terminals initiate requests through service links to NTN platforms, such as satellites, HAPs, or LAPs. These platforms can relay signals through inter-satellite/inter-aerial links across multiple nodes, until the signals pass through feeder links to a terrestrial Access Point (AP) for core network integration. To incorporate the respective advantages and characteristics of different types of satellites or platforms, NTNs combine all types of them and enhance the overall resilience performance of a communication system \cite{3}. If one component fails, others can sustain connectivity. In the following, the fundamental components of NTNs and their respective functions will be detailed.

\subsubsection{Aerial/Satellite Terminals} 
Aerial/satellite terminals can receive signals from NTNs and TNs to accomplish various tasks. According to latest researches as well as 3GPP standard, UAVs have been utilized to serve as new terminals for logistics transportation, data collection \cite{32}, environmental sensing \cite{33}, remote-controlled measurement, mobile edge computing \cite{34}, wireless power transfer \cite{35} and so on. HAPs have significant endurance and resource-carrying capabilities compared to UAVs. They are expected to function as mobile data centers to support edge computing, real-time data processing, and storage \cite{36}. Additionally, they can serve as centralized collection and charging stations for UAVs, further optimizing resource allocation. Satellites can lighten the burden on TNs by gathering and sending data directly to and from space when treated as users served by other satellite terminals \cite{39}.

\subsubsection{Aerial/Satellite Relays} 
Equipped with transparent payloads, aerial/satellite relays contribute to signal enhancement and coverage extension, where radio frequency filtering, frequency conversion, and signal amplification are primary functions. This architecture can be used to link the base stations and the core networks, providing backhauling services \cite{40}. Deploying aerial relays can expand the coverage of TNs, thereby facilitating the realization of the Space-Air-Ground Integrated Networks (SAGINs) \cite{41}.

\subsubsection{Aerial/Satellite Base Stations} 
Equipped with regenerative payloads, aerial/satellite base stations perform functions like filtering, frequency conversion, amplification, encoding/decoding, modulation/demodulation, switching, and routing. Compared to terrestrial base stations, aerial base stations offer broader coverage areas, stronger LoS channels, and more flexible deployments \cite{42}. These capabilities enable temporary and reliable communication services to access remote areas or emergency communication support in disaster-stricken regions \cite{43}. In the context of aerial/satellite networks, aerial base stations can also serve as the command and control center for the relay network. It can also integrate with other network management systems to optimize the overall network performance.
 
\subsubsection{Radio Links} 
Radio links enable wireless signal communication between terrestrial users and NTN platforms. The radio links between NTNs and TNs involve several types, such as service links between users and satellite/aerial networks, feeder links between satellite/aerial networks and terrestrial gateways, inter-satellite/aerial links, and inter-platform links connecting satellite and aerial networks \cite{27}. Among these, the first two are classified as Air-to-Ground (A2G) channels, while the remainder fall into Air-to-Air (A2A) channels. The CSI of these service links directly impacts the performance of NTNs. Factors such as the Doppler effect, path loss, and atmospheric attenuation can lead to signal distortion or interruption.
 
\subsubsection{Gateways and Terrestrial Core Networks} 
As the bridge between TNs and NTNs, gateways are responsible for transmitting and receiving wireless signals from NTN nodes, converting these signals according to the specified protocols. After that, the converted signals are routed to the terrestrial core networks, which transmit and process ground-based signals.

\begin{table*}[htbp!]
\caption{Characteristics and Suitable NTN platforms of Different Types of FAs}
\label{tab:FA-class}
\centering
\resizebox{\textwidth}{!}{
\begin{tabular}{
p{2.8cm}|p{6cm}|p{5.1cm}|p{5.1cm}|p{2.4cm}
}
\hline
 \multicolumn{1}{c|}{\textbf{Types of FAs}} & \multicolumn{1}{c|}{\textbf{Features}} & \multicolumn{1}{c|}{\textbf{Advantages}} & \multicolumn{1}{c|}{\textbf{Disadvantages}}  & \multicolumn{1}{c}{\textbf{Suitable NTN Platforms}} \\

\hline
\hline

\makecell{Liquid-based FA \\ \cite{44}}
&
\makecell[l]{
A feeding network delivers/receives the
RF signal \\ to/from the fluidic radiating element contained in \\the reservoir.
}
&
\makecell[l]{
\textbullet\ High switching speeds;\\
\textbullet\ Simple mechanism actuation requiring \\(e.g., pump/valves);\\
\textbullet\ Support multi-dimensional reconfiguration \\(e.g., size/shape) in some designs.
}
&
\makecell[l]{
\textbullet\ Need space for liquid reservoir; \\
\textbullet\ Leakage risk; \\
\textbullet\ Temperature/material stability issues;\\
\textbullet\ Complex dynamic attitude and acceleration \\control.
}
&
\makecell{
Terrestrial users \& \\base stations, LAPs
}
\\
\hline

\makecell{Pixel Array Antenna \\ \cite{48, 49, 50}}
&
\makecell[l]{
This antenna consists of several static radiating \\
elements. By toggling switches on/off, the section \\utilized
for radiating can be controlled. Increasing\\
the number of switches in the system will increase\\ 
the range of different configurations available to the \\antenna.
}
&
\makecell[l]{
\textbullet\ Extremely fast switching;\\
\textbullet\ Better reliability potential of electronic \\control;\\
\textbullet\ Low maintenance cost of non-mechanical \\motion;\\
\textbullet\ Suitable for large-scale array deployment.
}
&
\makecell[l]{
\textbullet\ Restricted by number of switches;\\
\textbullet\ Large portions of the antenna are unused;\\
\textbullet\ Mutual coupling and Impedance matching \\need circuit-antenna co-design.
}&
\makecell{
Terrestrial users \& \\base stations, LAPs, \\HAPs, satellites
}
\\
\hline

\makecell{Movable Antenna \\ \cite{52, 53}}
&
\makecell[l]{
The CPU sends a control signal to two motors to \\
control the MA's location horizontally and vertically.\\
Accordingly, the MA adjust its position in a 2-D \\space. The received signal is then passed through \\RF chains before being fed back to the CPU for\\ processing.
}
&
\makecell[l]{
\textbullet\ Larger position search space; \\
\textbullet\ Greater spatial diversity potential; \\
\textbullet\ Support rotational reconfigurability; \\ 
\textbullet\ Can reuse conventional antenna and RF-\\chain designs.
}
&
\makecell[l]{
\textbullet\ Significant mechanical latency;\\
\textbullet\ Increased hardware size, weight, maintena-\\ nce cost, and energy consumption;\\
\textbullet\ Mechanical wear and vibration sensitivity.
}
&
\makecell{
Terrestrial users \& \\base stations, LAPs, \\HAPs
}
\\
\hline
\end{tabular}}
\end{table*}

\subsection{Fundamentals of FAs}
\begin{figure}[t]
    \centering
    \includegraphics[width=0.4\textwidth]{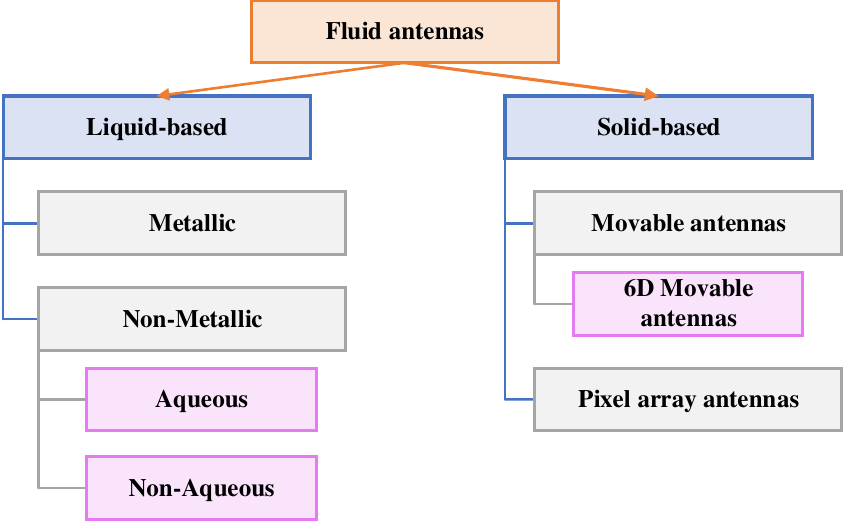}
    \caption{Classification of different types of FAs.}
    \label{fig:FA-class}
    \vspace{-2ex}
\end{figure}

FA is a flexible antenna concept that allows the antenna to be reconfigured to ensure optimal channel conditions for wireless communications. FAs include any movable or non-movable flexible antennas, such as fluid-based liquid antennas, Radio Frequency (RF) pixel array antennas, and mechanically MAs, which can adjust their positions and radiation characteristics in various ways. The different forms of FAs are summarized in Fig. \ref{fig:FA-class} and Table \ref{tab:FA-class} describes the features, advantages, disadvantages, and suitable NTN platforms.
-

\subsubsection{Liquid-Based Antennas}
\begin{figure}[t]
    \centering
    \includegraphics[width=0.4\textwidth]{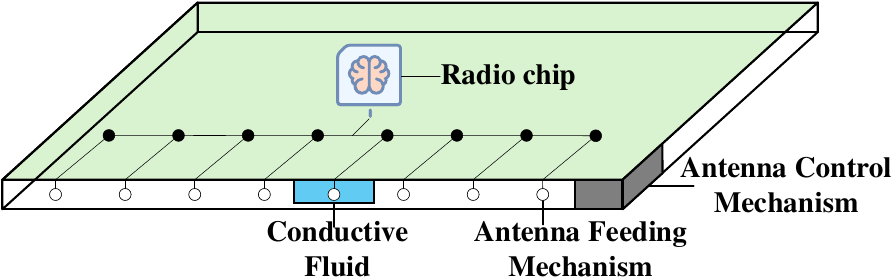}
    \caption{Liquid-based FA. The antenna feeding mechanism gathers CSI from different ports. The radio chip processes collected information to identify the optimal port. The antenna control mechanism directs the conductive fluid within the container to the desired location for optimal signal reception.}
    \label{fig:1-D FA}
    \vspace{-2ex}
\end{figure}

One idea to realize FAs is to utilize the flexibility of liquid or fluid as the radiating element of an antenna \cite{44}. Both metallic and non-metallic liquids can be used in designing the element. A classic one-Dimensional (1-D) FA for a receiver is shown in Fig. \ref{fig:1-D FA}. The key components of this system include the conductive fluid, antenna feeding mechanism, radio chip, and antenna control mechanism. The antenna feeding mechanism collects channel information from various ports and transmits it through the RF chains to the radio chip. The radio chip processes the collected data to identify the optimal port based on the maximum channel gain and then feeds this port selection information to the antenna control mechanism. Subsequently, the antenna control mechanism (e.g., micro-pump/valve) directs the conductive fluid within the container to the desired location for optimal signal reception. Unlike terrestrial networks, in the dynamic NTNs, we need to consider the fluid material's ability to cope with temperature and pressure differences caused by altitude changes, and we also need to comprehensively consider conductivity, control stability, and cost. Although most non-metallic liquids are inexpensive and easy to control, their conductivity and temperature adaptability often fail to meet the requirements of high-altitude and space environments. In recent years, gallium-based alloys, such as eutectic gallium-indium alloys (eGaIn) or Galinstan, have become the mainstream in the field of liquid-based antennas \cite{44}. These non-toxic and non-flammable alloys have high conductivity and show great potential in terms of thermal performance related to radio frequency applications. It is worth noting that their melting points are close to or below room temperature. This makes them suitable for most low-altitude environments.

To model the FA channels, we temporarily assume a 1-D FA with $ N $ preset positions (i.e., ports), uniformly distributed within a region of length $ W\lambda $, where $W$ refers to the normalized size of FA relative to the wavelength and $ \lambda $ denotes the wavelength. The FA can continuously analyze the fading envelope of the spatial channel to identify the position with the highest channel gain within the predefined movable range and adjust the antenna position accordingly. At a given port position, the FA is considered an ideal point antenna, and the position of the $ n $-th port is expressed as:
\begin{equation}
    d_n = \left( \frac{n - 1}{N - 1} \right) W\lambda, \quad \text{for } n = 1, 2, \ldots, N.
    \label{eq-1d-1}
\end{equation}
The normalized channel vector between the transmitter and a receiver equipped with $ N $ ports of a single FA can be expressed as $ \bm{h} = [h_1, \ldots, h_N]^\text{T}$, where $(\cdot)^\text{T}$ denotes the transpose operation and $ h_n \sim \mathcal{CN}(0, 1) $ represents the channel gain from the transmitter to the $ n $-th port, with a standard complex Gaussian distribution, i.e., a Rayleigh fading channel. As the small inter-port distance (less than $ \lambda/2 $), a significant correlation exists among the $ h_n $. Thus, the correlation coefficients between ports can be parameterized with the first port serving as the reference port. Specifically, the $n$-th port of $\bm{h}$ can be computed as:
\begin{equation}
	\begin{cases}
        \text{when } n = 1, h_1 = x_1 + j y_1 \\
	\text{when } n = 2, \ldots, N, \\h_n = \left( \sqrt{1 - \mu_n^2} x_n + \mu_n x_1 \right) + j \left(\sqrt{1 - \mu_n^2} y_n + \mu_n y_1 \right),
	\end{cases}
    \label{eq-1d-2}
\end{equation}
where $x_1, \dots, x_N $ and $ y_1, \dots, y_N$ represent independent and identically distributed (i.i.d) real Gaussian random variables with zero-mean and variance $ 0.5 $. The $ \mu_n $ is given by:
\begin{equation}
	\mu_{n} = J_0\left( \frac{2\pi (n - 1) W}{N - 1} \right),
    \label{eq-1d-3}
\end{equation}
where $ J_0(\cdot) $ denotes the zero-order Bessel function of the first kind. Based on the Jakes channel correlation model, $ \mu_{n,k} $ is used to characterize the spatial correlation of the channel between any $ n $-th port and $ m $-th port and computed as:
\begin{equation}
    \mu_{n,m} = J_0\left( \frac{2\pi (n - m) W}{N - 1} \right).
    \label{eq-1d-4}
\end{equation}
According to \cite{45}, to reduce the number of parameters for $ h_n $, we uses a $ \mu = \mu_{n,m} $ as the average correlation parameter for all FA's ports. The $\mu$ is given by: 
\begin{equation}
	\mu^2 = \frac{2}{N(N-1)} \sum_{n=1}^{N-1} (N - n) J_0 \left( \frac{2 \pi n W}{N-1} \right).
    \label{eq-1d-6}
\end{equation}
However, with the FA's ports becoming more numerous, this simplification may lead to a cost of accuracy. 
Recently, Ramírez-Espinosa et al. in \cite{46} introduced a new block spatial correlation model to balance manageability and accuracy. In \cite{47}, Khammassi et al. employed eigenvalue decomposition to reveal that the covariance matrix of the channels is mainly concentrated in some large eigenvalues, thus proposing a low-rank approximation method to simplify the correlation matrix.

\begin{figure}[!t]
\centering
\subfloat[]{\includegraphics[width=0.7\linewidth]{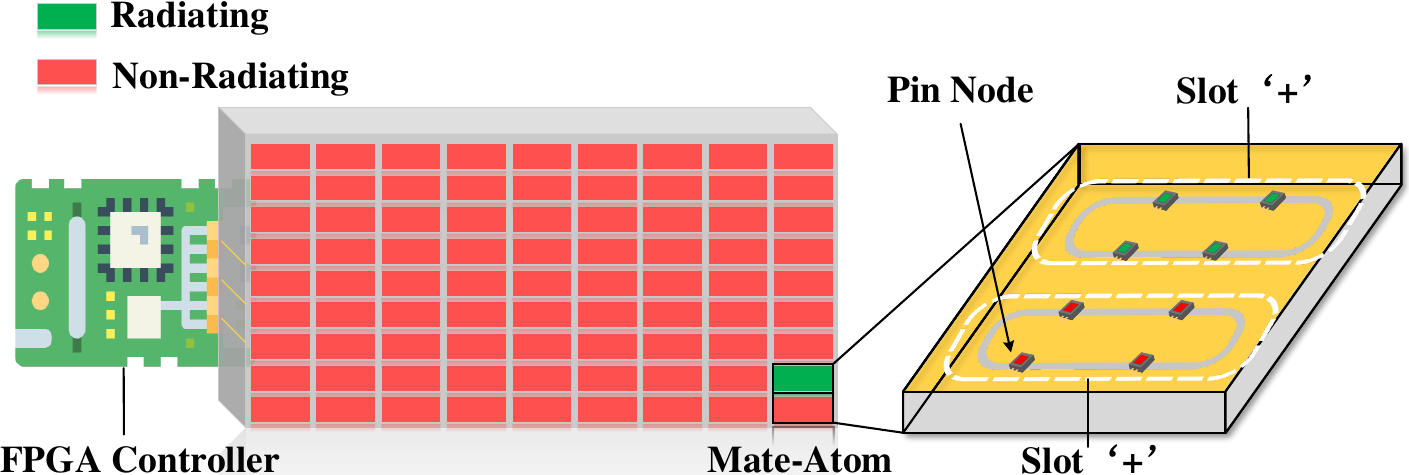}%
\label{meta-fa-1}}
\subfloat[]{\includegraphics[width=0.3\linewidth]{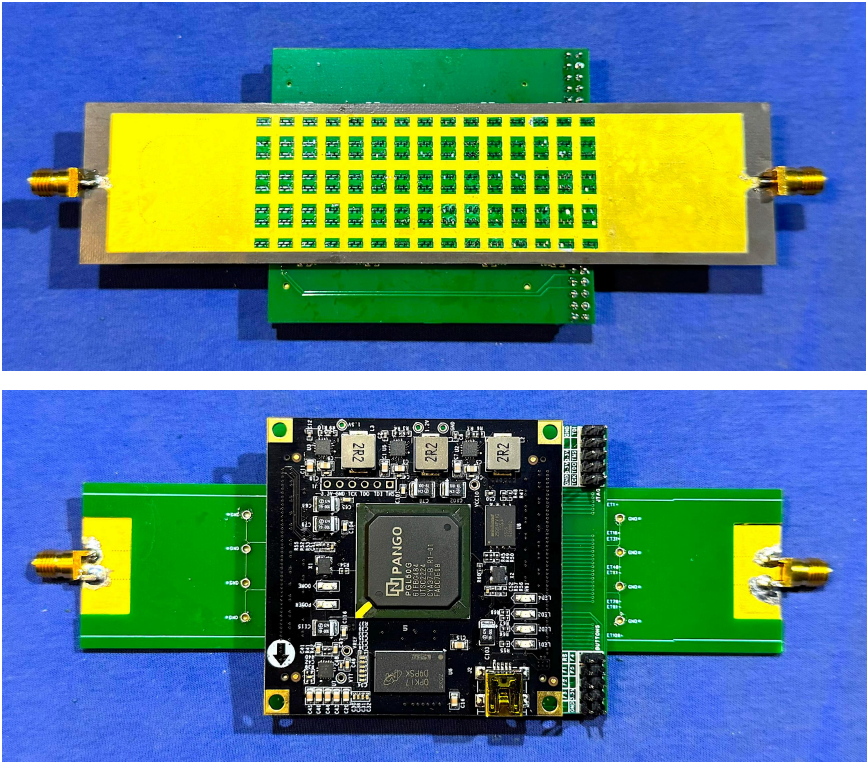}%
\label{meta-fa-2}}
\caption{Programmable Meta-FA in \cite{50}. (a) A potential architecture of Meta-FA. A 2-D plane deploys massive controllable basic units that contain a current-carrying slot and four PIN diodes. The FPGA controller can selectively activate the basic unit to adjust the amplitude and phase of the electromagnetic wave radiation. (b) Demo system of Meta-FA.}
\vspace{-1ex}
\label{fig:2-D FA}
\end{figure}

\subsubsection{Pixel Array Antennas}
Although liquid-based FAs have demonstrated promising results in theoretical simulations, their practical applications are restricted by long response times and challenges in achieving precise control. A potential alternative is the two-Dimensional (2-D) reconfigurable pixel FAs, which consist of a matrix of small controllable pixels \cite{48, 49}. By optimizing the connections between pixels, the radiation characteristics can be dynamically adjusted. As shown in Fig. \ref{fig:2-D FA}, Liu et al. in \cite{50} introduced a Meta-FA design with the basic unit (called an "element") which consists of a current-carrying slot and 4 Positive-Intrinsic-Negative (PIN) diodes. Through a Field-Programmable Gate Array (FPGA) controller, individual elements with four PIN diodes can be selectively activated, while other elements remain inactive. Each meta-atom is composed of two selectable elements, known as the `+' slot element and the `-' slot element. Each element functions as a magnetic dipole, fed by a waveguide that extracts energy from it and radiates electromagnetic waves into space. The slot states are controlled by the PIN diodes, which can adjust the amplitude and phase of the radiation. Compared to liquid-based FAs, the core advantage of meta-FAs lies in their ability to adjust radiation positions through feed control, which significantly reduces response latency and simplifies implementation complexity. Pixel FAs are one of the most used antennas in FA-assisted NTNs especially in the satellite scenarios, because they are easy to fabricate, have a low profile and low cost, and are easy to integrate.

We consider a 2-D uniform pixel FA with $ N = N_1 \times N_2 $ ports, where $N_1$ and $N_2$ indicate the numbers of horizontal and vertical FA elements, respectively. The size of FA is given by $ W_1\lambda \times W_2\lambda $, where $W_1$ and $W_2$ are the normalized lengths of FA's horizontal and vertical sides. The ports $(n_1, n_2)$ are indexed in a left to right and bottom to top order, with the index allocation $n_{n_1, n_2}$ defined as:
\begin{equation}
    n_{(n_1, n_2)} = (n_2 - 1) N_1 + n_1.
    \label{eq-2d-1}
\end{equation}
The spatial correlation among all ports at the receiver can be characterized by the correlation matrix $ \bm{\Sigma} \in \mathbb{C}^{N_1 \times N_2} $. The element $(\bm{\Sigma})_{n_{(n_1, n_2)},m_{(m_1, m_2)}}$ represents the spatial correlation between the $(n_1, n_2)$-th and $(m_1, m_2)$-th ports, which is given by:
\begin{equation}
    \begin{aligned}
    &(\bm{\Sigma})_{n_{(n_1, n_2)},m_{(m_1, m_2)}} =\\ 
    &j_0 \left( 2\pi \sqrt{ \left( \frac{n_1 - m_1}{N_1 - 1} W_1 \right)^2 + \left( \frac{n_2 - m_2}{N_2 - 1} W_2 \right)^2 } \right),
    \end{aligned}
    \label{eq-2d-2}
\end{equation}
where $j_0(\cdot)$ is the spherical Bessel function of the first kind. In \cite{51}, New et al. equipped FA at both transmitter and receiver side with the eigen-decomposition $\bm{\Sigma} = \bm{Q} \bm{\Lambda} \bm{Q}^\text{H}$, where $\bm{Q}$ consists of the eigenvectors of $\bm{\Sigma}$, $\bm{\Lambda}$ is a eigen matrix, and $(\cdot)^\text{H}$ denotes the conjugate transpose operation. As such, we can model the channel of 2-D FA:
\begin{equation}
    \bm{h} = \bm{Q} \bm{\Lambda}^{\frac{1}{2}} \bm{g},
\end{equation}
where $g = [g_1, \dots, g_N]^\text{T}$ and $g_n \sim \mathcal{CN}(0, 1)$. Furthermore, we can consider the complex channel where both the transmitter and receiver are equipped with a 2-D FA:
\begin{equation}
    \bm{H} = \bm{Q}_\text{rx} \bm{\Lambda}_\text{rx}^{\frac{1}{2}} \bm{G} \left( \bm{\Lambda}_\text{tx}^{\frac{1}{2}} \right)^\text{H} \bm{Q}_\text{tx}^\text{H},
    \label{eq-2d-channel}
\end{equation}
where $ \bm{G} \in \mathbb{C}^{N_\text{rx} \times N_\text{tx}} $ with each entry being i.i.d. and following $\mathcal{CN}(0, 1)$. 

In general, using this 2-D FA channel model to evaluate performance is more complex than the 1-D FA model due to the heightened interdependence between rows and columns. But when both the transmitter and receiver are equipped with 2-D FAs and multiple ports are activated, the proposed model can leverage spatial diversity and adaptability to enhance performance.

\begin{figure}[!t]
\centering
\subfloat[]{\includegraphics[width=0.46\linewidth]{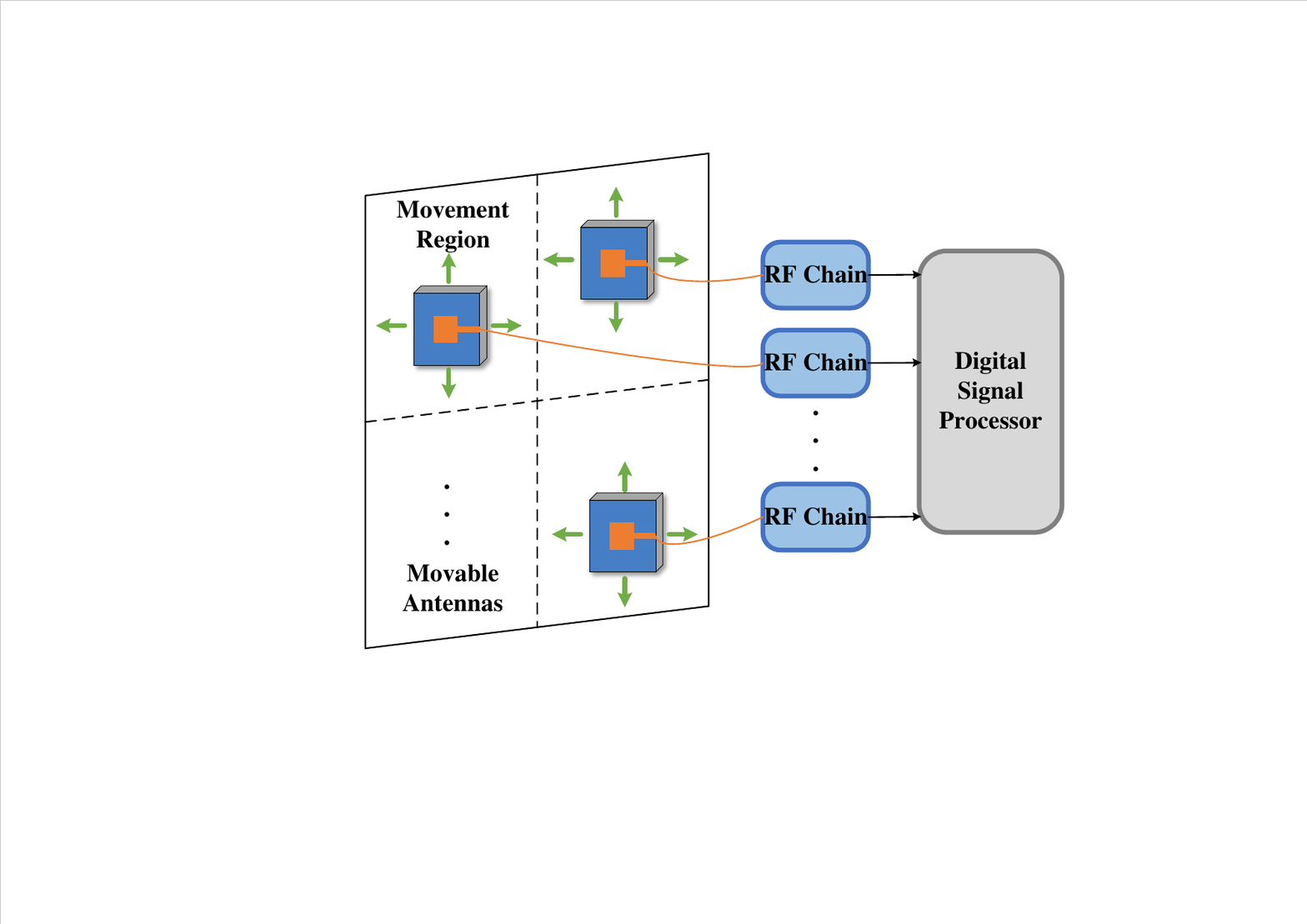}%
\label{ma-1}}
\subfloat[]{\includegraphics[width=0.45\linewidth]{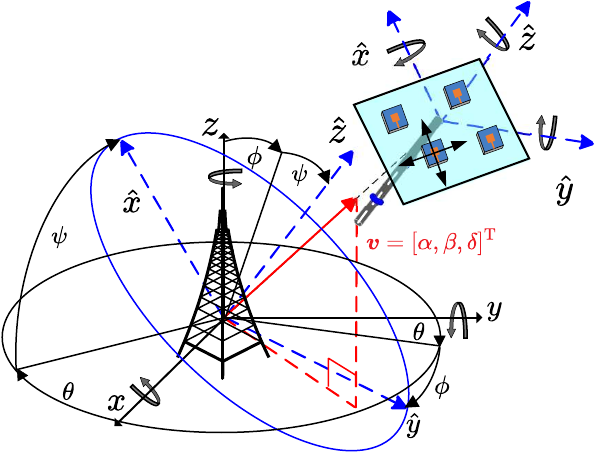}%
\label{ma-2}}
\caption{Mechanically MA. (a) A potential architecture of MA. The digital signal processor can perform signal processing and antenna position calculations, then move the antenna units through the stepper motor. (b) Illustration of the geometric coordinate system of the 6DMA.}
\label{fig:MA}
\vspace{-2ex}
\end{figure}

\subsubsection{Mechanically MA/6DMAs}
Another idea to realize FAs is to leverage stepper motor-driven radiating elements for position adjustment, known as mechanically MAs. Fig. \hyperref[ma-1]{7(a)} illustrates a potential architecture of MAs, which consists of a communication module and an antenna positioning module \cite{52}. The MA is connected to the RF chain via a flexible cable to supply power. A digital signal processor such as Central Processing Unit (CPU) can perform signal processing and antenna position computing. In \cite{53}, antennas are mounted on a 3-D mechanical slider driven by a stepper motor which, upon receiving control signals from the CPU, adjusts antennas to the desired location with the required accuracy. Additionally, the movable planar can be connected to the CPU transceiver through an extendable rod. Equipped with motors at both ends of the rod, the CPU can flexibly control 3-D spatial positions and rotation angles of antennas, forming a 6DMA. All 6DMA surfaces can independently control positions and rotations to adapt to the time-varying channel distribution in wireless networks. Compared to liquid-based FAs and reconfigurable pixel antennas, mechanically MAs offer higher movement freedom and lower hardware costs. However, their movement response time and spatial coupling are issues that cannot be neglected. In existing research, MAs and 6DMAs are more suggested for use in LAPs and HAPs due to the satellite-side MAs face several practical limitations. For example, the integration of movable hardware on satellites increases system complexity and is subject to strict payload constraints in terms of power and size. In addition, the dynamic adjustment of satellite-side MAs would require complex control mechanisms and continuous recalibration under fast orbital dynamics, which may compromise stability and reliability. Moreover, the hardware maintenance or replacement is extremely high in satellite networks.

Without loss of generality, we directly discuss the channel modeling of 6DMA, which covers the channel modeling for MA. We assume that the receiver is equipped with a stretchable 2-D square planar of side length $2L$ consisting of $N$ MAs, as shown in Fig. \hyperref[ma-2]{7(b)}. The planar can freely rotate in 3-D space to redefine the angles of departure. 
Consider a global 3-D Cartesian coordinate system $X$-$Y$-$Z$. The positions of transmitter and receiver are defined as $\bm{p}_\text{tx} = [x_\text{tx}, y_\text{tx}, z_\text{tx}]^\text{T}$ and $\bm{p}_\text{rx} = [x_\text{rx}, y_\text{rx}, z_\text{rx}]^\text{T}$.
The positions of the MAs are defined in a local coordinate system $\hat{X}$-$\hat{Y}$-$\hat{Z}$ centered at the origin and represented by $\hat{\mathcal{X}} = \{\hat{\bm{x}}_1, \hat{\bm{x}}_2, \ldots, \hat{\bm{x}}_N\}$, where $\hat{\bm{x}}_n = [\hat{x}_n, \hat{y}_n, \hat{z}_n]^\text{T}$, with $\hat{x}_n \in [-L, L]$, $\hat{y}_n \in [-L, L]$. The antenna rotation vector is defined as $\bm{a}_R = [\phi, \psi, \theta]^\text{T}$, where $\phi \in [-\pi, \pi]$, $\psi\in [-\pi, \pi]$, and $\theta\in [-\pi, \pi]$ represent the rotation angles of $\hat{X}$-$\hat{Y}$-$\hat{Z}$ with respect to the $X$-$Y$-$Z$ system. The rotation matrices can be defined as:

\begin{equation}
\bm{R}_x(\phi) = 
\begin{bmatrix}
1 & 0 & 0 \\
0 & \cos \phi & -\sin \phi \\
0 & \sin \phi & \cos \phi
\end{bmatrix},
\end{equation}

\begin{equation}
\bm{R}_y(\psi) = \!
\begin{bmatrix}
\!\cos \psi & 0 & \sin \psi \\
\!0 & 1 & 0 \\
\!-\sin \psi & 0 & \cos \psi
\end{bmatrix},
\end{equation}
and
\begin{equation}
\bm{R}_z(\theta) = 
\begin{bmatrix}
\cos \theta & -\sin \theta & 0 \\
\sin \theta & \cos \theta & 0 \\
0 & 0 & 1
\end{bmatrix},
\end{equation}
respectively. The steering vector from the transmitter to the 6DMA receiver is $\tilde{\bm{v}} = \bm{p}_\text{tx} - \bm{p}_\text{rx}$.
The steering vector from the transmitter to the 6DMA is given by $\hat{\bm{v}} = \bm{U}^\text{T} \tilde{\bm{v}}$, where $\bm{U} = \bm{R}_x(\phi) \bm{R}_y(\psi) \bm{R}_z(\theta)$. Thus, $\bm{v} = \frac{\hat{\bm{v}}}{\|\hat{\bm{v}}\|} \equiv [\alpha, \beta, \delta]^\text{T}$. The array steering vector $\bm{a}$ is:

\begin{equation}
\bm{a} = \left[ e^{j \frac{2\pi}{\lambda}\rho_{1}}, \ldots, e^{j \frac{2\pi}{\lambda}\rho_{n}}, \ldots, e^{j \frac{2\pi}{\lambda} \rho_{N}} \right]^\text{T},
\label{eq-rotation-steering-vector}
\end{equation}
where $\rho_{n} = \hat{x}_n \alpha + \hat{y}_n \beta$ represents the angular difference in signal propagation direction between the transmitter and the 6DMA receiver.

\subsection{Advantages of FAs}
FAs enable the reconfiguration of radiating element properties (e.g., size, position, and rotation) to redefine spatial diversity within a predefined space. They can achieve better multiplexing and diversity gains compared to traditional FPAs. By employing appropriate antenna position selection algorithms, FAs achieve parallel performance comparable to that of traditional FPA-based XL-MIMO and antenna selection techniques, with fewer hardware components and lower power consumption. Furthermore, their configuration flexibility facilitates additional enhancements, including adaptive beamforming and adjustable coverage range. The benefits of FAs are summarized as follows:

\subsubsection{Positional/Rotational Multiplexing Gain}
FAs can continuously analyze the fading envelope of spatial channels to identify the position of the highest channel gain within a predefined space, i.e., the movable range of the antenna, then adjust the position of FAs accordingly. As mentioned in \cite{9, 54}, regarding this flexibility as an additional DoF, FAs can enhance the adaptability of the communication system to dynamic environments by effectively exploiting spatial channel variations. For example, the FA position/rotation adjustment can change the relationships between the steering vectors of desired signals and the interference vectors in the LoS path and selects the best channel gain in the Non-Line-of-Sight (NLoS) random channel. As the movable range of FAs expands, analyzing a larger number of channel paths and a broader reception area significantly enhances overall system performance \cite{51, 55}. Moreover, this positional adaptability helps mitigate interference power by statistically determining the position of minimal channel power gain from interfering transmitters \cite{56}. In FA-assisted MIMO systems, all channel variations can be utilized to reconfigure the steering vectors. FAs can optimize their positions/rotations to reshape the singular values of the MIMO channel matrix, thereby reducing the probability of signal interruption and power consumption, even within a wavelength smaller than half \cite{51, 57, 58}. In rich scattering environments with severe multipath fading, adjusting the position and rotation angles of FAs to align the dominant channel with the LoS path can increase average signal power and reduce the receiver’s outage probability \cite{59}.

\subsubsection{Interference Mitigation}
Furthermore, FA arrays enable flexible beamforming strategies by jointly optimizing the position, rotation angle, and beamforming weights of multiple antennas \cite{60, 61}. Traditional FPA arrays suffer from reduced array gain in multi-beamforming due to their fixed geometry, which limits the correlation between steering vectors and beamforming vectors when they are orthogonal in the beam space. In contrast, FAs can dynamically reconfigure their geometry to enhance the correlations, thereby improving multi-beamforming performance. Besides, FA arrays can reduce the correlation between desired steering vectors and interfering vectors, potentially achieving full array gain in the signal direction while suppressing interference. Beyond beamforming, FAs can enhance wireless sensing by leveraging distributed positioning and orientation adjustments.

\subsubsection{Geometric Gain}
In Rotatable FAs such as 6DMAs, the antenna rotation affects the channel gain of each propagation path by modifying both the radiation pattern and polarization response. Specifically, rotating the antenna changes the impinging angle of the electromagnetic wave in the local coordinate system, which modifies the effective antenna gains. In addition, the rotation of the transmitting or receiving antenna alters its polarization orientation through the polarization response matrix, further reshaping the path gains. As a result, antenna rotation provides an additional gains for channel adaptation by adjusting the relative geometric differences between the receiver and the transceiver \cite{25}.

Beyond communication enhancement, FAs can also improve sensing accuracy by exploiting the geometric flexibility of position/rotation-adjustable antenna surfaces. In sensing tasks such as localization, the achievable accuracy depends strongly on the relative geometry between the target and the transmitting/receiving antennas. By optimizing the antenna position and rotation, the system can create a more favorable sensing geometry and thereby achieve a geometric gain in localization accuracy \cite{61-n}.

\begin{figure}[t]
    \centering
    \includegraphics[width=0.5\textwidth]{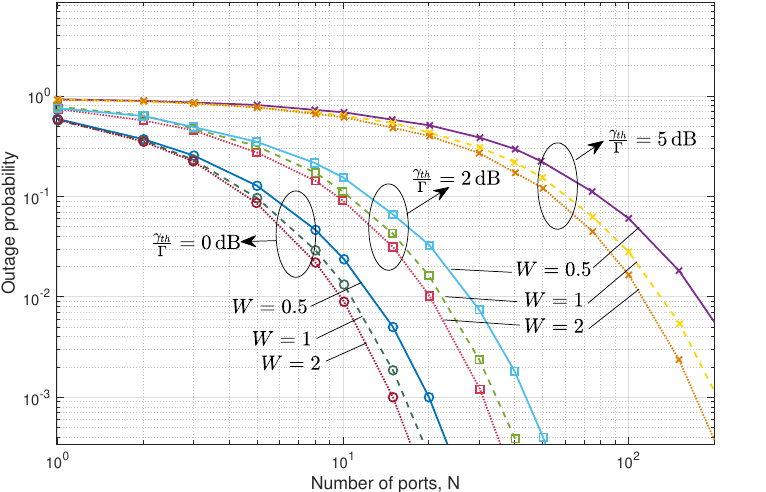}
    \caption{Outage probability against N for the FAs. As the $\frac{\gamma_{th}}{\Gamma}$ becomes more ambitious, the outage probability rises \cite{9}. If the FAs have larger $W$ and $N$, the outage probability will be reduced with no outage floor.}
    \label{fig:gain}
    \vspace{-2ex}
\end{figure}

\subsubsection{Diversity Gain}
FAs continuously analyze the fading envelope of channels to identify the maximum channel gain. Can we enhance diversity gain by increasing the number of channel ports? To address this question, we shed light from the perspective of outage probability as follows. Consider a 1-D FA with $N$ ports equipped at the receiver, where the motion range of the FA is $W\lambda$. Assuming a rich scattering environment (e.g., Rayleigh fading) and treating each port radiator as an ideal omnidirectional point source. The upper bound of outage probability $Pr_{\text{out}}(\gamma_{\text{th}})$ is given by \cite{9}:
\begin{equation}
    Pr_{\text{out}}(\gamma_{\text{th}}) < 
    \left( 1 - e^{-\frac{\gamma_{\text{th}}}{\Gamma}} \right) \prod_{n=2}^{N} \left( 1 - \frac{\varrho}{\sqrt{|\mu_n|}} e^{-\frac{\kappa}{1-\mu_n^2} \left( \frac{\gamma_{\text{th}}}{\Gamma} \right)} \right),
    \label{eq3}
\end{equation}
where $\gamma_{\text{th}}$ is the normalized target Signal-to-Noise Ratio (SNR), $\Gamma$ is the average received SNR at each port, $\kappa>1$, $0<\varrho<0.5$, and $\mu_n$ can follow the definition in (\ref{eq-1d-3}). For a given SNR target $\gamma_{\text{th}}$ and average SNR $\Gamma$, an FA with any dimension, $W\lambda$, can achieve an arbitrarily small outage probability when $N \to \infty$ and $|\mu_k| \neq 1$. In Fig. \ref{fig:gain}, we study how the outage probability performance of FAs scales with the number of ports, $N$, for various sizes, $W$, and SNR targets, $\frac{\gamma_{th}}{\Gamma}$. As expected, as the $\frac{\gamma_{th}}{\Gamma}$ becomes more ambitious, the outage probability rises. In addition, if the FAs have larger $W$ and $N$, the outage probability will be reduced with no outage floor. The above analysis demonstrates that FAs with extensive ports can yield significant diversity gains. 
\section{Basic Principles and Dynamic CSI Estimation of FA-Assisted NTNs}
\label{csi}
Accurate channel modeling and estimation are crucial for FA to achieve network throughput improvements. In this section, we first further investigate the FA-assisted NTN channel models to consider the key influential challenges in dynamic NTN environments. Then we provide a detailed review of model-based and AI-based CSI estimation methods.

\begin{figure}[t]
    \centering
    \includegraphics[width = 0.85\linewidth]{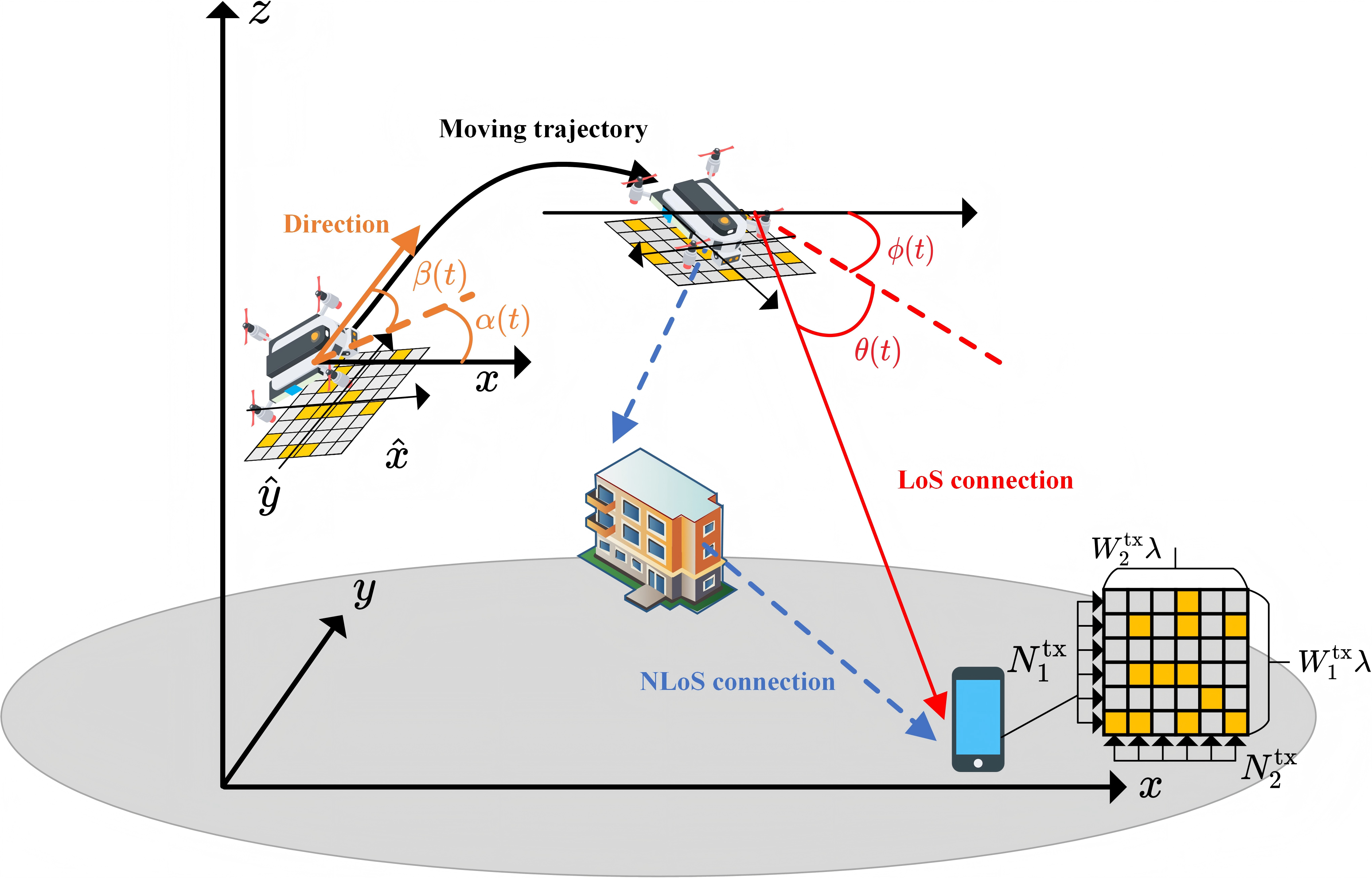}
    \caption{The FA-assisted aerial network channel model.}
    \label{fig:aerial channel model}
    \vspace{-2ex}
\end{figure}

\subsection{Channel Modeling of FA-Assisted NTNs}
An FA-assisted NTN structure can be modeled as a layered, functionally integrated system architecture consisting of three core subsystems. The terrestrial layer includes terrestrial base stations and edge nodes integrated with FAs, serving both aerial and terrestrial terminals. The aerial layer consists of UAVs or other aerial platforms equipped with FAs to support A2G and A2A links. The FA system setups comprise Single-Input Single-Output (SISO), Multiple-Input Single-Output (MISO), Single-Input Multiple-Output (SIMO), and MIMO. The control and management layer is centered on a software-defined networking controller and network orchestrator, enabling centralized control and global resource allocation, FA position movement, beamforming design, and task joint scheduling.

\subsubsection{FA-Assisted Aerial Channel Modeling}
In modeling the general MIMO A2G channels in FA-assisted aerial networks \cite{62}, we assume the aerial platform, such as the UAV, is equipped with a 2-D FA transmitter with a size of $ W_1^{\text{tx}}\lambda \times W_2^{\text{tx}}\lambda$ and $ N_{\text{tx}} = N_1^{\text{tx}} \times N_2^{\text{tx}} $ ports, where $\lambda$ is the carrier wavelength. Accordingly, the terrestrial receiver has a 2-D FA with a size of $ W_1^{\text{rx}}\lambda \times W_2^{\text{rx}}\lambda$ and $ N_{\text{rx}} = N_1^{\text{rx}} \times N_2^{\text{rx}} $ ports.

Considering a 3-D Cartesian coordinate system $X$-$Y$-$Z$ in Fig. \ref{fig:aerial channel model}. With a motion time index $t$, the positions of aerial platform and terrestrial receiver are defined as $\bm{p}_\text{tx}(t)$ and $\bm{p}_\text{rx}(t)$. The aerial platform keeps moving with a speed of $v(t)$, the azimuth and elevation angles in the direction of the motion are $\alpha(t)$ and $\beta(t)$ , respectively. The positions of transmitter FA ports are defined in a local coordinate system $\hat{X}$-$\hat{Y}$-$\hat{Z}$ centered at the origin and represented by $\hat{\bm{x}}_n^{\text{tx}} = [\hat{x}_n^{\text{tx}}, \hat{y}_n^{\text{tx}}, \hat{z}_n^{\text{tx}}]^\text{T}$, with $\hat{x}_n^{\text{tx}} \in [0, W_1^{\text{tx}}\lambda]$, $\hat{y}_n^{\text{tx}} \in [0, W_2^{\text{tx}}\lambda]$. Similarly, the positions of receiver FA ports are represented by $\hat{\bm{x}}_n^{\text{rx}} = [\hat{x}_n^{\text{rx}}, \hat{y}_n^{\text{rx}}, \hat{z}_n^{\text{rx}}]^\text{T}$.

We assume that the transmitter FAs and receiver FAs activated $L_{\text{tx}} \leq N_{\text{tx}}$ and $L_{\text{rx}} \leq N_{\text{rx}}$ ports, respectively, and the signal transmitted by the aerial platform reaches the terrestrial receiver through a rich scattering environment. Thus, the A2G channel matrix for the proposed channel model can be represented by:
\begin{equation}
\begin{aligned}
        \bm{H}^{\text{aerial}}_\text{A2G}(t,\tau) = & PL_{\text{LoS}}(t)  Pr_{\text{LoS}}(t) \bm{h}_{\text{LoS}}(t) \delta\left(\tau - \tau_{\text{LoS}}(t)\right) \\ 
         + & PL_{\text{NLoS}}(t) Pr_{\text{NLoS}}(t) \bm{h}_{\text{NLoS}}(t) \delta \left(\tau - \tau_{\text{NLoS}}(t)\right) ,
\end{aligned}
\end{equation}
where $PL_{\text{LoS}}(t)$ and $PL_{\text{NLoS}}(t)$ are LoS and NLoS pathloss, respectively, $Pr_{\text{LoS}}(t)$ is the probability of LoS path, $Pr_{\text{NLoS}}(t) = 1-Pr_{\text{LoS}}(t)$, $\bm{h}_{\mathrm{LoS}}(t) \in \mathbb{C}^{L_{\text{rx}} \times L_{\text{tx}}}$ and $\bm{h}_{\mathrm{NLoS}}(t) \in \mathbb{C}^{L_{\text{rx}} \times L_{\text{tx}}}$ are the complex channel response of LoS and NLoS path, respectively, $\delta(\tau)$ is the impulse function, $\tau_{\text{LoS}}$ and $\tau_{\text{NLoS}}$ are the propagation delay for the LoS and NLoS path, respectively. The LoS pathloss and NLoS pathloss in dB from aerial platform to the terrestrial receiver are given by, respectively:
\begin{equation}
    PL_{\text{LoS}}(t) = 20\log(\frac{c}{\lambda}) + 20\log(\frac{4\pi}{c}) + 20 \log \big(d(t)\big) + \eta_{\text{LoS}},
    \label{pl_los}
\end{equation}
\begin{equation}
    PL_{\text{NLoS}}(t) = 20\log(\frac{c}{\lambda}) + 20\log(\frac{4\pi}{c}) + 20\log \big(d(t)\big) + \eta_{\mathrm{NLoS}},
    \label{pl_nlos}
\end{equation}
where $c$ is the speed of light, $d(t) = || \bm{p}_\text{tx} - \bm{p}_\text{rx} ||$ is the distance between aerial platform and terrestrial receiver, $\eta_{\text{LoS}}$ and $\eta_{\text{NLoS}}$ are additional attenuation factors due to the LoS and NLoS connections. The probability of LoS path $Pr_{\text{LoS}}(t)$ is given by:
\begin{equation}
    Pr_{\text{LoS}}(t)=\frac{1}{1+a\,\exp\!\big(-b(\theta(t)-a)\big)},
\end{equation}
where $a$ and $b$ are constants depending on the environment, $\theta(t)$ is the elevation angle between aerial platform and terrestrial receiver. The complex channel response of LoS path is expressed as:
\begin{equation}
    \bm{h}_{\mathrm{LoS}}(t) = \widetilde{D}(t) e^{j\rho_{\text{LoS}}(t)} \bm{S}_\text{rx}^{\text{T}}(t) \bm{a}_\text{rx}(t) \bm{a}_\text{tx}^{\text{H}}(t) \bm{S}_\text{tx}(t),
\end{equation}
where $\widetilde{D}(t)$ is the dynamic Doppler phase shift, $\rho_{\text{LoS}}(t)$ is the phase of the LoS component, $\bm{S}_\text{rx}(t) \in \mathbb{C}^{L_{\text{rx}} \times N_{\text{rx}}}$ and $\bm{S}_\text{tx}(t) \in \mathbb{C}^{L_{\text{tx}} \times N_{\text{tx}}}$ are port
selection matrices, $\bm{a}_\text{rx}(t) \in \mathbb{C}^{N_{\text{rx}} \times 1}$ and $\bm{a}_\text{tx}(t) \in \mathbb{C}^{N_{\text{tx}} \times 1}$ are array steering vectors, which can follow the definition of (\ref{eq-rotation-steering-vector}). The Doppler phase shift $\widetilde{D}(t)$ can be expressed as:
\begin{equation}
    \begin{aligned}
    \widetilde{D}(t) = & e^{j \frac{2\pi}{\lambda} v(t)t
     \cos\left(\phi(t)-\alpha(t)\right)
     \cos\theta(t) \cos\beta(t)}
\qquad \\
     \times & e^{j \frac{2\pi}{\lambda} v(t)t
     \sin\theta(t) \sin\beta(t)},
    \end{aligned}
\end{equation}
where $\phi(t)$ is the azimuth angle between aerial platform and terrestrial receiver.
The complex channel response of NLoS path is expressed as:
\begin{equation}
    \bm{h}_{\mathrm{NLoS}}(t) = \widetilde{D}(t) \bm{S}_\text{rx}^{\text{T}}(t)\bm{H}(t)\bm{S}_\text{tx}(t),
\end{equation}
where $\bm{H}(t) \in \mathbb{C}^{N_{\text{rx}} \times N_{\text{tx}}}$ is the spatial correlated channels of FAs, which can follow the definition in (\ref{eq-2d-channel}). These formulas accurately reflect the influence of relevant propagation parameters, including transmission paths, FA spatial correlation, transmit-receive angles, and UAV movement, on the characteristics of the proposed FA-assisted NTN channel model.

\begin{figure}[t]
    \centering
    \includegraphics[width = 0.8\linewidth]{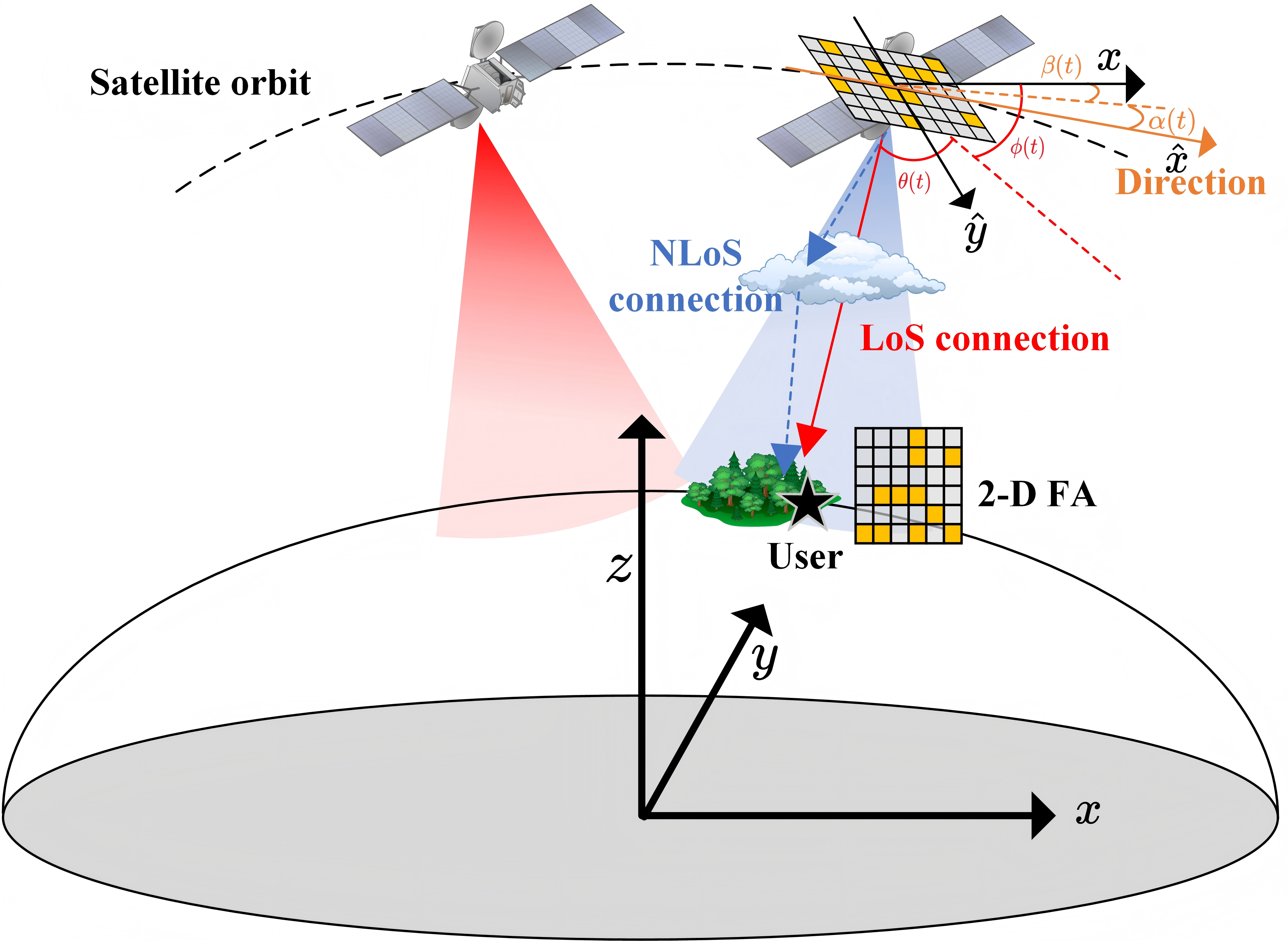}
    \caption{The FA-assisted satellite network channel model.}
    \label{fig:satellite channel model}
    \vspace{-2ex}
\end{figure}

\subsubsection{FA-Assisted Satellite Channel Modeling}
We consider a general MIMO A2G channels for FA-assisted satellite networks as shown in Fig. \ref{fig:satellite channel model}. The satellite utilizes a 2-D FA with the size of $ W_{\text{tx}} \lambda = W_1^{\text{tx}}\lambda \times W_2^{\text{tx}}\lambda$ and $ N_{\text{tx}} = N_1^{\text{tx}} \times N_2^{\text{tx}} $ ports, where $\lambda$ is the carrier wavelength. Accordingly, the terrestrial user has a 2-D FA with the size of $ W_1^{\text{rx}}\lambda \times W_2^{\text{rx}}\lambda$ and $ N_{\text{rx}} = N_1^{\text{rx}} \times N_2^{\text{rx}} $ ports. Without loss of generality, the terrestrial user is located in an urban/remote environment with several natural obstacles like mountains and trees, which generate a small number of
paths, while maintaining a dominant LoS path. Thus, the A2G channel matrix for the proposed channel model can be represented by:
\begin{equation}
\small
\begin{aligned}
        & \bm{H}^{\text{sat}}_\text{A2G}(t,\tau) =  \sqrt{\frac{K}{K+1}}PL^{\text{sat}}_{\text{LoS}}(t) \bm{h}^{\text{sat}}_{\text{LoS}}(t) \delta\left(\tau - \tau_{\text{LoS}}(t)\right) \\ 
         & + \sqrt{\frac{1}{L_p(K+1)}}\sum_{l=1}^{L_p} PL^{\text{sat}(l)}_{\text{NLoS}}(t) \bm{h}^{\text{sat}(l)}_{\text{NLoS}}(t) \delta \left(\tau - \tau^{(l)}_{\text{NLoS}}(t)\right) ,
\end{aligned}
\end{equation}
where $K$ is the Rice factor, $L_p$ is the total number of NLoS paths, $PL^{\text{sat}}_{\text{LoS}}(t)$ and $PL^{\text{sat}(l)}_{\text{NLoS}}(t)$ are LoS and $l$-th NLoS pathloss, respectively, $\bm{h}^{\text{sat}}_{\mathrm{LoS}}(t) \in \mathbb{C}^{L_{\text{rx}} \times L_{\text{tx}}}$ and $\bm{h}^{\text{sat}(l)}_{\mathrm{NLoS}}(t) \in \mathbb{C}^{L_{\text{rx}} \times L_{\text{tx}}}$ are the complex channel response of LoS and $l$-th NLoS path, respectively. Considering free space loss and rainfall attenuation, the LoS pathloss and NLoS pathloss in dB from satellite to the terrestrial receiver are given by, respectively:
\begin{equation}
    PL^{\text{sat}}_{\text{LoS}}(t) = PL_{\text{LoS}}(t) + PL_g(t)+PL_s(t)+PL_e(t),
    \label{pl_sat_los}
\end{equation}
\begin{equation}
    PL^{\text{sat}}_{\text{NLoS}}(t) = PL_{\text{NLoS}}(t)  +PL_g(t)+PL_s(t)+PL_e(t),
    \label{pl_sat_nlos}
\end{equation}
where $PL_g(t)$ is the attenuation due to atmospheric gases, e.g. rainfall attenuation, $PL_s(t)$ is the attenuation due to either ionospheric or tropospheric scintillation, and $PL_e(t)$ is building entry loss. The above formula is for signal attenuation as defined by 3GPP in \cite{27,28,29}. In the LoS complex channel response vector, we incorporate the randomness caused by shadow Rician model into the spatial correlation of the FAs for a unified representation:
\begin{equation}
    \bm{h}^{\text{sat}}_{\mathrm{LoS}}(t) = \widetilde{D}(t) e^{j\rho_{\text{LoS}}} \bm{S}_\text{rx}^{\text{T}}(t) \bm{a}_\text{rx}(t) \bm{H}(t)\bm{a}_\text{tx}^{\text{H}}(t) \bm{S}_\text{tx}(t).
\end{equation}
The complex channel response of NLoS
path is expressed as:
\begin{equation}
    \bm{h}^{\text{sat}(l)}_{\mathrm{NLoS}}(t) = \widetilde{D}(t) e^{j\rho_{\text{NLoS}}} \bm{S}_\text{rx}^{\text{T}}(t) \bm{a}^{(l)}_\text{rx}(t) \bm{H}(t)\left(\bm{a}_\text{tx}^{(l)}(t)\right)^{\text{H}} \bm{S}_\text{tx}(t),
\end{equation}
where $\rho_{\text{NLoS}} \sim \mathcal{U}(0, 2\pi)$ with a uniform distribution, models the random phase in multipath environments, $\bm{a}^{(l)}_\text{rx}(t)$, and $\bm{a}_\text{tx}^{(l)}$ are array steering vectors of the $l$-th NLoS path.

Current studies mostly focus on joint processing in the signal domain. To cope with Doppler shifts and latency, practical receivers typically adopt a “prediction-precompensation-tracking” mechanism. Specifically, satellite ephemeris/orbital information and terminal positioning are exploited to align the timing for the propagation delay and to precompensate the Doppler before the data transmission. Moreover, session-management timers should increase the duration to mitigate the latency introduced by satellite links. However, such a mechanism is hardware-incompatible and has limited performance. Recently, Orthogonal Time Frequency Space (OTFS) modulation offer a promising solution which can convert the domain of the signal into a Delay-Doppler domain to mitigate severe time-delay Doppler impairments in mobile satellite networks. Yang et al. in \cite{101} proposed a novel FA-OTFS satellite channel model for enhancing both the energy and computational efficiency in high-mobility environments. Note that the time index $t$ is omitted because a block-fading assumption is adopted. Hence, the Delay-Doppler-domain OTFS FA-assisted satellite channel can be expressed as:
\begin{equation}
\small
\begin{aligned}
        & \bm{H}^{\text{OTFS}}_\text{A2G}(\tau, \nu) =  \sqrt{\frac{K}{K+1}}PL^{\text{sat}}_{\text{LoS}} \hat{\bm{h}}^{\text{sat}}_{\text{LoS}} \delta\left(\tau - \tau_{\text{LoS}}\right)\delta\left(\nu - \nu_{\text{LoS}}\right) \\ 
         & + \sqrt{\frac{1}{L_p(K+1)}}\sum_{l=1}^{L_p} PL^{\text{sat}(l)}_{\text{NLoS}} \hat{\bm{h}}^{\text{sat}(l)}_{\text{NLoS}} \delta \left(\tau - \tau^{(l)}_{\text{NLoS}}\right)\delta\left(\nu - \nu^{(l)}_{\text{NLoS}}\right) ,
\end{aligned}
\end{equation}
where $PL^{\text{sat}}_{\text{LoS}}$ and $PL^{\text{sat}(l)}_{\text{NLoS}}$ are LoS and $l$-th NLoS pathloss, which can follow the definition in (\ref{pl_sat_los}) and (\ref{pl_sat_nlos}),  respectively, $\hat{\bm{h}}^{\text{sat}}_{\text{LoS}}\in \mathbb{C}^{L_{\text{rx}} \times L_{\text{tx}}}$ and $\hat{\bm{h}}^{\text{sat}(l)}_{\text{NLoS}}\in \mathbb{C}^{L_{\text{rx}} \times L_{\text{tx}}}$ are LoS and $l$-th NLoS complex channel response in Delay-Doppler-domain, respectively, $\nu_{\text{LoS}} = \cos\left(\phi(t)-\alpha(t)\right)\cos\theta(t) \cos\beta(t) + \sin\theta(t) \sin\beta(t)$ and $\nu^{(l)}_{\text{NLoS}} = \nu_{\text{LoS}} + \delta_{\nu}$ are LoS and $l$-th NLoS Doppler shift, respectively. Herein $\delta_{\nu} \sim \mathcal{U}(-\nu_{\text{max}},\nu_{\text{max}})$, where $\nu_{\text{max}}$ is the maximum relative Doppler shifts. The complex channel response is expressed as:
\begin{equation}
    \hat{\bm{h}}^{\text{sat}}_{\text{LoS}} = e^{j\rho_{\text{LoS}}} \bm{S}_\text{rx}^{\text{T}}(t) \bm{a}_\text{rx}(t) \bm{H}(t)\bm{a}_\text{tx}^{\text{H}}(t) \bm{S}_\text{tx}(t),
\end{equation}
\begin{equation}
    \hat{\bm{h}}^{\text{sat}}_{\text{NLoS}} = \bm{S}_\text{rx}^{\text{T}}(t) \bm{a}_\text{rx}(t) \bm{H}(t)\bm{a}_\text{tx}^{\text{H}}(t) \bm{S}_\text{tx}(t).
\end{equation}
It is worth noting that in satellite networks, the Delay-Doppler domain becomes more sparse due to the limited number of paths between ground terminals and satellites, thus reducing the complexity of OTFS detection even further.

\begin{table*}[htbp!]
\caption{A Summary of Model-Based CSI Estimation Methods}
\label{tab:param_model_csi}
\centering
\setlength{\tabcolsep}{6pt}
\resizebox{\textwidth}{!}{
\begin{tabular}{p{2cm}|p{5.2cm}|p{5cm}|p{5cm}|p{3.3cm}}
\hline
 \multicolumn{1}{c|}{\textbf{Methods}} & \multicolumn{1}{c|}{\textbf{Principles}} & \multicolumn{1}{c|}{\textbf{Strengths}} & \multicolumn{1}{c|}{\textbf{Weaknesses}}  & \multicolumn{1}{c}{\textbf{Pilot overhead}}  \\
\hline\hline

\makecell{LMMSE\\ \cite{73, 74-n}}
&
\makecell[l]{Exploit second-order channel statistics to \\perform MMSE estimation; reduce training \\burden through pilot reuse, pilot allocation \\and port dependency.}
&
\makecell[l]{\textbullet\ Statistically optimal under correct \\covariance/statistics;\\
\textbullet\ Pilot reuse is potential for NTN channels;}
&
\makecell[l]{\textbullet\ Requires reliable statistical CSI;\\
\textbullet\ Matrix inversion and joint processing can\\ be heavy for payload-limited satellites;\\
\textbullet\ Sensitive to model mismatch and resi-\\dual Doppler/delay errors.}
&
\makecell[l]{High pilot overhead}
\\
\hline

\makecell{Compressed \\ Sensing\\ \cite{76,77, 77-n}}
&
\makecell[l]{
Based on the channel angular domain or dir-\\ectional sparsity assumption, angles and path\\ gains are recovered from observed positions\\ through compressed sensing methods.}
&
\makecell[l]{\textbullet\ Significant overhead reduction;\\
\textbullet\ Suitable for NTNs where channels are\\ sparse in angle (LoS-dominant).}
&
\makecell[l]{\textbullet\ Off-grid leakage and quantization loss;\\
\textbullet\ Strongly depends on observable-position\\ design;\\
\textbullet\ Iterative recovery can be latency-sensitive\\ in fast mobility.\\
\textbullet\ Sensitive to model mismatch and resi-\\dual Doppler/delay errors.
}
&
\makecell[l]{Pilot overhead is significantly \\reduced compared to complete \\LMMSE estimation.}
\\
\hline

\makecell{Successive \\ Bayesian \\ Estimation \\ \cite{78}}
&
\makecell[l]{Model the spatial channel function as a \\continuous stochastic process and reconstruct \\full CSI from successive sampling;}
&
\makecell[l]{\textbullet\ Robust to varying model parameters;\\
\textbullet\ Exploit spatial correlation of FA ports to \\reduce pilot overhead;\\
\textbullet\ High performance under appropriate priors.\\
\textbullet\ Low estimation latency.
}
&
\makecell[l]{\textbullet\ Covariance matrix choice matters;\\
\textbullet\ Posterior updates can be heavy as sample \\size grows;\\
\textbullet\ Require strong correlation structure.}
&
\makecell[l]{Pilot overhead is reduced co-\\mpared to complete LMMSE \\and Compressed sensing.}
\\
\hline

\makecell{Tensor \\Decomposition \\ \cite{79-n, 80-n}}
&
\makecell[l]{Organize received pilots across space-time-\\coding into a low-rank tensor. Exploit tensor\\decomposition to recover factor matrices \\containing angles, gains, and Doppler shifts.}
&
\makecell[l]{\textbullet\ Training inserted only in header slots,\\ then predict subsequent slots;\\
\textbullet\ Multi-dimensional coupling improves \\accuracy; \\
\textbullet\ Suitable for high-mobility scenarios.}
&
\makecell[l]{
\textbullet\ High computational complexity;\\
\textbullet\ Sensitive to model mismatch and resi-\\dual Doppler/delay errors.}
&
\makecell[l]{Pilot overhead is significantly \\reduced compared to complete \\LMMSE estimation.}
\\
\hline

\makecell{Moving-port\\prediction\\ \cite{80}}
&
\makecell[l]{
Predict the optimal FA port location where \\the deviation between the time-varying \\channel and the static channel is minimized, \\making the effective channel approach static.}
&
\makecell[l]{\textbullet\ Convert fast time-varying CSI estimation\\ into parameter estimation and port selection\\, mitigating channel aging;\\
\textbullet\ Asymptotically near-zero prediction error \\under strong LoS with high port density;\\
\textbullet\ Reduces repeated precoding burden;}
&
\makecell[l]{
\textbullet\ Sensitive to model mismatch under \\scattering environments;\\
\textbullet\ Require sufficient FA port density and \\feasible switching speed;}
&
\makecell[l]{Pilot overhead is significantly \\reduced compared to complete \\LMMSE estimation.}
\\
\hline
\end{tabular}}
\end{table*}

\subsection{CSI Estimations of FA-Assisted NTNs}
The long propagation delay and high mobility in NTN channels exacerbate channel estimation errors, while large number of FA ports further increase the training and feedback overhead. Since effective FA optimization depends on accurate CSI, research on robust and efficient channel estimation is a central topic for FA-assisted NTNs.

\subsubsection{Model-Based CSI Estimations}
Model-based estimation methods exploit the intrinsic temporal and structural correlation of wireless channels. They characterize the channel response through a geometric, parametric representation. From a channel-feature perspective, model-based CSI estimation in FA-assisted NTNs can be formulated as a high-dimensional extrapolation problem: infer the full CSI across tens to hundreds of ports from pilots observed on only a limited subset of ports. The feasibility relies on: (i) strong spatial correlation induced by densely sampled FA apertures; (ii) low-dimensional channel structures such as angular sparsity in LoS-dominant links or low-rank couplings in space-time-frequency domains; (iii) assume that the channel evolves smoothly and remains stable over a time scale. Under these assumptions, the full CSI can be inferred from historical observations or from CSI measured on a subset of FA ports.

The mainstream channel estimation methods for FA-assisted NTN systems can be systematically categorized into five representative technical paradigms: Linear Minimum Mean-Square Error (LMMSE) methods, compressed sensing methods, successive Bayesian estimation, tensor decomposition, and moving-port prediction. Table \ref{tab:param_model_csi} summarizes the principles, strengths, weaknesses and pilot overhead of the model-based estimation methods. In \cite{73}, Skouroumounis et al. proposed a LMMSE-based approach to estimate the CSI from a limited number of ports, using the estimated CSI to approximate other CSIs in its neighborhood. This method introduces estimation and approximation errors. However, the problems of pilot overhead and training burden are still difficult to eliminate as the number of FA ports increases. In \cite{74-n}, Li et al. investigated the MIMO orthogonal frequency division multiplexing channel estimation for low-earth-orbit satellite communication systems. After pre-compensation for Doppler and delay at terminals, the estimation complexity can be significantly reduced by performing pilot signal reuse and LMMSE estimation in the angle-delay domain channels.

Compressed sensing can reconstruct the CSI for all ports from sparse channel parameters \cite{76,77, 77-n}. In \cite{76}, a compressed sensing-based channel estimator was proposed under the assumption of angular-domain sparsity in FA channels. By leveraging the orthogonal matching pursuit algorithm for sparse recovery, the proposed approach jointly estimates key channel parameters, including the angles and the corresponding path gains of multipath components. On the basis, Xiao et al. in \cite{77} further investigated the sparse channel characteristics and compressed sensing estimator under different FA location configurations (e.g., uniform setup, edge of region setup, and cross-shape setup). To jointly reconstruct both the rotation angles and positions of the 6DMAs, Shao et al. in \cite{77-n} introduced a directional sparsity channel representation, assuming that signals arrive at the 6DMA from only a few dominant directions in the 3-D environments. Building upon this property, a compressed sensing estimator is developed to estimate both statistical and instantaneous CSI with less pilot overhead while achieving better estimation accuracy.

Inspired by Bayesian linear regression, Zhang et al. in \cite{78} generally modeled the channel as a stochastic process to perform successive Bayesian estimation. Leveraging the spatial correlation of FA ports, they select informative ports for sampling and combine with the prior covariance matrix for accurate channel estimation. This non-parametric estimation method is robust in the variable NTN environment, and the pilot overhead can be reduced by the correlation of the FA port. In \cite{79}, New et al. introduced an oversampling strategy considering FA dimensions. They revealed the trade-off between sampling density and estimation accuracy.

Tensor decomposition is a powerful tool for processing high-dimensional channel data, such as MIMO FAs, long-term extrapolation, and block-fading channel models. In \cite{79-n}, a tensor-driven channel estimation framework was developed for FA-MIMO systems, where the received pilot signals are modeled as a third-order low-rank tensor (transmit domain × receive domain × path-gain domain). By performing tensor decomposition, the resulting factor matrices enable joint estimation of angles, and path gains. Considering fast time-varying uplink MU-MIMO channels, Wang et al. in \cite{80-n} organized multi-slot pilot signals into a third-order low-rank tensor (space domain × time domain × coding domain). Tensor decomposition then jointly estimates angles, path gains, and Doppler shifts. For high-mobility scenarios, only a few initial training slots are required to estimate the channel parameters, after which Doppler phase evolution is exploited to predict subsequent slots, mitigating channel aging and reducing pilot overhead.

Unlike traditional CSI prediction for addressing Doppler and latency issues in time-varying channels, Li et al. in \cite{80} utilize the position Reconfigurability of FAs to transform dynamic channels into static channels. Specifically, the base station estimates parameters including paths, angles, Doppler shifts, and gains within the coherence time and reconstructs the reference channel for each port. Then, the user side select the optimal FA ports to make the future channels approximate the static reference channels, significantly alleviating CSI aging and reducing the burden of continuous training and repetitive precoding computation.

\begin{table*}[htbp!]
\caption{A Summary of AI-Based Channel Estimation Methods}
\label{tab:ai_channel_extrapolation}
\centering
\resizebox{\textwidth}{!}{
\begin{tabular}{p{2cm}|p{5cm}|p{5cm}|p{5cm}|p{2cm}}
\hline
 \multicolumn{1}{c|}{\textbf{Methods}} & \multicolumn{1}{c|}{\textbf{Principles}} & \multicolumn{1}{c|}{\textbf{Strengths}} & \multicolumn{1}{c|}{\textbf{Weaknesses}}  & \multicolumn{1}{c}{\textbf{Complexity}} \\
\hline\hline


 \makecell{LSTM-based\\\cite{64,65,66}}
 & \makecell[l]{Model the channel extrapolation as a sequ-\\ence-to-sequence problem, process the input \\CSI sequence serially.}
 & \makecell[l]{\textbullet\ Suitable for sequential data, effectively \\ capture the spacial and temporal correlations \\ inherent in channels;\\
  \textbullet\ Variable sequence length handling.}
 & \makecell[l]{ \textbullet\ Error accumulation due to sequential extr-\\apolation;\\  \textbullet\ Sequential extrapolation induced long-late-\\ncy for large historical CSI sequence, unfit \\for high-speed communication.\\
\textbullet\ Not suitable for parallel computing.
 } 
 & \makecell{$ \mathcal{O}( T \tilde{t} d_h (d_h + $\\$ d_{\text{in}}^{\text{LSTM}})) $}\\
\hline

\makecell{Transformer-\\based \\ \cite{69, 70, 71}}
 & \makecell[l]{Model the CSI extrapolation as a sequ-\\ence-to-sequence problem, use self-attention \\mechanism to obtain global coupled fea-\\tures across time/ports/subcarriers.}
 & \makecell[l]{\textbullet\ Suitable for variable sequential data, eff-\\ectively capture the spacial and temporal \\correlations inherent in channels;\\
\textbullet\ Parallel computation effectively mitigate \\channel aging in high-mobility scenarios..}
 & \makecell[l]{ \textbullet\ Rely on extensive datasets and 
precise\\ data preprocessing;\\ 
\textbullet\ When deploying on the edge, model \\compression is essential.\\
\textbullet\ Positional encoding need improvements \\for CSI prediction.
 } 
 & \makecell{$\mathcal{O}(T^2 d_{\text{Trans}})$}\\
\hline

 \makecell{Diffusion-\\based \\ \cite{72}}
 & \makecell[l]{Formulate CSI estimation as conditional\\ generation via iterative denoising,\\ progressively recovering full CSI distribu-\\tion from noisy observations.}
 & \makecell[l]{\textbullet\ Strong distribution modeling for nonli-\\near, non-Gaussian, multi-modal dynamics;\\
 \textbullet\ Supports CSI completion/extrapolation\\ under low pilot regimes and noisy CSI;\\
 \textbullet\ Suitable for statistical CSI Estimation.}
 & \makecell[l]{\textbullet\ Slow inference due to multi-step sampling,\\ may yield outdated CSI in fast fading;\\
 \textbullet\ High training cost and data dependency,\\ sensitive to domain shift;\\
 \textbullet\ Limited physical interpretability without\\ explicit constraints.}
 & \makecell{$\mathcal{O}(TN_{\text{denoise}}$\\$ d_{\text{denoise}})$}\\
\hline
\end{tabular}%
}
\end{table*}

\subsubsection{AI-Based CSI Estimations}
In highly dynamic and heterogeneous NTN channels, model-based CSI estimation remains severely challenged. Firstly, platform mobility and intermittent blockage can rapidly reshape the propagation environment (e.g., evolving multipath clusters and the emergence/disappearance of dominant reflectors), thus altering the assumed geometric parameters of the NTN channel (e.g., LoS/NLoS probability, path-loss model, and user distribution). During the long-term CSI extrapolation, this can result in model mismatch and degrade estimation performance. Traditional CSI estimation methods based on static parametric models and short-term stability assumptions are not suitable for long-term CSI extrapolation. Secondly, severe Doppler effects with estimation and feedback latency effectively shorten the channel coherence time, so CSI can easily become outdated. Thirdly, the large number of FA ports imposes substantial pilot overhead and repeated joint processing within each coherence interval, which is inefficient in NTNs \cite{63}. In contrast, AI-based methods can reconstruct channel statistics from data and exploit strong spatial-temporal correlations across FA ports to reconstruct and predict the full CSI from a limited set of observed ports, reducing pilot overhead and estimation latency in time-varying NTNs. AI-based channel estimation methods are categorized into Long Short-Term Memory (LSTM)-based \cite{64,65,66}, Convolutional Neural Network (CNN)-based \cite{67}, Generative Adversarial Networks (GAN)-based \cite{68}, transformer-based \cite{69, 70, 71}, and diffusion-based \cite{72}, to exploit the spatial and temporal correlations among ports where model-based models are difficult to establish. Table \ref{tab:ai_channel_extrapolation} summarizes the principles, strengths, weaknesses and complexity of the AI-based estimation methods.

\begin{figure}[!t]
\centering
\subfloat[]{\includegraphics[width=1\linewidth]{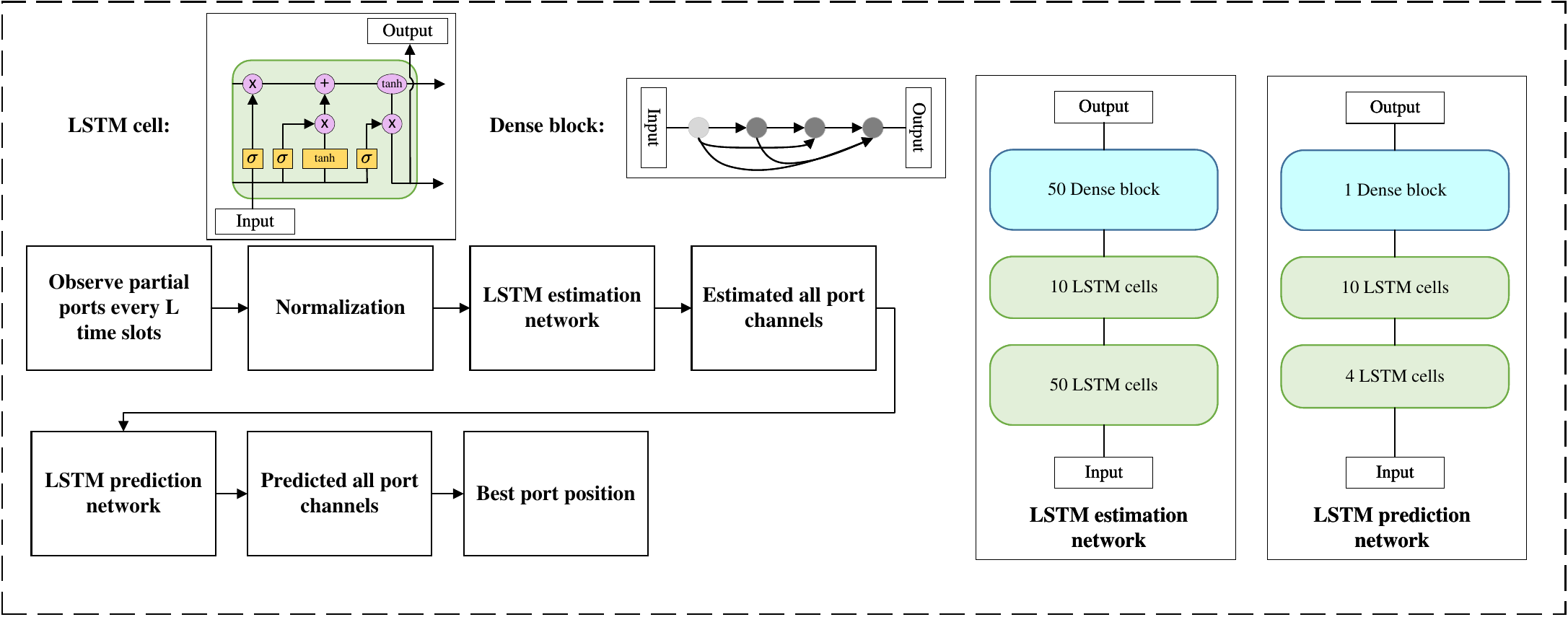}%
\label{a-3-1}}
\quad
\subfloat[]{\includegraphics[width=1\linewidth]{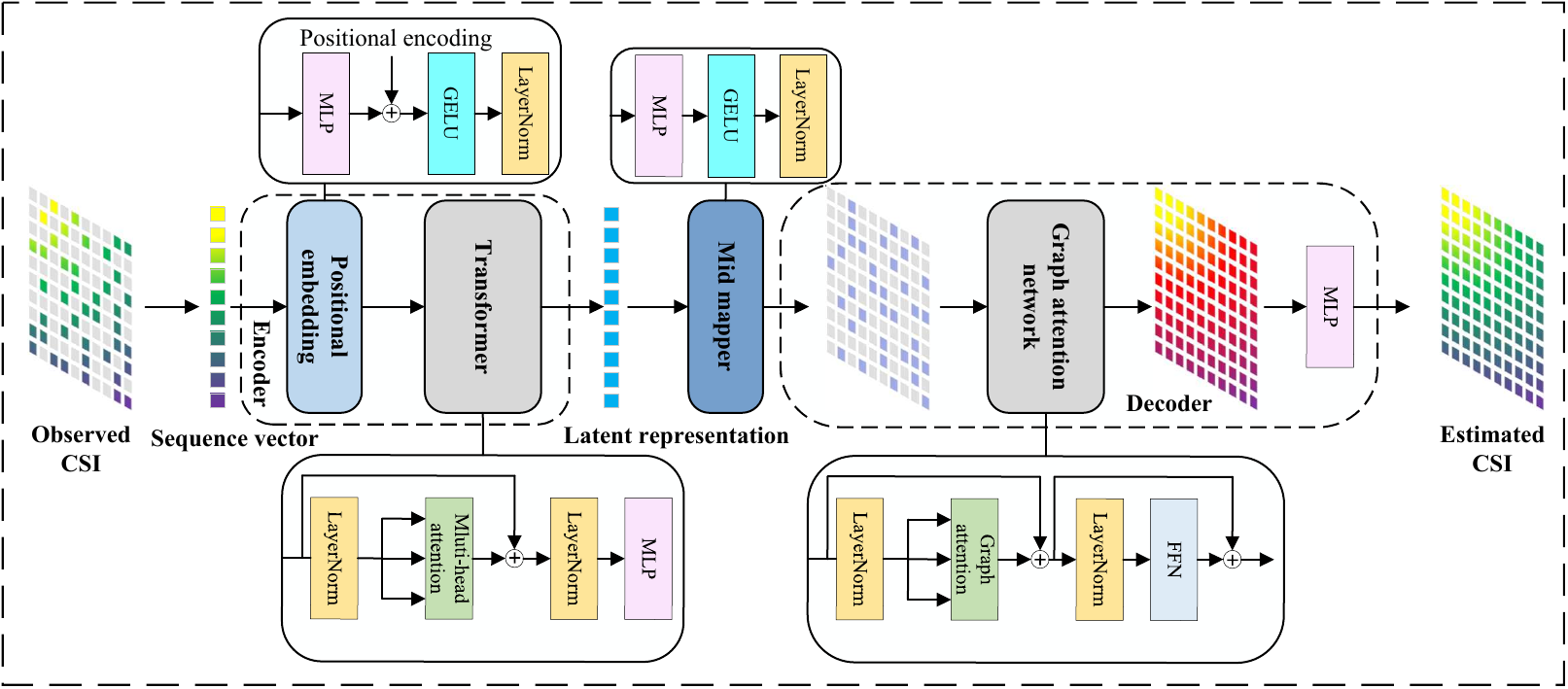}%
\label{a-3-2}}
\quad
\subfloat[]{\includegraphics[width=1\linewidth]{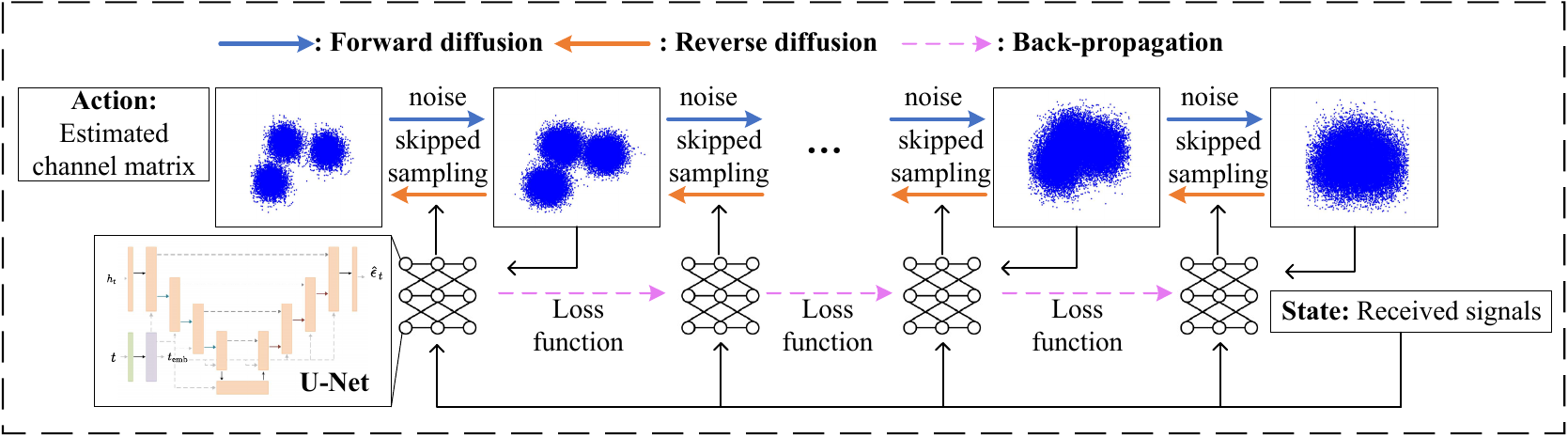}%
\label{a-3-3}}
\caption{The model architecture of (a) LSTM-based, (b) transformer-based, and (c) diffusion-based for FA CSI estimation.}
\label{a-3}
\vspace{-2ex}
\end{figure}

For example, Zhang et al. in \cite{65} expected to find a channel mapping function of an FA with $N$ ports: $\{\bm{h}_\mathcal{U}, \bm{H}_{t \in T}\} = f_{\text{net}}(\bm{h}_\mathcal{O}, \bm{\Theta})$, where $\bm{h}_\mathcal{U}$, $\bm{h}_\mathcal{O}$ respectively represent the estimated CSI and observable CSI, $\bm{H}_{t \in \mathcal{T}}$ denotes the CSI of all ports the predicted CSI of all ports at the next $T$ time indexes, $f_\text{net}(\cdot)$ functions as the neural network operation, and $\bm{\Theta}$ is the set of learnable parameters. For the above channel estimation and prediction problems, the LSTM architecture in \cite{65} can be used. An LSTM cell consists of three gating mechanisms: the forget gate regulates how much past information is retained, the input gate controls how new information is written, and the output gate determines how much of the internal state is exposed as the output. The LSTM cell acts as a long-term memory pathway that preserves and propagates information across time steps. This architecture can Model the CSI estimation as a sequential problem, processing the input CSI sequence serially. The loss function of the estimation network can take the form of Mean Square Error (MSE):
\begin{equation}
    \mathcal{L}_{\text{est}} = \text{E} \left[ \frac{1}{|\mathcal{U}|} \sum_{l \in \mathcal{U}} (|h_l| - | \widetilde{h}_l|)^2 \right],
\end{equation}
where $\widetilde{h}_l$ is the estimated CSI of unknown FA ports through the LSTM estimation network, $h_l$ is the known channel gain in the dataset. Similarly, the prediction network can use the loss function:
\begin{equation}
    \mathcal{L}_{\text{pred}} = \text{E} \left[ \frac{1}{N|\mathcal{T}|} \sum_{n \in N} \sum_{\nu \in \mathcal{T}} (|h_{n,\nu}| - |\widetilde{h}_{n,\nu}|)^2 \right],
\end{equation}
where $\mathcal{T}$ includes the present time index $t$ and the $\tilde{t}$ previous time indices $t-1,t-2, \cdots, t-\tilde{t}$, $\widetilde{h}_{n, \nu}$ is the predicted CSI of all ports through the LSTM prediction network, and $h_{n, \nu}$ is the known channel gain in the dataset. 
The per-layer complexity is
$
\mathcal{O}\!\left( T \tilde{t} d_h\left(d_h+d_{\mathrm{in}}^{\mathrm{LSTM}}\right)\right),
$
where $T$ is the sequence length, $d_h$ is the hidden-state dimension, and $d_{\mathrm{in}}^{\mathrm{LSTM}}$ is the input dimension. The LSTM-based network is shown in Fig. \hyperref[a-3-1]{11(a)}.

Most simple models such as LSTM-based \cite{64,65,66}, CNN-based \cite{67} framework have achieved good results in some fixed scenarios and lightweight tasks, but they still have some shortcomings, such as: (i) error accumulation due to sequential extrapolation, challenging to deal with a variable number of CSI inputs due to their weak long-range information feature capture; (ii) limited performance improvement, cannot achieve enough precise estimation accuracy for practical applications; (iii) poor generalization and high computational complexity, unable to adjust dynamic and complex channel environments. 
To overcome these problems, Zhang et al. in \cite{70} employed a unified asymmetric masked autoencoder architecture to better obtain the spatiotemporal nonlinear characteristics of each FA port. The model introduces a multi-head self-attention operation to capture the spatial correlations between observable ports and surrounding ports and designs an asymmetric encoder-decoder architecture to balance computational complexity. To enhance input flexibility and model generalization, the training process is based on 800,000 channel data samples and adopts a random masking strategy. The model architecture of the transformer-based network is shown in Fig. \hyperref[a-3-2]{11(b)}.
In an OTFS-based FA-assisted LEO satellite system with high latency, severe Doppler, and strong port correlation, Yang et al. in \cite{71} argued that channel prediction is essential for joint optimization under overhead-limited operation. Accordingly, they input a compressed channel vector that retains delay-Doppler features into a transformer-based model and conduct time-series prediction by explicitly leveraging the correlation across FA ports.
The loss function of the transformer used is the MSE. 
The per-layer complexity is $
\mathcal{O}\!\left(T^2 d_{\text{Trans}}\right),
$, where $T$ is the sequence length and $d_{\text{Trans}}$ is the
feature dimension. Although transformer-based models outperform in various information feature extraction and generation tasks, their training relies on extensive datasets and precise data preprocessing. With the development of expert models and model compression, training overhead is gradually decreasing, which is crucial for the practical deployment in NTNs. 

Recently, a new class of generative model architectures, diffusion models, has demonstrated superior channel estimation performance under conditions of substantial interference and thermal noise. Tang et al. in \cite{72} trained a diffusion model to learn complex data distributions as prior knowledge, employing posterior sampling for channel estimation from observable ports. The denoising network employs a U-Net structure to strike a balance between performance and computational complexity. The denoising network updates its parameters via a designated loss function:
\begin{equation}
    \begin{aligned}
    \mathcal{L}_{\text{est}} = 
    \frac{1}{\mathcal{B}}\sum_{i=1}^{\mathcal{B}} \left\| \bm{\epsilon}^{(i)} - \bm{\epsilon}_{\bm{\theta}} \left( \sqrt{\bar{\alpha}_{t(i)}} h_0^{(i)} + \sqrt{1 - \bar{\alpha}_{t(i)}} \bm{\epsilon}^{(i)}, t^{(i)} \right) \right\|_2^2,
    \end{aligned}
\end{equation}
where $\mathcal{B}$ is the batch size, $\bm{\epsilon}^{(i)}$ is the corresponding noise sample, $\bm{\epsilon}_{\bm{\theta}}(\cdot)$ is the trained denoising network, $\bm{\theta}$ is the parameters of the denoising network, 
$\bar{\alpha}_{t(i)}$ is the preset weight coefficient,
$\bm{h}_{0}^{(i)}$ is the original channel data in the dataset and $t(i) \sim \text{Unif}\{1, \ldots, N_{\text{denoise}}\}$ indexes the hierarchical level of the latent variables, where $N_{\text{denoise}}$ is the number of denoising layers.
The sampling trajectory $\{ \tau_i\}_{i=1}^{N'}$ with $N'\ll N_{\text{denoise}}$ is introduced in the generation process to accelerate the sampling process. The calculation formula for the ground truth of the channel estimation is as follows:
\begin{equation}
    \hat{\bm{h}}_{\bm{\theta}}(\bm{h}_{\tau_i}, \tau_i) = \frac{1}{\sqrt{\bar{\alpha}_{\tau_i}}} \left( \bm{h}_{\tau_i} - \sqrt{1 - \bar{\alpha}_{\tau_i}} \, \bm{\epsilon}_{\bm{\theta}}(\bm{h}_{\tau_i}, \tau_i) \right),
\end{equation}
where $\hat{\bm{h}}_{\bm{\theta}}(\bm{h}_{\tau_i}, \tau_i)$ is the updated CSI estimation in timestep $\tau_i$ and $\bm{h}_{\tau_i}$ is the latent variable sampled by the forward diffusion process. The model architecture of the diffusion-based network is shown in Fig. \hyperref[a-3-3]{11(c)}. The complexity of diffusion model is
$\mathcal{O}(TN_{\text{denoise}} d_{\text{denoise}})$, where $T$ is the input size and $d_{\text{denoise}}$ is the denoise network dimension. Although diffusion models stand out for their high-quality sample generation and environmental alignment through conditional injection for complex NTN environments, their relatively slow inference speed is a barrier for further applications in dynamic scenarios.

$\textbf{Summary and Insights: }$
Beyond single-architecture designs, hybrid AI models combine complementary strengths from multiple models to tackle especially challenging scenarios. A promising direction is to leverage diffusion models for high-fidelity statistical CSI estimation to provide informative priors for channel prediction, and then employ fast inference models such as transformers to perform real-time instantaneous CSI estimation and forecasting.

The CSI estimation in FA-assisted NTNs still face several challenges, such as data scarcity, resource load constraints, and robustness against complex and dynamic environments. Moreover, customizing loss functions for the unique signal characteristics of FAs can enhance system performance, such as incorporating complex-valued architectures. For higher levels in the future, the function of AI models will scale to a more general framework \cite{81}, integrating CSI estimation and prediction, 3-D UAV positioning, as well as signal detection and classification, which furnish prior information for decision-making. Furthermore, the development of continual learning and adaptive learning algorithms can enhance robustness and adaptability, while the adoption of FL techniques contributes to protecting data privacy and security.
\vspace{-1ex}
\section{Joint Optimizations of FA-Assisted NTNs}\label{advance}
In this section, we present a comprehensive overview of the joint optimization strategies of FA-assisted NTNs, focusing on better data transmission and interference mitigation. Existing works can be mainly investigated in three scenarios: hovering UAVs with static scenarios, mobile UAVs with dynamic scenarios and satellite networks. The influencing parameters include FA activated ports, FA positions, adaptive beamforming, power consumption, and motion modes of NTN platforms. Some representative works are summarized in Table \ref{joint optimization}.

\begin{table*}[htbp!]
\caption{A Summary of Representative Works on Joint Optimizations of FA-Assisted NTNs
}
\centering
\resizebox{\textwidth}{!}{
\begin{tabular}{p{1cm}|p{2cm}|p{2.5cm}|p{1.5cm}|p{1cm}|p{8cm}|p{4cm}}
\hline
\multicolumn{1}{c|}{\textbf{Ref.}} & \multicolumn{1}{c|}{\textbf{NTN Platform}} & \multicolumn{1}{c|}{\textbf{System Setup}} & \multicolumn{1}{c|}{\textbf{Transmitter}} & \multicolumn{1}{c|}{\textbf{Receiver}} & \multicolumn{1}{c|}{\textbf{Variables}} & \multicolumn{1}{c}{\textbf{Indicators}} \\ 
\hline
\hline
\makecell{\cite{82}} & \makecell{Hovering UAV} 
& \makecell{Uplink, SIMO} & \makecell{FA} & \makecell{FA} 
& \makecell{Transmit \& receive FA positions, transmit power} & \makecell{Sum capacity}
\\ \hline
\makecell{\cite{83}} & \makecell{Hovering UAV} 
& \makecell{Uplink, SIMO} & \makecell{FPA} & \makecell{FA} 
& \makecell{Receive FA positions, transmit power, receive beamforming} & \makecell{Minimum rate}
\\ \hline
\makecell{\cite{84}} & \makecell{Hovering UAV} 
& \makecell{Uplink, SIMO} & \makecell{FPA} & \makecell{FA} 
& \makecell{Receive FA positions, receive beamforming} & \makecell{Minimum rate}
\\ \hline
\makecell{\cite{85}} & \makecell{Hovering UAV} 
& \makecell{Uplink, MIMO} & \makecell{FA} & \makecell{FA} 
& \makecell{Transmit \& receive FA positions, transmit beamforming} & \makecell{Achievable rate}
\\ \hline
\makecell{\cite{57}} & \makecell{Hovering UAV} 
& \makecell{Uplink, MIMO} & \makecell{FA} & \makecell{FA} 
& \makecell{Transmit \& receive FA positions, transmit beamforming} & \makecell{Achievable rate}
\\ \hline
\makecell{\cite{51}} & \makecell{Hovering UAV} 
& \makecell{Uplink, MISO} & \makecell{FA} & \makecell{FA} 
& \makecell{Transmit \& receive FA positions, transmit \& receive\\beamforming, transmit power} & \makecell{Achievable rate}
\\ \hline

\makecell{\cite{87}} & \makecell{Hovering UAV} 
& \makecell{Downlink, MISO} & \makecell{FA} & \makecell{FPA} 
& \makecell{Transmit FA positions, transmit beamforming} & \makecell{Sum-rate}
\\ \hline
\makecell{\cite{88}} & \makecell{Hovering UAV} 
& \makecell{Downlink, MISO} & \makecell{FA} & \makecell{FA} 
& \makecell{Transmit \& receive FA positions, transmit beamforming} & \makecell{Minimum rate}
\\ \hline
\makecell{\cite{89}} & \makecell{Hovering UAV} 
& \makecell{Downlink, MISO} & \makecell{FA} & \makecell{FPA} 
& \makecell{Transmit FA positions, transmit beamforming} & \makecell{Sum-rate}
\\ \hline
\makecell{\cite{61}} & \makecell{Hovering UAV} 
& \makecell{Downlink, MIMO} & \makecell{FA} & \makecell{FA} 
& \makecell{Transmit \& receive FA positions, transmit beamforming} & \makecell{Achievable rate}
\\ \hline
\makecell{\cite{90}} & \makecell{Hovering UAV} 
& \makecell{Downlink, MIMO} & \makecell{FA} & \makecell{FA} 
& \makecell{Transmit \& receive FA positions, transmit beamforming} & \makecell{Achievable rate}
\\ \hline
\makecell{\cite{58}} & \makecell{Hovering UAV} 
& \makecell{Downlink, MIMO} & \makecell{FA} & \makecell{FA} 
& \makecell{Transmit \& receive FA positions} & \makecell{Channel capacity}
\\ \hline
\makecell{\cite{91}} & \makecell{Hovering UAV} 
& \makecell{Downlink, MIMO} & \makecell{FA} & \makecell{FA} 
& \makecell{Transmit \& receive FA positions, transmit beamforming} & \makecell{Sum-rate}
\\ \hline

\makecell{\cite{94}} & \makecell{Mobile UAV} 
& \makecell{Downlink, MISO} & \makecell{FA} & \makecell{FPA} 
& \makecell{UAV position, transmit FA positions, transmit beamforming} & \makecell{Minimum beamforming gain}
\\ \hline
\makecell{\cite{21-n}} & \makecell{Mobile UAV} 
& \makecell{Downlink, MISO} & \makecell{FA} & \makecell{FPA} 
& \makecell{UAV position, transmit FA positions, transmit beamforming} & \makecell{Minimum rate}
\\ \hline
\makecell{\cite{95-n}} & \makecell{Mobile UAV} 
& \makecell{Relaying Network} & \makecell{FPA} & \makecell{FPA} 
& \makecell{UAV position, relaying FA rotation angles, beamforming} & \makecell{Minimum SNR}
\\ \hline
\makecell{\cite{96}} & \makecell{UAV swarm} 
& \makecell{Downlink, MISO} & \makecell{FA} & \makecell{FPA} 
& \makecell{UAV positions, transmit FA positions, transmit beamforming} & \makecell{Minimum rate}
\\ \hline
\makecell{\cite{18}} & \makecell{Mobile UAV} 
& \makecell{Downlink, SIMO} & \makecell{FPA} & \makecell{FA} 
& \makecell{Receive FA positions, receive beamforming, \\base station selection} & \makecell{Maximum SINR}
\\ \hline
\makecell{\cite{18-n}} & \makecell{Mobile UAV} 
& \makecell{Uplink, SIMO} & \makecell{FPA} & \makecell{FA} 
& \makecell{UAV positions, receive FA positions, \\transmit \& receive beamforming} & \makecell{Sum-rate}
\\ \hline
\makecell{\cite{17}} & \makecell{Mobile UAV} 
& \makecell{Downlink, MISO} & \makecell{FA} & \makecell{FPA} 
& \makecell{UAV positions, transmit FA positions, transmit beamforming} & \makecell{Sum-rate}
\\ \hline
\makecell{\cite{99}} & \makecell{Mobile UAV} 
& \makecell{Relaying Network} & \makecell{FPA} & \makecell{FPA} 
& \makecell{UAV positions, relaying FA positions,\\transmit \& receive beamforming of UAV, time slot allocation} & \makecell{Sum-rate}
\\ \hline
\makecell{\cite{100}} & \makecell{Mobile UAV} 
& \makecell{Downlink, SISO} & \makecell{FA} & \makecell{FPA} 
& \makecell{UAV positions, transmit FA rotation} & \makecell{Data collection time}
\\ \hline

\makecell{\cite{101}} & \makecell{Satellite} 
& \makecell{Downlink, SISO} & \makecell{FPA} & \makecell{FA} 
& \makecell{Receive FA position} & \makecell{Receive SNR}
\\ \hline
\makecell{\cite{102}} & \makecell{Satellite} 
& \makecell{Downlink, MISO} & \makecell{FA} & \makecell{FPA} 
& \makecell{Transmit FA positions, transmit beamforming} & \makecell{Interference leakage power}
\\ \hline
\makecell{\cite{23-n}} & \makecell{Satellite} 
& \makecell{Uplink, SIMO} & \makecell{FPA} & \makecell{FA} 
& \makecell{Receive FA positions, Receive beamforming} & \makecell{Achievable rate}
\\ \hline
\makecell{\cite{103}} & \makecell{Satellite} 
& \makecell{Uplink, SIMO} & \makecell{FPA} & \makecell{FA} 
& \makecell{Receive FA positions, activated FA ports} & \makecell{Receive SNR}
\\ \hline
\makecell{\cite{103-n}} & \makecell{Satellite} 
& \makecell{Frequency Division \\Duplex, MIMO} & \makecell{FA} & \makecell{FA} 
& \makecell{Common rate allocation, downlink transmit FA positions,\\ downlink beamforming} & \makecell{Minimum ergodic user-rate}
\\ \hline
\makecell{\cite{104}} & \makecell{Satellite} 
& \makecell{FD system, MIMO} & \makecell{FA} & \makecell{FA} 
& \makecell{Uplink receive \& downlink transmit FA positions,\\uplink transmit power, downlink beamforming} & \makecell{Total downlink \& uplink \\ transmit power}
\\ \hline
\end{tabular}}
\label{joint optimization}
\end{table*}

\subsection{FA-Assisted Hovering UAV Networks}
Hovering UAVs can serve as a aerial base station at a fixed height to provide stronger connections for terrestrial users, where FAs can be selectively deployed on the transceiver to offer higher DoFs and diversity gains for interference mitigation. However, there are two critical challenges in the joint optimization of FA positions, beamforming vectors, and power consumption. First, FA positions introduce new coupling among transmit power, transmit beamforming, and the receive beamforming. Optimizing FA positions may cause a mismatch between the channel vectors and the given receive beamforming, leading to channel gain loss and interference. A general solution is to formulate the beamforming vector as a function of FA positions for the derivation of closed-form solutions. The second one is the poor convergence properties of the conventional three-variable Alternating Optimization (AO) algorithm.

\subsubsection{Uplink Transmission}
\begin{figure}[t]
    \centering
    \includegraphics[width=0.85\linewidth]{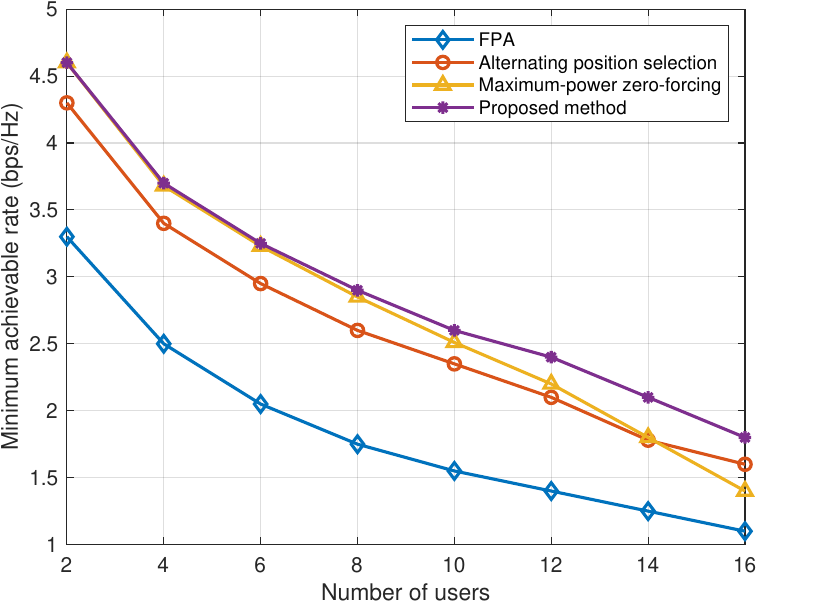}
    \caption{
    Experimental result in \cite{83}: Minimum achievable rates versus number of users in a multiuser uplink SIMO system. The base station is equipped with FAs.}
    \label{a-2-1}
    \vspace{-3ex}
\end{figure}

In uplink transmission, Tang et al. in \cite{82} investigated the multiuser-SIMO capacity gains by optimizing FA positions at the transceiver sides, where exhaustive search, majorization-minimization, and gradient descent algorithms are employed. However, beamforming, power allocation, and multiuser interference are not discussed. In \cite{83}, Xiao et al. conducted joint optimizations of the receive FA positions, the beamforming vector, and the transmit power to maximize the minimum rate in the multiuser-SIMO system. To address the poor convergence of traditional AO algorithms, a dual-loop iterative algorithm based on Particle Swarm Optimization (PSO) can efficiently derive a suboptimal solution. Figs. \ref{a-2-1} shows minimum achievable rates for different schemes versus the number of users. We can observe that the proposed scheme in \cite{83} outperforms all other benchmark schemes, including FPA, maximum-power zero-forcing, and alternating position selection schemes. 
In reality, we initially access instantaneous CSI, while the second is the real-time adjustment of FA positions. However, the updates of FA positions in every channel coherence time result in rapid instantaneous CSI changes and high mechanical latency. In \cite{84}, Hu et al. investigated the two-timescale transmission design, which consider the use of statistical CSI for FA positions adjustment in the large timescale, and instantaneous CSI for beamforming design in the small timescale.
With the activation of multiple ports at the transmitter and the receiver, FA-enabled MIMO systems can flexibly reshape the channel matrix according to varying requirements and SNRs. Specifically, under low SNR conditions, FAs reconfigure the MIMO channel matrix by adjusting FA positions and transmit covariance matrix to maximize singular values \cite{85}. In \cite{57}, Ma et al. derived an asymptotic expression for FA-MIMO channel capacity under low SNR conditions, revealing that position optimization enhances channel power while reducing the condition number of the MIMO channel matrix. In contrast, for high SNR scenarios, FAs balance these singular values to optimize power allocation across multiple eigenmodes. In \cite{51}, New et al. proposed a joint optimization algorithm based on Rank-Revealing QR (RRQR) factorization, effectively maximizing the system rate under high SNR conditions. Additionally, they rigorously established the Diversity-Multiplexing Tradeoff (DMT) for FA-enabled MIMO systems, demonstrating its superior efficiency compared to conventional MIMO systems.

\subsubsection{Downlink Transmission}
In downlink transmission, Cheng et al. in \cite{87} proposed an FA-enabled multiuser-MISO system. They presented a pair of efficient algorithms, leveraging the principles of AO, gradient descent, and backtracking line search methods. The aforementioned studies, while presenting the superior aspects of FAs, often assume ideal hardware at both the transmitters and the receivers. However, both transmitters and receivers are inevitably subject to hardware impairments that introduce additional distortion components. In \cite{88}, Yao et al. highlighted the potential of FAs to mitigate hardware impairments and improve the achievable rate with a low transmit power. In \cite{89}, Zhang et al. outlined the high costs of equipping mobile motors at each antenna and digital signal processing unit. They proposed a sub-connected FA array scheme driven by a single motor, enabling independent movement of each sub-array within a predefined area. By jointly optimizing hybrid beamforming and sub-connected array positions, the sum-rate can be further improved than fully-connected FPA arrays. Considering FA-enabled MIMO systems, Chen et al. in \cite{61} proposed a Constrained Stochastic Successive Convex Approximation (CSSCA) algorithm and two simplified antenna port switching schemes (i.e., linear and planar) to optimize port selections and beamforming. In \cite{90}, quantum computing tools were employed to maximize the SNR through joint optimization of FA positions at both ends. In \cite{58}, a simple FA-enabled MIMO architecture was introduced, with joint convex relaxation and AO of transmitter and receiver FA positions to maximize capacity. To simplify the optimization of highly coupled variables introduced by FA-enabled MIMO system, Weng et al. in \cite{91} proposed a DRL framework based on heterogeneous Multi-Agent Deep Deterministic Policy Gradient (MADDPG) to learn beamforming and transceiver FA movement strategies. Moreover, when there are multiple obstacles during propagation, FAs can be deployed as relay nodes on the UAV or both sides of the obstacles to improve the throughput \cite{92, 93}.

$\textbf{Summary and Insights: }$
Current research primarily centers on optimizing FA positions, beamforming vectors, and power allocation with a fixed number of antennas. The number and placement of FAs are critical for system performance: the number of FAs can shape the dimension of the channel matrix and the diversity gain, whereas their positions can redefine the channel eigenvalues and fading states. In NTNs, determining the optimal number, positions, and rotations of activated FAs is highly complex due to diverse user requirements, resource overhead, and varying channel conditions. A DRL framework can be leveraged to establish a general paradigm for addressing the aforementioned optimization problems. On the other hand, a large movable region requires effective FA movement strategies. Nevertheless, there is less literature that considers designs of FA structure and mobility schemes.

\begin{figure}[!t]
\centering
\subfloat[]{\includegraphics[width=0.85\linewidth]{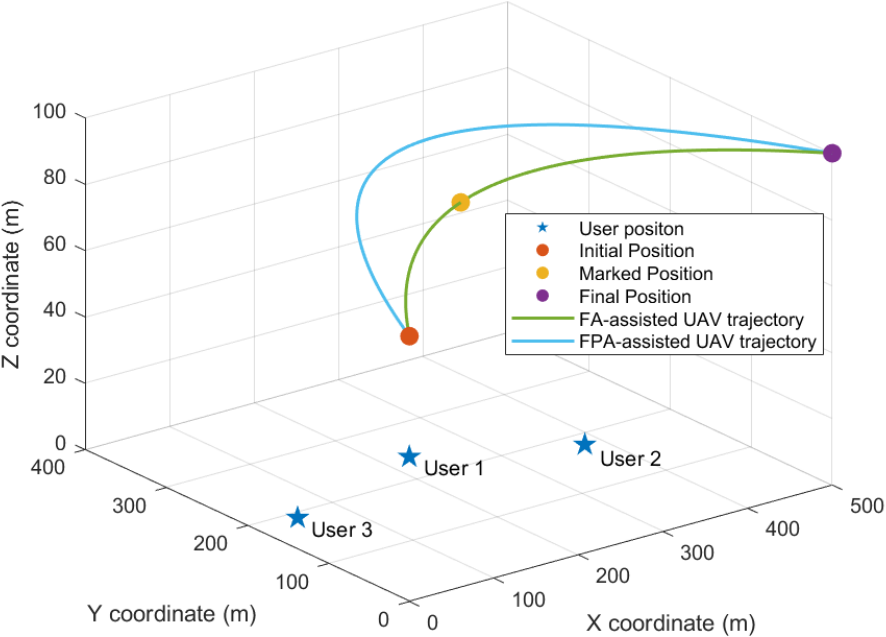}%
\label{uav-1}}
\quad
\subfloat[]{\includegraphics[width=0.85\linewidth]{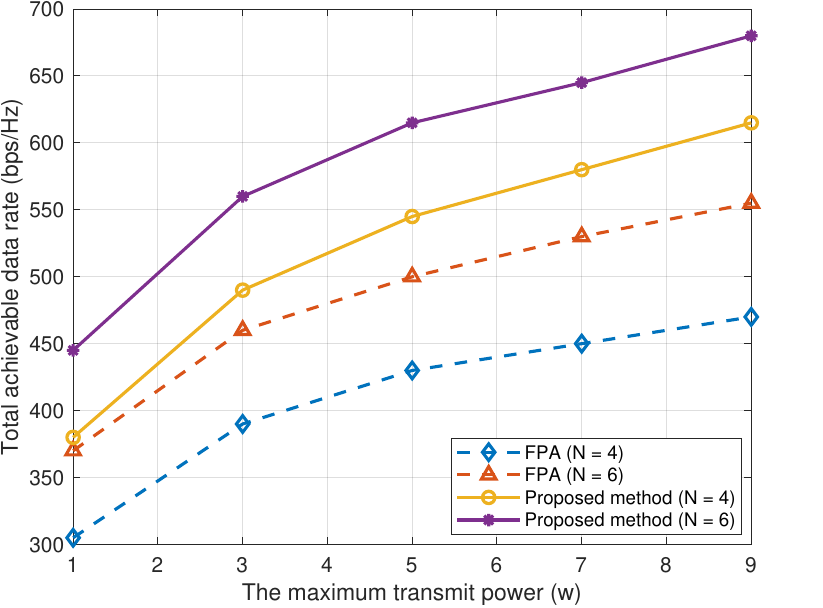}%
\label{uav-2}}
\quad
\subfloat[]{\includegraphics[width=0.82\linewidth]{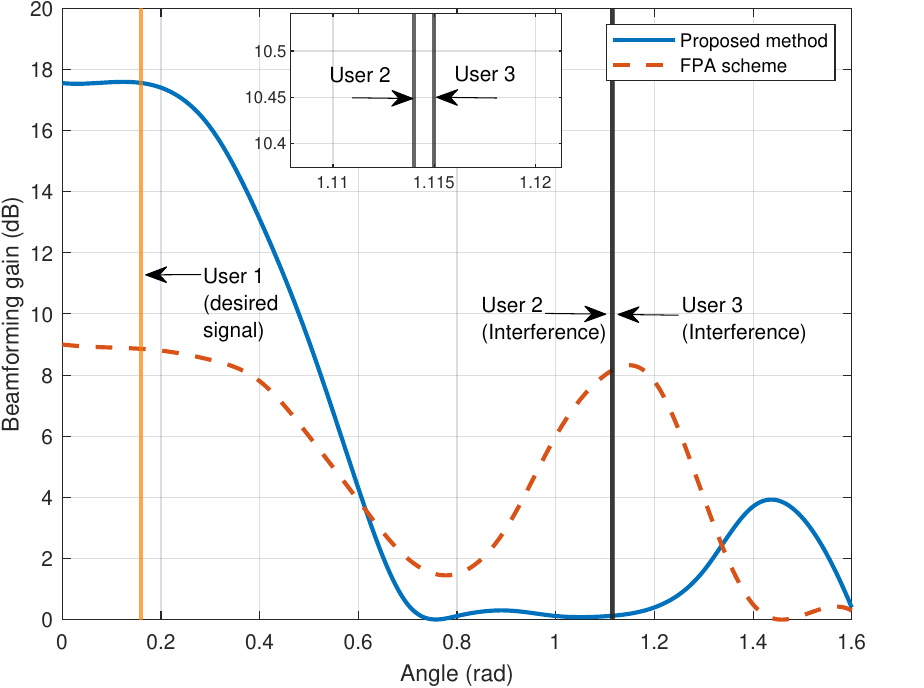}%
\label{uav-3}}
\caption{Experimental results in \cite{17}: (a) 3D UAV trajectory of the proposed algorithm; (b) The achievable data rate versus transmit power. (c) Comparison of beamforming gains with FA and FPA array.}
\vspace{-2ex}
\label{fig:uav}
\end{figure}

\subsection{FA-Assisted Mobile UAV Networks}
Mobile UAV networks can provide a broader optimization space but result in more complex system models and variable coupling, including FA positions, beamforming vectors, power allocations and the positions of the UAVs. In \cite{94}, Tang et al. assumed that an FA-assisted UAV base station can adjust its altitude in vertical height, and design a three-variable AO algorithm to optimize the UAV altitude, FA positions, and beamforming vectors to reduce inter-user interference. In addition to reducing distance-dependent path loss, UAV altitude and FA positions are designed to enhance phase-sensitive beamforming. 
In \cite{21-n}, Mao et al. targeted the uncertainty in AoD caused by UAV jitter and proposed a robust framework assisted by FAs for multiuser-downlink transmission to maximize the minimum rate. The approach consists of two stages: first, iteratively optimizing the deployment position and power under ideal beam patterns; then, combining cutting-set with Successive Convex Approximation (SCA) to iteratively optimize the beamforming and FA positions. In \cite{95-n}, Liu et al. studied a 6DMA-assisted UAV relaying system and jointly optimized the UAV placement, 6DMA rotation angles, and beamforming to maximize the minimum received SNR across users. They develop an AO framework combining SCA with multidimensional search, and incorporate Gibbs sampling to mitigate convergence to poor local optima. Considering the potential of FA-assisted UAV swarms, Lu et al. in \cite{96} integrated FAs into a UAV swarm for a two-level FA system, where each UAV has a local FA array and the swarm collaborates to form a large-scale distributed FA system. The SCA algorithm is used to alternately optimize the UAV positions, the FA local position, and the receive beamforming for the minimum rate maximization.
When operating as a terminal in a large-scale cellular network, UAVs can jointly optimize 6DMA positions and rotations, as well as UAV positions to maximize the SINR at each available base station \cite{18}. After obtaining the maximum SINR from all accessible base stations, the UAV may select the optimal base station with the highest SINR for communication, thus mitigating multi-cell interference.

Moreover, due to nearly unrestricted 3-D movement of UAVs, UAV trajectories may lead to complex and dynamic channel environments. Optimizing the positions of UAVs across different time slots can achieve trajectory optimization over a specified period. In \cite{18-n}, Zhang et al. proposed a optimization framework of FA-assisted UAVs to maximize the sum rate in the uplink data collection. They addressed UAV trajectory through SCA, updated beamforming and power through AO, and applied PSO for optimal FA positioning. Simulations demonstrate that the proposed algorithm achieves faster convergence and higher sum rate than the fixed UAV trajectory and FPA scheme. In downlink data transmission, Liu et al. in \cite{17} utilized an FA-assisted UAVs to maximize the sum rate. The FA positions, transmit beamforming, and the UAV trajectory are alternatively optimized through SCA. Fig. \hyperref[uav-1]{13(a)} shows the optimized FA-assisted UAV trajectory between users to achieve better energy efficiency compared to the FPA scheme. Fig. \hyperref[uav-2]{13(b)} illustrates the total achievable data rate versus the maximum transmit power when UAV is deployed at the marked position in \hyperref[uav-1]{13(a)}. Fig. \hyperref[uav-1]{13(c)} shows the beamforming gain of FAs and FPAs in different directions. It can be observed that FPAs cannot completely mitigate interfering signals and may lose array gain over the desired signal due to the fixed positions of radiating elements. In contrast, FAs can achieve full array gain of the target signal while maintaining suppression of interfering users.
In \cite{99}, Zhou et al. proposed an FA-assisted UAV relaying framework to maximize the throughput by optimizing beamforming, time slot allocation, FA positions, and UAV trajectory. A penalty-based AO algorithm is used to solve the coupled optimization variables and the rank-one constraints.
From the perspective of ML methods, Bai et al. in \cite{100} minimize the data collection time of FA-assisted UAVs through a DRL framework. The state space includes the UAV position, data source label, and relative positions between the data source and the UAV, while the action space comprises the UAV trajectory and FA rotation angles. A soft actor-critic network is employed for stable training under constraints of communication rate, FA angle, and energy.

$\textbf{Summary and Insights: }$
Current research mainly focuses on optimizing existing communication systems. For seamless access to NTNs, it is worthwhile to investigate the potentials of FAs in enabling UAVs across diverse tasks. For example, during a flight trajectory from a given point to a destination, an FA-assisted UAV performs different tasks, such as communication, sensing, and computation. Moreover, The UAV power consumption models should be considered in more detail, such as basic endurance, UAV mobile consumption, transmit power, hardware impairments, and FA mobile consumption. 
Besides, delay-sensitive models in FA-assisted NTNs are important for practical deployment including the signal propagation delay, CSI estimation delay, and antenna movement delay. Furthermore, FAs can promote collaboration among UAV swarms, satellite networks, and diverse NTN platforms at varying altitudes for overall reliability and performance improvement.


\begin{figure}[t]
    \centering
    \includegraphics[width=0.9\linewidth]{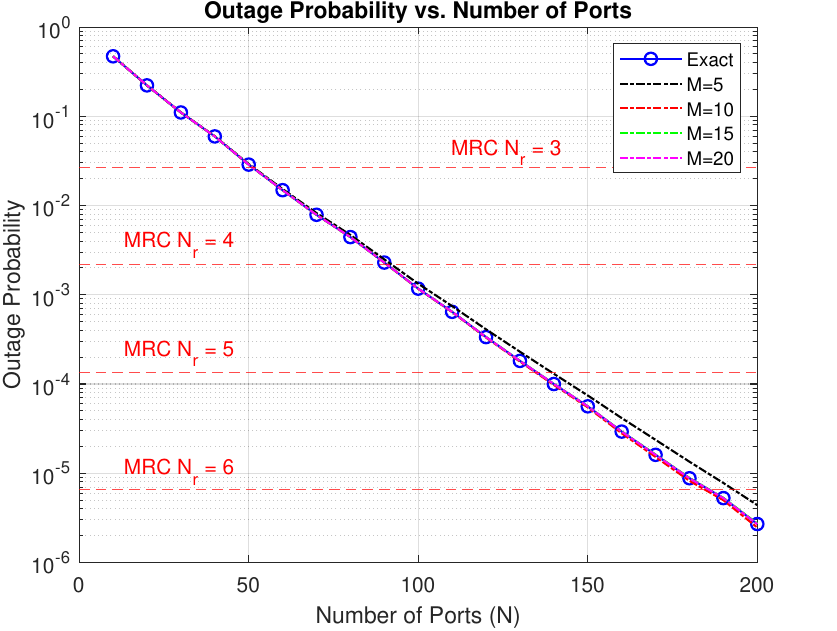}
    \caption{
    Experimental results in \cite{101}: Outage probability versus the number of FA ports for the FA-OTFS system in the single-LoS channel model (exact vs. Gauss-Hermite quadrature with $M$ nodes).}
    \label{fig:satellite-2}
    \vspace{-2ex}
\end{figure}

\subsection{FA-Assisted Satellite Networks}
The main challenges of satellite networks include highly dynamic Doppler shifts, propagation delays, scarce spectrum, significant path loss, and complex fading. To mitigate Doppler impairments, Yang et al. in \cite{101} proposed an FA-OTFS system for high-mobility LEO IoT links, where a satellite with an FPA transmits the OTFS signal to a terrestrial user with a single FA. They first developed an A2G delay-Doppler domain FA-OTFS channel model with a dominant LoS component and a few scattered paths. Based on this model, the outage probability was derived in closed form. Fig. \ref{fig:satellite-2} demonstrates the outage probability versus the number of FA ports in the LoS-only satellite channel. As expected, increasing $N$ significantly reduces the outage probability, confirming substantial spatial diversity gains provided by multiple ports. Furthermore, FAs can have a better performance compared to the benchmark maximal-ratio combining (MRC) scheme using $N_r = 6$ receive FPA. These findings reinforce the effectiveness of FA-OTFS in high-mobility satellite environments.


In LEO satellite networks, FPAs are affected by limited beam directivity and signal sidelobe leakage, resulting in inter-satellite interference and reduced capacity. This challenge is further amplified by orbital motion, as the coverage and interference regions evolve continuously, making static beamforming of FPAs insufficient for interference suppression. To address this issue, Zhu et al. in \cite{102} proposed an FA-assisted beam-coverage model based on a typical Walker-Delta constellation and developed an AO algorithm of FA positions and beamforming designs, which achieves target direction array gain while minimizing average leakage power and hardware overhead. In contrast to satellite-side FA structures, deploying FAs at the ground station can also mitigate inter-satellite interference. Wang et al. in \cite{23-n} constructed an A2G channel based on a ground-station-centric Cartesian coordinate system, where the FA-equipped ground station receives concurrent transmissions from multiple satellites. The model includes a dominant LoS component and accounts for directional antenna gains. They apply an SCA-based AO scheme to jointly optimize FA positions and beamforming vectors for maximizing the achievable rate of the ground station.

\begin{figure}[t]
    \centering
    \includegraphics[width=0.85\linewidth]{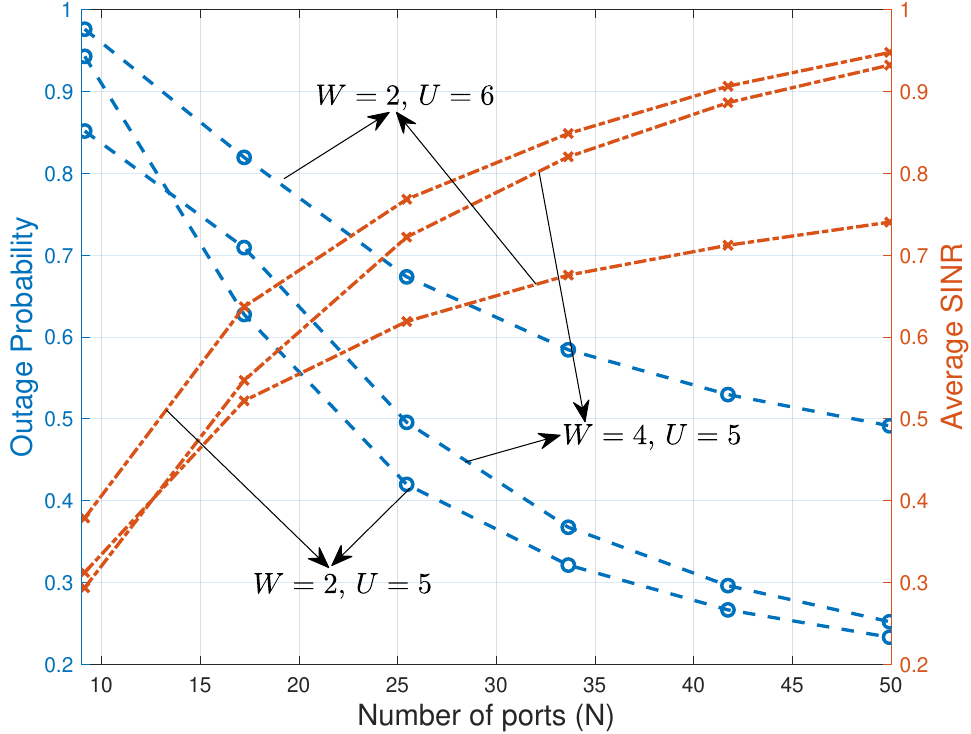}
    \caption{
    Experimental result in \cite{103}: Outage probability and average SINR against the number of FA ports $K$ under different FA size $W$ and user number $U$.}
    \label{satellite-3}
    \vspace{-2ex}
\end{figure}

In \cite{103}, Han et al. introduced a novel multiuser uplink reception scheme through a compact FA array at the satellite driven by a single RF chain while each user is equipped with a FPA. Unlike selecting FA ports with the strongest channel gains, satellite activates multiple channel-aligned ports to coherently combine the desired signal, while interference signals treated as noise are added randomly. Simulations reveal that the compact FA array can better identify port sets with specific signs of in-phase and quadrature components under LoS satellite channels, achieving lower outage probability and higher ergodic rate than the MRC scheme. Fig. \ref{satellite-3} illustrates that as the number of FA ports $K$ increases, the average SINR improves and the outage probability decreases. Furthermore, for a fixed number of FA ports $K$, increasing FA size $W$ leads to a decrease in SINR, resulting in degraded performance. This shows that FAs must have a sufficient number of ports in a limited space to achieve significant diversity gain.

In satellite communications, time synchronization errors disrupt pilot and symbol alignment, increasing CSI estimation errors and bit error rates. Orbital uncertainties introduce propagation delays and Doppler prediction errors, leading to Doppler pre-compensation bias at the terminal and exacerbating the CSI inaccuracy \cite{27}. To improve the robustness of FA-satellite networks using imperfect CSI, Zhang et al. in \cite{103-n} proposed an FA-Rate-Splitting Multiple Access (RSMA)-assisted satellite network that leverages FA position reconfigurability and RSMA’s robustness to imperfect CSI for interference management. The channel is modeled by a Shadowed-Rician process with the imperfect CSI induced by high mobility and orbital uncertainties. A Conditional Mamba solver with unsupervised transfer learning is developed to jointly optimize common-rate allocation, FA port positions, and beamforming for maximum ergodic user rate in dynamic environments. Moreover, FD technology acts as a promising approach to deal with the spectrum scarcity issue in satellite communications. Nevertheless, it poses new in-band self-interference challenges. In \cite{104}, Lin et al. introduced an FA-assisted LEO FD communication system, where the receive signals suffer from co-channel interference from multiple users and self-interference from the FD system. They apply a PSO-based AO algorithm of FA positions, beamforming, and power allocation under imperfect CSI to reduce total downlink and uplink transmit power.

$\textbf{Summary and Insights: }$
Most existing studies focus on communication between a single satellite and terrestrial users. However, future work should extend to multi-satellite cooperative transmission and dual-FA deployments at both satellite and terrestrial segments. On the one hand, multi-satellite cooperative transmission can form a large-scale distributed FA array to better utilize the diversity gain of the space channel and reduce the hardware load of each satellite. Moreover, FAs present the potential to enhance the stability of inter-satellite links and increase throughput, motivating a systematic investigation of FA-assisted satellite constellations. On the other hand, deploying FA on both the terrestrial and satellite sides can provide stable beam coverage in dynamic environments according to user requirements. Meanwhile, RIS can further refine the coverage accurately.

\section{Compatibility with Emerging B5G Technologies}
\label{b5g}
Leveraging unique flexibility and compatibility, FA-assisted NTNs present great potential in combination with advanced B5G technologies for better performance. In this section, we comprehensively investigate the technologies assisted by FA-assisted NTNs to achieve joint performance enhancement on the original basis, including CF-mMIMO, FD communication, NGMA, and RIS.

\subsection{CF-mMIMO}
UAV networks are the core scenarios of NTNs and CF-mMIMO. In existing terrestrial cellular networks, UAV uplink signals are easily received by multiple neighboring cells. However, this can also interfere with the UAV's received signals. Furthermore, the low-altitude sidelobe coverage triggers frequent handovers and overshoot issues, increasing the complexity of network management. The key enabling technology for UAV networks to overcome these challenges is the CF-mMIMO system. In CF-mMIMO networks, each user is treated as a service-centric entity, with coordinated distributed antenna units for stable signals and personalized services \cite{105, 105-n3, 106-n3}. This approach promises significant enhancements in spectral efficiency, interference mitigation, and energy efficiency. 

Despite the significant potential of CF-mMIMO to enhance communication performance in UAV networks, it also faces significant challenges. In CF-mMIMO systems, resource allocation must consider distributed antenna deployment, complex precoding, user mobility, and different service demands, making a complex optimization algorithm design. For multiuser scenarios, the APs require timely and accurate signal detection and joint processing, but insufficient uplink rates may induce interference, degrading system performance. Also, during the downlink transmission, APs need to transmit data symbols of different users over the same time-frequency resource through beamforming design. Traditional FPAs struggle to address these issues, as they are constrained by limited channel spatial flexibility. FA-assisted UAV networks exploit flexible FA-assisted APs and aerial base stations with antenna position optimization, beamforming design, UAV trajectory coordination, and power allocation to maximize achievable data rates. 

For instance, Olyaee et al. in \cite{106} proposed a joint optimization algorithm for FA positions and uplink power allocation to enhance transmission rates and signal quality. Due to the assumptions of perfect CSI and discrete FA positions, the sub-problem can be solved applying the bisection method and exhaustive search algorithms.
In \cite{107}, Shi et al. adjusted the rotation angles of 6DMA antennas in distributed APs to maximize the average user sum rate. They developed a Bayesian optimization-based algorithm to tackle the non-convex and highly nonlinear nature of rotation angle optimization, flexibly adapting to user spatial distributions. 
In \cite{22-n}, Wang et al. proposed a novel aerial 6DMA-enabled CF-mMIMO network where multiple UAVs serve as APs and jointly optimize 3D UAV placement and 6DMA rotation for uplink reception. UAV deployment enhances coverage and LoS conditions, while 6DMA rotation enables fine-grained beam alignment and interference suppression. To reduce complexity and CSI exchange, they adopt a team MMSE combiner with partial CSI sharing and use multi-agent RL to learn the positioning and rotation policies. Simulations demonstrate significant uplink rate gains over FPA baselines with stable convergence.
Furthermore, FA-assisted UAVs can enhance CF-mMIMO by releasing the urgent need for perfect CSI and simplifying the precoding scheme at the transmitter with FAMA in multiple access channels.

$\textbf{Summary and Insights: }$
In the future, providing high data rates to the users with different mobility conditions  (e.g., terrestrial mobile users and aerial UAV swarms) is one of the major challenges in CF-mMIMO systems. In dynamic scenarios, the relative motion between users and APs induces temporal variations of channel coefficients and Doppler shifts, leading to channel estimation errors and outdated in uplink/downlink transmission. It is worthwhile to study obtaining accurate and up-to-date CSI for FAs in CF-mMIMO systems. FAs present huge potential in channel hardening and Doppler effect mitigation by adjusting FA positions for channel self-adaptation and flexible beamforming. Another challenge is realtime massive access under constrained network capacity. The traditional CF-mMIMO systems require that all APs share identical time-frequency resources to serve all users, while the fronthaul network capacity between the CPU and APs is constrained. Besides, data transmission is often bursty and demands ultra-low latency in NTNs. Therefore, it is a promising research direction to jointly optimize FA configurations and resource allocation of APs for fronthaul network capacity improvement with low computational complexity.

\subsection{FD Communication}
Facing the scarcity of spectrum resources for non-terrestrial networks, FD communication is a promising technology that enhances network data rates and spectral efficiency. Based on non-paired spectrum, FD allows simultaneous downlink and uplink resource usage, achieving double spectrum utilization and reduced latency. However, the self-interference cancellation is the core topic of FD communications. For conventional FD systems based on FPAs, existing self-interference cancellation techniques can be broadly classified into passive suppression and active suppression. Passive suppression is typically implemented at the antenna/hardware design without relying on accurate CSI, and reduces self-interference through spatial separation, polarization diversity, directional antennas, and beam alignment \cite{107-n}. Specifically, increasing the distance between transmit and receive ports, optimizing their relative orientations, or exploiting polarization mismatch can improve isolation to reduce self-interference. However, such approaches often require larger antenna footprints, higher hardware cost, or more complex structures. On the other hand, active suppression operates after CSI acquisition and minimize the total transmit power through the joint optimization of beamforming, power allocation, and multiple-access strategies. Nevertheless, FPAs lack spatial flexibility, which limits the interference mitigation capability of analog and digital processing modules to cancel interference signals.

The main advantage of FAs is to exploit spatial DoFs within a constrained space. By reconfiguring FA positions/rotations, FAs can improve transmit-receive isolation in limited space while preserving the diversity gains of multi-antenna operation. In addition, antenna rotation can modify the relative polarization between transmitting and receiving links, thereby reducing interference coupling \cite{25}. Furthermore, FA reconfigurability introduces new optimization variables into system design, enabling the joint optimization of transmit/receive FA positions, beamforming, and power control in FD systems to provide greater flexibility for interference management. Specifically, Lin et al. in \cite{104} propose a two loop PSO algorithm based on a multiobjective optimization framework for FA Array-assisted FD satellite communication systems to minimize transmit power under imperfect CSI, while also decrease the burden on self-interference cancellation. In \cite{108}, a point-to-point FA-enabled Co-frequency Co-time FD (CCFD) system was introduced to jointly reduce self-interference. To solve the suboptimal solutions of traditional AO algorithm, Ding et al. in \cite{108} focused on the Projected Particle Swarm Optimization (PPSO)-based solution. In \cite{109}, Skouroumounis et al. evaluated the performance of FA-assisted FD systems in large-scale cellular networks under channel estimation errors and residual loop interference. They also analyzed the impact of key parameters, including port number, spatial density, and transmit power. In \cite{110}, a cooperative communication strategy with a stochastic geometry framework was proposed to enhance the performance of multi-user FD-Non-Orthogonal Multiple-Access (NOMA) networks, where the users are divided into near and remote users with FAs. In \cite{111}, Ding et al. exploit the spatial DoFs of FA-assisted FD base station to optimize beamforming and antenna positions at FD base stations for sum of secrecy rates maximization in secure communication. A Multi-Velocity Particle Swarm Optimization (MVPSO) is proposed, which is an improved version of the standard PSO, to simultaneously optimize all MA positions.

In recent years, FD-ISAC systems have attracted attention due to their efficient utilization of spectrum resources. Especially in the NTNs, the FD-ISAC system not only transmits data signals for uplink and downlink users, but also realizes the sensing functions such as positioning, speed measurement, and collision warning, making it highly promising for efficient UAV communication and positioning. In \cite{112}, Li et al. introduced a novel FA-assisted FD-ISAC base station where the transmit and receive FAs operate simultaneously: the transmitter sends data/sensing signals to downlink users, while the receiver collects uplink data and target echoes, thereby achieving higher spectral efficiency than half-duplex or simplex operation. Under minimum SINR constraints for uplink communication, downlink communication, and radar sensing, the authors minimize the total base-station transmit power by jointly optimizing the transmit/receive FA positions, beamforming, sensing waveform, and uplink transmit power through binary particle swarm optimization. Looking forward, FD-ISAC systems with a unified beam for ISAC remain underexplored, where the joint optimization is highly complex, and AI-based methods may provide a promising solution.

$\textbf{Summary and Insights: }$
In the future multiuser FD communication systems, the interference challenges encompass self-interference, sub-band interference, and adjacent-channel interference. Considering the analytical complexity, current research mainly focuses on designing mitigation algorithms for a single interference type. In NTN-FD systems, the joint optimization of UAV positions, transceiver FA positions, transmit beamforming, receive combining matrices, and power allocation can improve multi-type interference mitigation under dynamic channels. Beyond multiuser systems that employ beamforming to suppress interference, FD systems require higher power than received signal power to mitigate self-interference, so that the operational overhead in NTNs is an essential constraint. Furthermore, FD systems are scaling from single-function toward multifunction, such as FD-ISAC systems and FD-AirComp systems. However, the multiple optimization objectives and constraints engender profound algorithmic complexity, where DRL exhibits a promising solution for addressing optimization decisions in dynamic environments.

\subsection{NGMA}
To facilitate massive communications in 6G, the adoption of NGMA is evidently crucial. One avenue of exploration involves examining the interplay between existing multiple access methods such as NOMA and FAMA.

\subsubsection{NOMA}
NOMA allows data transmission on the same frequency at different power levels, which can maximize the utilization efficiency of subcarrier resources at receivers. With superposition coding and effective Serial Interference Cancellation (SIC), NOMA achieves higher spectrum sharing efficiency than Orthogonal Multiple Access (OMA) when CSI is available. However, several challenges remain in the dynamic NTN channel conditions: the accurate CSI estimation, the optimal SIC order, efficient power allocation schemes, fast and reliable user pairing, as well as interference and fading mitigation. Furthermore, due to the limited DoFs and channel flexibility of FPAs, NOMA cannot effectively distinguish the power differences between different users.

FA-assisted NTNs significantly enhance the performance of NOMA. Firstly, NTN platforms such as LAPs and HAPs can be used to establish robust LoS links which offer NOMA a promising opportunity for massive connectivity and larger coverage. Secondly, if FA is deployed at the receivers, it may be more efficient to allocate equal DoFs and power allocation for different users when the CSI at the transmitter is unavailable. Specifically, In \cite{110}, Tlebaldiyeva et al. derived the outage probability for the FA receiver in the FD cooperative NOMA system. Nevertheless, the impact of FA on NOMA capability remains unexplored. In \cite{56}, New et al. compared FA-OMA and FA-NOMA frameworks by analyzing the influence of SNR, port numbers, and user numbers on performance. They also discussed that FA without CSI at the transmitter may outperform the optimal scheme of traditional antennas with CSI. In \cite{113}, an FA-assisted UAV framework was proposed to mitigate small-scale and large-scale fading for NOMA system, where the UAV position, FA port selection, and power allocation are optimized to maximize the total rate of terrestrial NOMA users. In \cite{114}, Gao et al. introduced an FA-assisted wireless-powered communication networks utilizing NOMA for uplink data transmission. To maximize the throughput of NOMA systems, the continue/discrete MA positions, the time allocation, and the uplink power allocation were jointly optimized. Notably, the optimum for wireless power and wireless information transmission is achieved using identical MA positions.

\subsubsection{FAMA}
In \cite{16}, Wong et al. proposed a novel multiple access scheme unique to FAs called FAMA to meet the requirements for massive access within the limited spectrum. During signal propagation in free space, FAs are expected to exploit the deep fading of interference signals by FA position selections of the maximum SINR. Unlike traditional multiple access techniques, FAMA does not require CSI at the transmitters for complex precoding and SIC at the receivers. Instead, it directly analyzes the CSI of both desired and interfering signals at the receiver, adjusting the channel conditions to reduce interference.

\begin{figure}[t]
    \centering
    \includegraphics[width=0.85\linewidth]{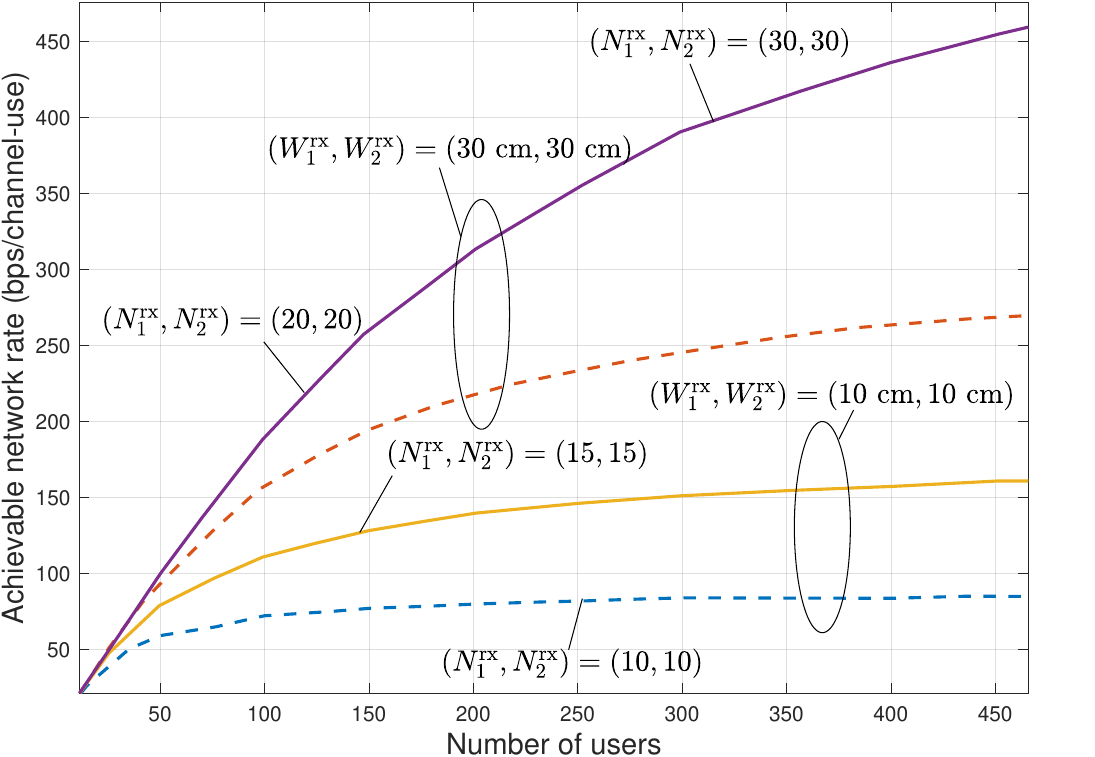}
    \caption{Achievable network rates of f-FAMA against the number of users with varying sizes $W_1^\text{rx} \times W_2^\text{rx}$ of FAs at each user \cite{16}. Each user is equipped with a 2-D FAs offering $N_1^\text{rx} \times N_2^\text{rx}$ positions or ports.}
    \label{fig:fama-1}
    \vspace{-1ex}
\end{figure}

FAMA is classified into two types: fast-FAMA (f-FAMA) and slow-FAMA (s-FAMA) based on their dependency on the receiver's CSI. The implementation of f-FAMA requires the receiver to acquire and analyze instantaneous CSI. This method enables rapid port switching within a symbol-level time. In a f-FAMA system, we assume $M$ users with a 2-D FA and a base station with $M$ FPAs and the $m$-th FPA is assigned to send the symbol to the $m$-th user. The size and number of ports of the FAs are $W_1^\text{rx}\lambda \times W_2^\text{rx}\lambda$ and $N_1^\text{rx} \times N_2^\text{rx}$. For the $m$-th user, the best FA port $n^*$ can be selected for every symbol instance which is given by:
\begin{equation}
\begin{aligned}
        n^* &= \underset{n \in \{1, \dots, N\}}{\arg\max}\ \text{SINR}_n \\
        &= \underset{n \in \{1, \dots, N\}}{\arg\max}\ \frac{\left| h_{m,m}^{(n)}(t) s_m \right|^2}{\left| \sum_{k'\ne k}^K h_{m,m'}^{(n)}(t) s_{m'} + \zeta_m^{(n)}(t) \right|^2},
\end{aligned}
\label{optimal-f-fama}
\end{equation}
where $h_{m,m'}^{(n)}(t) \triangleq [\bm{h}_{m, m'}(t)]_n$ is the time-varying channel gain of the $m$-th user's $n$-th FA port and $\zeta^{(n)}_{m}(t) \sim \mathcal{CN}(0, N_0)$ is the Additive White Gaussian Noise (AWGN). Notably, the insight of the (\ref{optimal-f-fama}) is that interference and noise are correlated, thus it is possible to completely suppress interference and noise by instantaneous port switching. Simulations of f-FAMA in Fig. \ref{fig:fama-1} demonstrate its extreme spectral efficiency and capability to support simultaneous access for hundreds of users. 
The results investigate the average network rate performance of f-FAMA under finite scattering channels with Rician factor $K = 7$ and $2$ scattered paths. The results show that f-FAMA can handle 300 users (with smaller FAs), and serving $500$ users is possible if larger FAs are allowed.

\begin{figure}[t]
    \centering
    \includegraphics[width=0.842\linewidth]{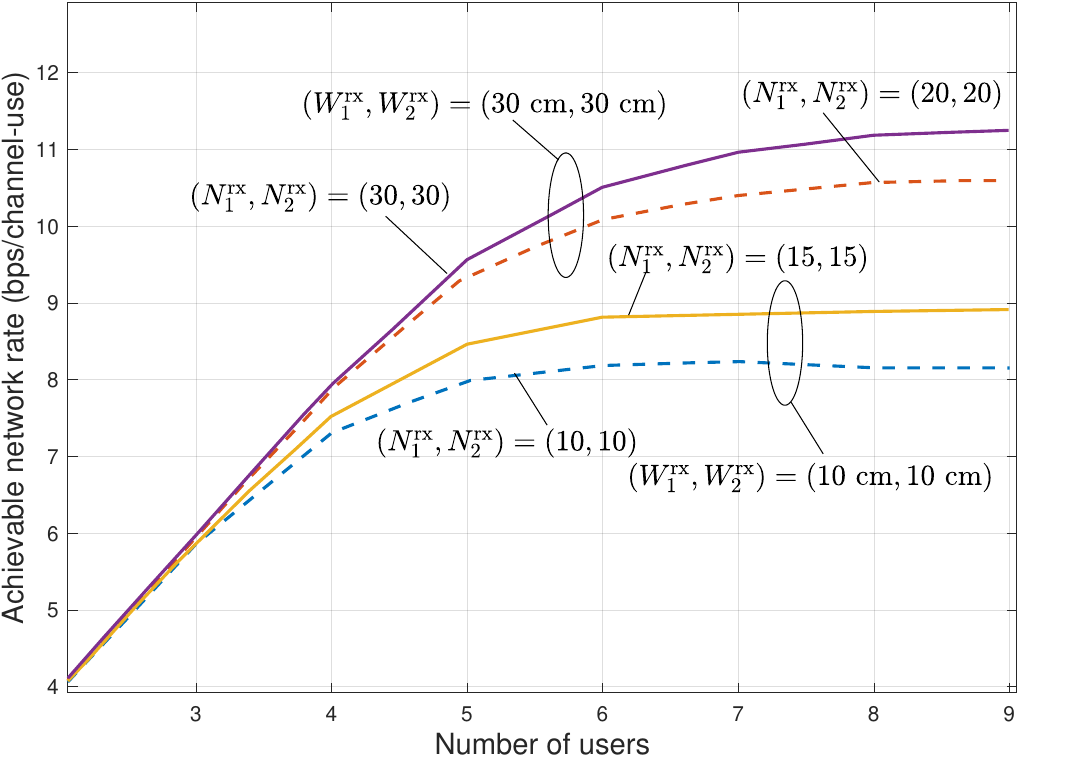}
    \caption{Achievable network rates of s-FAMA against the number of users with varying sizes $W_1^\text{rx} \times W_2^\text{rx}$ of FAs at each user \cite{102}. Each user is equipped with a 2-D FAs offering $N_1^\text{rx} \times N_2^\text{rx}$ positions or ports.}
    \label{fig:fama-2}
    \vspace{-2ex}
\end{figure}

Although f-FAMA demonstrates impressive performance gains, it relies on instantaneous CSI and symbol-level port switching, which currently confines it to the theoretical simulation stage. In contrast, s-FAMA, built upon the fundamental principles of FAMA, offers a practical multiple access approach \cite{102}. s-FAMA operates under a scenario where each user equipped with an FA selects the optimal port based on the positioning principle of FAMA, adjusting only when the statistical channel variation exceeds the threshold. To simplify the analysis, we assume the system setup is similar to the f-fama and we can compute the best FA port $n^*$ for the $m$-th user:
\vspace{-2ex}
\begin{equation}
\begin{aligned}
    n^* &= \underset{n \in \{1, \dots, N\}}{\arg\max}\ \text{SINR}_n \\
        &= \underset{n \in \{1, \dots, N\}}{\arg\max}\ \frac{P\left| h_{m,m}^{(n)}\right|^2} {P\sum_{m'\ne m}^M \left|h_{m,m'}^{(n)} \right|^2 + N_0} ,\\
\end{aligned}
\label{optimal-s-fama}
\end{equation}
where $P$ is the transmitted power, $h_{m,m'}^{(n)}\triangleq [\bm{h}_{m, m'}]_n$ is the statistical channel gain and $N_0$ is the noise power. The results in Fig. \ref{fig:fama-2} investigate the average network rate performance of s-FAMA under finite scattering channels with Rician factor $K = 7$ and $2$ scattered paths. Simulation results show that s-FAMA can support more than $10$ users simultaneously \cite{115}. In the future, we can achieve greater spectrum efficiency by combining FAMA technology with other multiple access technologies such as NOMA.

As previously discussed, s-FAMA enables practical deployment while suffering limited multiple-access capabilities. Researchers aim to enhance the performance of s-FAMA while preserving its practicality, promoting the development of Compact Ultra Massive Antenna Array (CUMA), an advanced variant of s-FAMA \cite{116}. Instead of activating a single optimal port, CUMA activates multiple appropriate ports of a set of selected ports $\mathcal{N}$ to receive signals in the analog domain for analysis. In a CUMA system of $M$ users, The $m$-th receiver can separate the received signal $\bm{r}_m$ into in-phase (real part) $r^{\text{real}}_m = \underset{n \in \mathcal{N}}{\sum} \text{real} \left( r_m^{(n)} \right)$ and quadrature (imaginary part) $r^{\text{imag}}_m = \underset{n \in \mathcal{N}}{\sum} \text{imag} \left( r_m^{(n)} \right)$ components, where $r_m^{(n)} \triangleq [\bm{r}_m]_n$. By denoting the transmitted signal as $s_m = s_m^\text{real} + \text{j}s_m^\text{imag}$, $r_m^{\text{real}}$ can be expressed as:
\begin{equation}
\begin{aligned}
    r_m^{\text{real}} &= \left[ \sum_{n \in \mathcal{N}} \text{real} \left( h_{m,m}^{(n)} \right) \right] s_m^{\text{real}} + \left[ -\sum_{n \in \mathcal{N}} \text{imag} \left( h_{m,m}^{(n)} \right) \right] s_m^{\text{imag}}\\
    & + \sum_{n \in \mathcal{N}} \text{real} \left(\sum_{m'= 1, m' \ne m}^M h_{m, m'}^{(n)} s_{m'} + \zeta_m^{(n)} \right),
\end{aligned}
\end{equation}
and similarly, $r_m^{\text{imag}}$ is given by:
\begin{equation}
\begin{aligned}
    r_m^{\text{imag}} &= \left[ \sum_{n \in \mathcal{N}} \text{real} \left( h_{m,m}^{(n)} \right) \right] s_m^{\text{imag}} + \left[ \sum_{n \in \mathcal{N}} \text{imag} \left( h_{m,m}^{(n)} \right) \right] s_m^{\text{real}}\\
    & + \sum_{n \in \mathcal{N}} \text{imag} \left(\sum_{m'= 1, m' \ne m}^M h_{m, m'}^{(n)} s_{m'} + \zeta_m^{(n)} \right).
\end{aligned}
\end{equation}
Obviously, if we selectively activate FA ports to maximize $\left| \underset{n \in \mathcal{N}}{\sum} \text{real} \left( h_{m,m}^{(n)} \right) \right|$ or $\left| \underset{n \in \mathcal{N}}{\sum} \text{imag} \left( h_{m,m}^{(n)} \right) \right|$, we can easily detect the transmitted signal through matrix inverse. Notably, CUMA only requires CSI for the desired signal and not for the interference users, reducing the complexity of the receiver design. In summary, the efficiency of CUMA lies in selecting the correct ports to enhance the estimated desired signal while weakening interfering signals through random superposition. As the increasing number of activated ports and RF chains, the interference resilience also increases \cite{117, 118, 119}. Even under conditions limited to LoS-only precoding, CUMA outperforms massive MIMO.

$\textbf{Summary and Insights: }$ 
Current research mainly assumed the stable channel conditions for the joint optimization of FA-NOMA techniques. However, in the NTNs, dynamic channels pose several challenges for NOMA, encompassing CSI estimation errors and outdated, dynamic user pairing, and constrained resource allocation. For instance, when UAVs apply NOMA, their movement introduces errors in channel estimation and uses outdated CSI for superposition coding, while also requiring continuous detection of new user pairings due to the coverage changing. Additionally, the limited transmit power of UAVs may diminish the distinguishability between stronger and weaker users. To address these problems, FA-NOMA should leverage effective allocation of multiple resource blocks to support multiuser in NTNs, including spatial dimensions, slot configurations, frequency band partitioning, coding selections, and power control. This approach promises to significantly enhance connectivity, spectral efficiency, energy efficiency, and low-latency, reliable performance in NTNs. Furthermore, it is important to investigate how FA can benefit from improved CSI accuracy and SIC.

\subsection{RIS}
RIS, as a passive/active relay node, has emerged as a critical solution for restoring connectivity between base stations and users, thereby reducing the need for costly network densification. The key benefits of RIS include its ability to enhance the overall SINR without affecting the hardware, reflect the signals with optimal phase shifts, and reduce the design complexity at the transmitter and receiver \cite{120}. However, due to obstacles or long distances between transmitters and receivers, RIS are struggling from severe channel blockage and double fading losses in hostile radio environments. Although RIS can focus beamforming through phase matrix optimization, the received signal power typically remains weak. 

The integration of FAs with RIS offers an inherent complement to each other, and they collaborate from several perspectives. In particular, Ghadi et al. in \cite{121} proposed a RIS-assisted FA receiver framework, derived closed-form theoretical expressions for outage probability and delay outage rate under rich scattering environments. Compared to MISO systems based FPAs, simulations demonstrated that, under perfect CSI conditions, activating a single FA port alone can obtain better system performance.
In \cite{122}, the outage probability expression of the RIS-assisted FA system was approximated using a block correlation model. Simulations demonstrate that, with a large number of RIS elements, this approximation closely aligns with the simulation outcomes, thus reducing computation complexity.
In \cite{123}, Lai et al. jointly optimized the spatial positions of the FA and RIS for passive beamforming to maximize the user's SNR.

In RIS-FA systems, CSI acquisition plays a crucial role in the design of optimization algorithms, necessitating a trade-off between performance and computational complexity. 
In \cite{124}, Yao et al. investigated the upper bounds of outage probability of the RIS-FA systems and CSI-based/free optimization algorithms to enhance network throughput. Moreover, Yao et al. in \cite{125} employed a block diagonal matrix approximation to characterize the spatial correlation of RIS elements, leveraged statistical CSI to design RIS phase-shift configurations, and derived the upper and approximate lower bounds of the outage probability. This method not only reduces the burden on base stations for instantaneous CSI and the feedback overhead but also simplifies the calculations of adjusting RIS phase-shift, since the adjustments are performed only within the intervals of statistical CSI updates. 
The combination of instantaneous CSI and statistical CSI for optimization is expected to achieve a balance between performance and overhead.
In \cite{126}, Zhang et al. proposed a two-timescale design for multi-user RIS-FA systems. This approach utilizes statistical CSI to optimize FA port selection, regularization factors, and RIS phase-shift configurations. In parallel, instantaneous CSI can be used to design zero-forcing or regularized zero-forcing precoding. Notably, owing to the use of two-timescale CSI, the optimization objective targets ergodic sum rate rather than instantaneous performance.

\begin{figure}[t]
    \centering
    \includegraphics[width=0.48\textwidth]{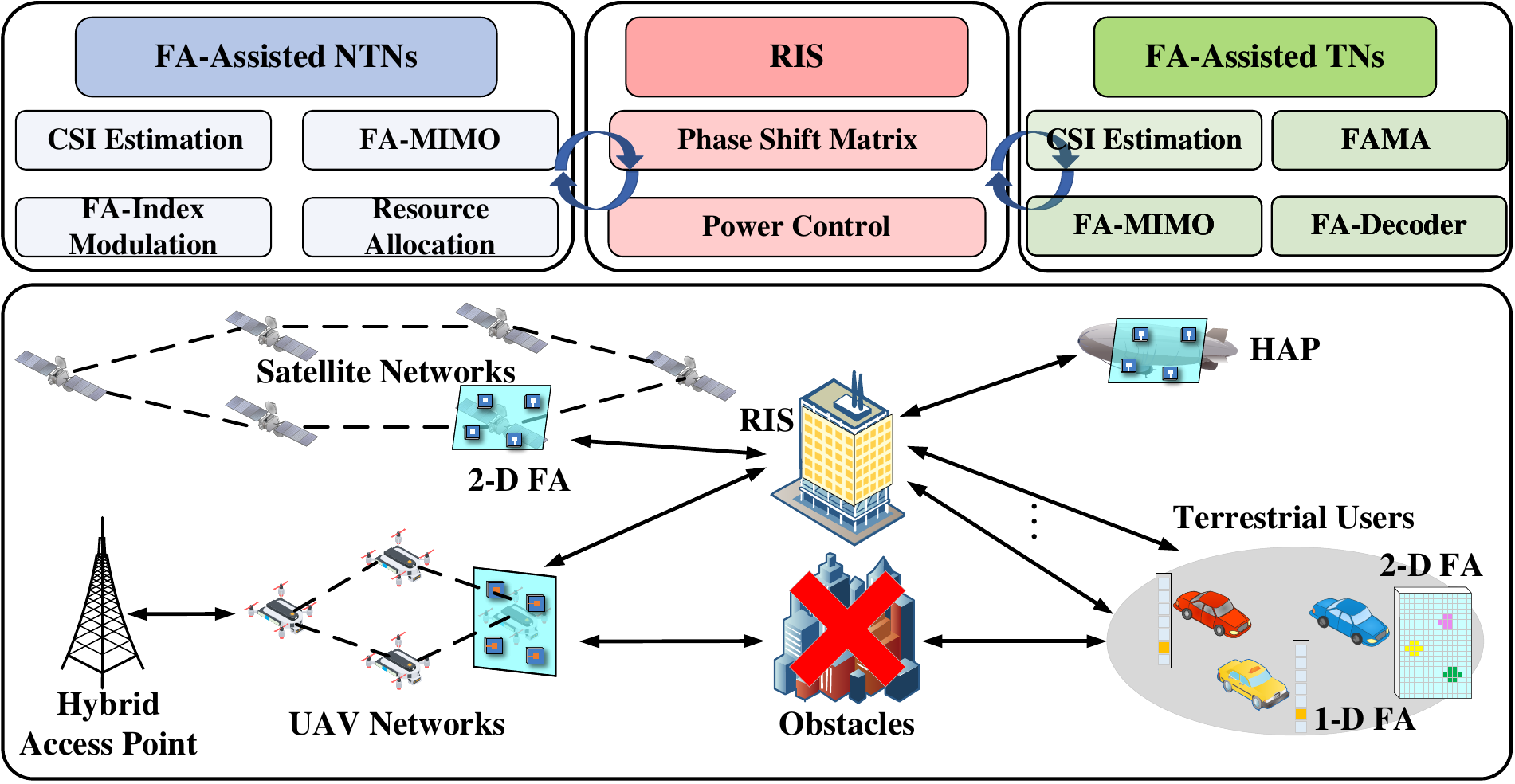}
    \caption{A potential RIS-FA-NTN system. In satellite networks, RIS-FA systems deployed at satellites and TNs can improve the stability of inter-satellite links and the throughput, respectively. In UAV/HAP networks, RIS-FA systems and UAVs can be jointly optimized for the network throughput, coverage region, and energy consumption.}
    \label{fig:ris-fa}
    \vspace{-2ex}
\end{figure}

Moreover, RIS-FA has some promising architectures, such as RIS-FA-NTN and RIS-CUMA. A potential RIS-FA-NTN system is shown in Fig. \ref{fig:ris-fa}. In satellite networks, RIS-FA systems deployed at satellites and TNs can improve the stability of inter-satellite links and the throughput, respectively. In UAV/HAP networks, RIS-FA systems and UAVs can be jointly optimized for the network throughput, coverage region, and energy consumption. Specifically, the work in \cite{127} proposed a RIS-FA-UAV networks, where UAV deployment mitigates large-scale path loss, the FA positions and beamforming counteract small-scale fading, and RIS regulates phase shifts to address transmission path blockages and enhance service coverage.
The simulations show that more FA antennas provide additional adjustable positions to combat small-scale fading in the UAV-RIS link. Similarly, increasing RIS elements enhances the rate by enabling more phase-shift adjustments for different users, mitigating channel fading in the RIS-user link. Also, with more FA antennas, the requirement for RIS elements can be reduced. Additionally, the random deployment and configuration of RIS enrich the scattering environment, enhancing the multiple access capability of CUMA. This idea is easily achievable, as it eliminates the need for additional RIS optimization \cite{117}.

$\textbf{Summary and Insights: }$
The joint deployment of RIS and FAs enables flexible reconfiguration of propagation environments across transceivers and cascaded channels, making it suitable for the dynamic channels of NTNs. Although joint FA-RIS deployment in satellite networks remains unexplored, these structures are promising for enhancing link stability, beam coverage, interference mitigation, and improving user rates. However, there are several practical challenges in the RIS-FA-NTNs. Due to the dynamic and outdated nature of the A2G channels, it is essential to explore more accurate and lower-latency CSI estimations. Besides, RIS-FA systems need to develop corresponding optimization schemes under different CSI conditions, or leverage additional DoFs to fortify the channel stability. Moreover, there is a broad research space for the optimization algorithm design in multi-RIS-assisted FA enabled-MIMO systems. Besides, RIS and FAs can provide additional DoFs and benefit from each other, they can also collaborate from a different perspective. For instance, FAMA performance hinges on the diversity of multiple access channels, which relies on rich-scattering environments. In this case, RIS can construct artificial scattering settings to assist FAMA, where its performance bounds and complexity analyses are worth investigating.
\section{Intelligent Function Integration}
\label{integration}
The evolution of 6G imposes function integration demands during the intelligent transformation of NTNs, including integrated communication-computation systems and ISAC systems. FA-assisted NTNs play a central role in unified orchestration, scheduling, and coordination across diverse functional systems, driving the intelligent integration of 6G network resources. In this section, we review the mobile edge computing, over-the-Air Computation (AirComp), Federated Learning (FL), and ISAC systems in FA-assisted NTNs.

\begin{figure}[t]
    \centering
    \includegraphics[width=0.485\textwidth]{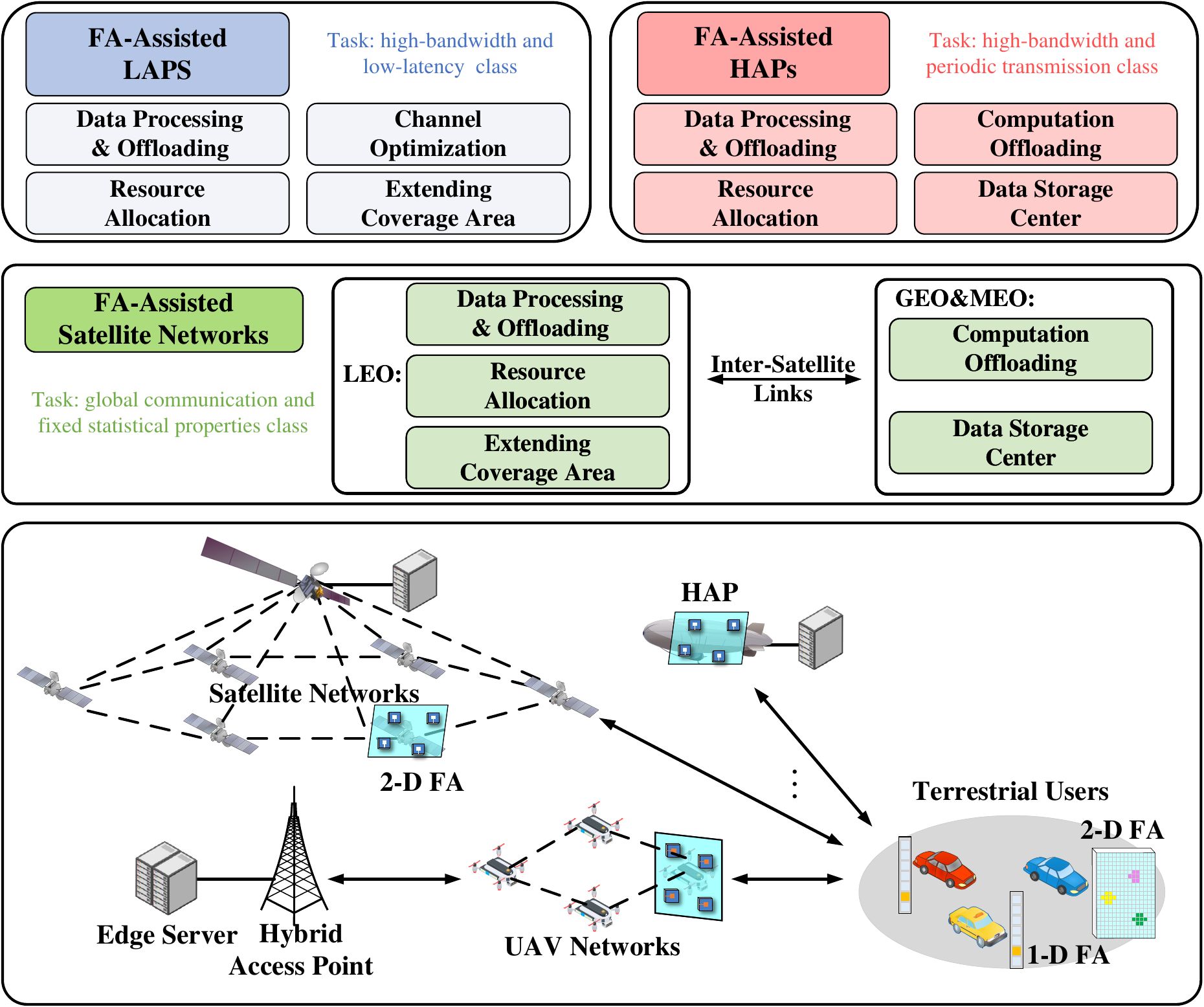}
    \caption{An FA-assisted NTNs integrated mobile edge computing systems. For high-bandwidth and low-latency tasks, FA-assisted LAPs (e.g., UAVs) can serve as relay nodes, performing data processing/offloading and channel optimization, then sending the processed results to the nearby AP edge servers. For high-bandwidth, but periodic transmission tasks, FA-assisted HAPs with edge servers can process computation offloading tasks for remote users. For tasks requiring global communication with fixed data statistical properties, FA-assisted satellite networks transmit the collected data from LEOs to the edge servers of GEOs/MEOs.}
    \label{fig:fa-MEC} 
    \vspace{-2ex}
\end{figure}

\subsection{Mobile Edge Computing}
The exploding growth in data processing and low-latency computation demands in 6G will exceed the capabilities of traditional cloud computing. Mobile edge computing, a real-time distributed computing technology, deploys computing resources closer to data sources for processing, computation, and storage. In this way, the computing tasks can be faster responded in lower network latency, energy consumption, and bandwidth burden. NTNs can function as mobile terminals, relay nodes, and edge servers in mobile edge computing systems, thereby expanding network performance, but also leading to congestion. Consequently, effective management and allocation of communication and computing resources are essential to support advanced 6G applications \cite{128, 129}. However, in traditional mobile edge computing systems, reducing system latency through resource allocation is much more difficult because of the reliance on FPA arrays at the APs. This configurations increase the system complexity and latency, and fail to fully leverage the channel spatial characteristics. 

FAs can enhance mobile edge computing system performance from several perspectives compared with FPA arrays: (i) Additional spatial diversity gains and DoFs enable the optimization of wireless channels, thus enhancing resource allocation efficiency. In \cite{130}, Zuo et al. jointly optimized FA positions, offloading ratios, and CPU frequencies to minimize system latency. In this way, FAs can adjust antenna positions to achieve better channel conditions and reduce transmission delay. On the other hand, different FA position selection strategies reconfigure offloading ratios and CPU frequencies, promoting more effective resource allocation.
(ii) fewer number of antennas in mobile edge computing systems, thereby lowering hardware costs and system complexity. In \cite{131}, FAs can play a crucial role in Interference suppression and system latency reduction in mobile edge computing systems. Attributed to the flexible position selection and beamforming strategies, FA-assisted mobile edge computing systems can obtain lower hardware costs and complexity. (iii) extending the coverage of base stations and relay nodes. In \cite{132}, FA-assisted UAVs, as relay nodes, combined with mobile edge computing systems, effectively broaden the coverage and task offloading capacity of the mobile edge computing systems.

Moreover, mobile edge computing systems based on FPAs typically require higher transmission and circuit power to enhance communication performance, making their deployment on lightweight NTN platforms (e.g., UAVs) impractical. As an improvement, FA-assisted NTNs can simplify the process of resource loading and allocation, thereby improving the efficiency of computation offloading tasks and resource distribution. In \cite{133}, Chen et al. proposed an FA-enhanced scheme for wireless-powered mobile edge computing systems. In this work, FAs deployed at hybrid access nodes adjust their positions to increase spatial DoFs for improving the efficiency of downlink wireless power transfer and uplink task offloading. Moreover, three FA configurations, including dynamic, semi-dynamic, and static, are introduced to balance performance gains and implementation complexity.

$\textbf{Summary and Insights: }$
To address the diverse computation demands in the future, we propose a promising architecture of FA-assisted NTNs for mobile edge computing systems, as illustrated in Fig. \ref{fig:fa-MEC}. For high-bandwidth, low-latency tasks, FA-assisted LAPs (e.g., UAVs) can serve as relay nodes, performing data processing/offloading and channel optimization, then sending the processed results to the nearby AP edge servers. This configuration improves coverage and efficiency while reducing latency. Under the latency and power constraints, the primary challenges include UAV deployment, resource allocation, FA configuration, offloading ratios, and CPU frequency. For high-bandwidth, but periodic transmission tasks, FA-assisted HAPs with edge servers can process computation offloading tasks for remote users due to their flexible deployment and substantial load-bearing capacity. For tasks requiring global communication with fixed data statistical properties, FA-assisted satellite networks transmit the collected data from LEOs to the edge servers of GEOs/MEOs. This approach reduces the reliance on terrestrial stations. Additionally, FAs can facilitate direct connections between terrestrial users and satellite networks to provide services.

\begin{figure}[t]
    \centering
    \includegraphics[width=0.485\textwidth]{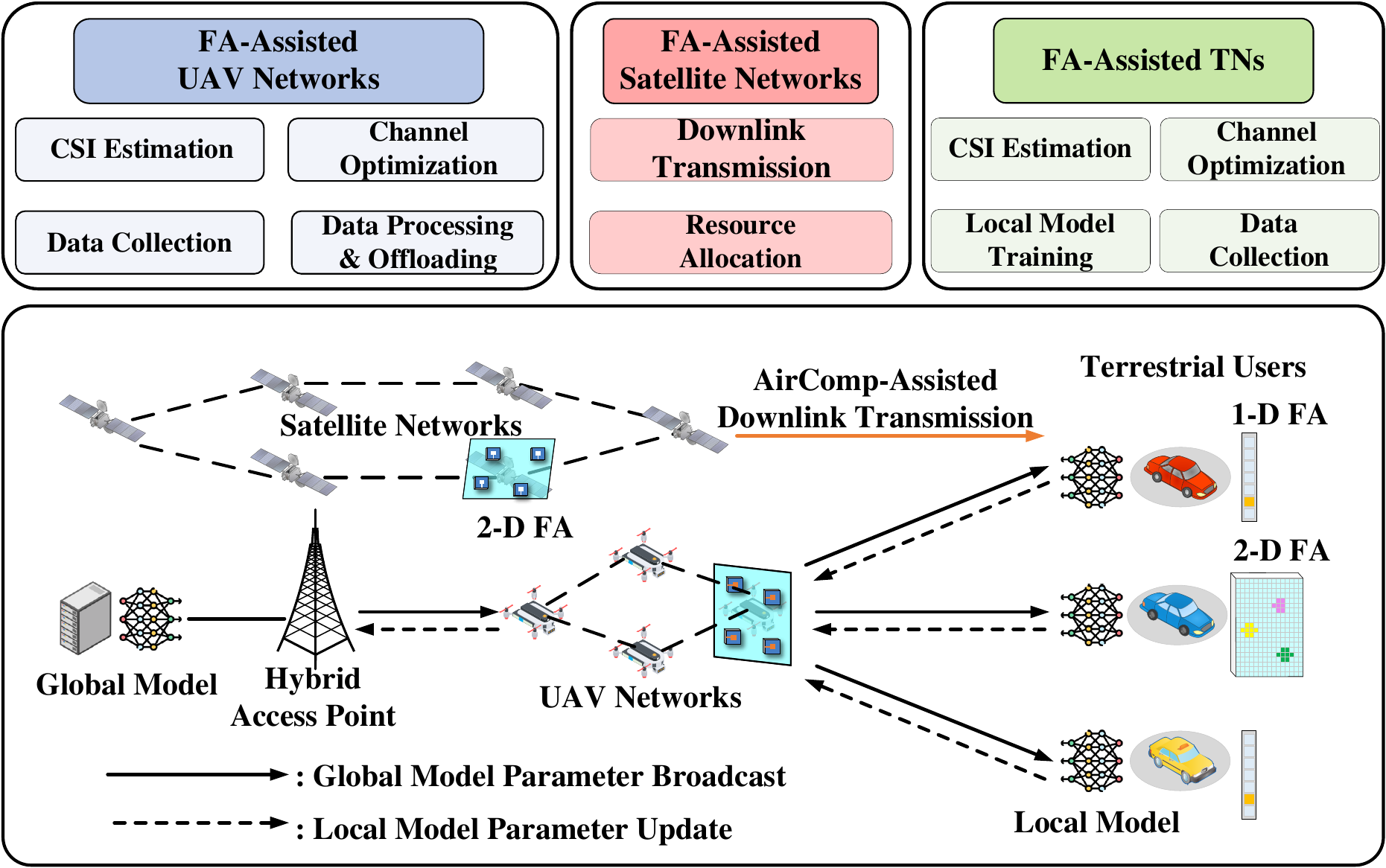}
    \caption{An FA-assisted NTNs integrated over-the-air FL systems. In the FA-assisted UAV networks, UAVs can serve as relay nodes for first-hop AirComp, update local model parameters, then upload the results to higher-level APs for second-hop AirComp and global model parameter update. In FA-assisted satellite networks, AirComp optimizes downlink data transmission. FAs can improve precision computation and data aggregation efficiency by reshaping the wireless channel gain.}
    \label{fig:fa-FL}
    \vspace{-2ex}
\end{figure}
 
\subsection{AirComp and FL}
To address computing limitations in massive node access, AirComp, a distributed intelligent technology integrated with communication and computation, emerges as a promising solution. By supporting the superposition of multiple devices' concurrent transmission for electromagnetic waves, AirComp enables rapid computation at receiving nodes with direct result transmission, thus reducing channel usage and computing latency. Undoubtedly, AirComp imposes strict requirements on channel consistency and distributed node management. In this context, cooperating FA-assisted NTNs with AirComp facilitates the fusion of dynamic adaptability and distributed intelligence. As illustrated in Fig. \ref{fig:fa-FL}, in the FA-assisted UAV networks, UAV swarms can serve as relay nodes to perform the first-hop air-computation, and then upload the results to higher-level APs for second-hop air-computation and result aggregation, which facilitates massive device access and energy efficiency improvement. On the other hand, FA-assisted satellite networks can use AirComp to optimize downlink data transmission. Furthermore, FAs can achieve high-precision computation by reshaping the wireless channel gain, and also can optimize resource allocation between devices and receiving nodes to enhance data aggregation efficiency. Such distributed intelligence significantly improves the energy and spectral efficiency in AirComp systems. In \cite{134}, Cheng et al. demonstrated that FAs can effectively reduce the MSE by flexibly reshaping wireless channels between terminal devices and receiving nodes. On this basis, Zhang et al. in \cite{135} investigated a 1-D antenna motion FA array-assisted AirComp system, where results showed that compared to traditional FPAs, FAs employing an AO approach have lower computation MSE. Extending 1-D FAs, Li et al. in \cite{136} presented a 2-D antenna motion FA array-enhanced AirComp system. To fully exploit DoFs in antenna positioning, a two-loop iterative algorithm based on PSO is proposed, outperforming the AO method in minimizing MSE.

The above studies indicated that FAs reduce the MSE in AirComp, making the integration of FAs and AirComp a viable approach to enhance the performance of wireless FL systems, thus minimizing communication overhead. In \cite{137}, Park et al. integrated FAs into an AirComp-FL system and optimized antenna port selection to provide dynamic spatial diversity, which mitigates channel variations and improves signal aggregation. Moreover, Ahmadzadeh et al. in \cite{140} employed UAVs as FL clients to support IoT devices, jointly optimizing antenna placement and beamforming vectors to minimize MSE. Therefore the dynamic environmental challenges can be addressed by redefining the problem as an MDP suitable for a Twin Delayed Deep Deterministic policy gradient (TD3) algorithm. In \cite{138}, Ahmadzadeh et al. derived the performance gap between actual and optimal losses in AirComp-FL to quantify the impact of beamforming vectors and FA positioning. They reformulated the performance enhancement as a non-convex optimization problem and a Markov decision process (MDP), which is solved using a DRL framework. Furthermore, in \cite{139}, LSTM layers are integrated into a DRL framework to learn temporal correlations during user dynamic movement, enabling adaptive decision-making under different channel conditions.

$\textbf{Summary and Insights: }$ 
Further research directions focus on coordinating the dynamic adjustments of the FA-assisted NTNs with the computation tasks performed by AirComp systems. Firstly, meticulously designed algorithms and communication protocols are needed to ensure that AirComp's computation resources remain synchronized with the constantly changing configuration of the FA-assisted NTNs. Additionally, resource constraints such as available CPU frequency and memory must be considered when optimizing AirComp's computations. Moreover, under dynamic conditions, both FA-assisted NTNs and AirComp need to make real-time adjustments rapidly to maintain optimal performance, requiring accurate system models and resource prediction algorithms. The optimization problem often becomes a non-convex optimization problem with complex dependencies between variables.

\subsection{ISAC System}
The ISAC system consolidates hardware and radio resources, unifying communication and sensing within a single frequency band and architecture. In ISAC systems based on FPAs, conventional self-interference mitigation and resource optimization are typically formulated as a tradeoff between communication and sensing: either maximizing sensing quality (e.g., radar sensing SINR) under communication-rate constraints, or maximizing communication performance (e.g., user receive SINR) subject to sensing requirements. 

\begin{figure}[t]
    \centering
    \includegraphics[width=0.45\textwidth]{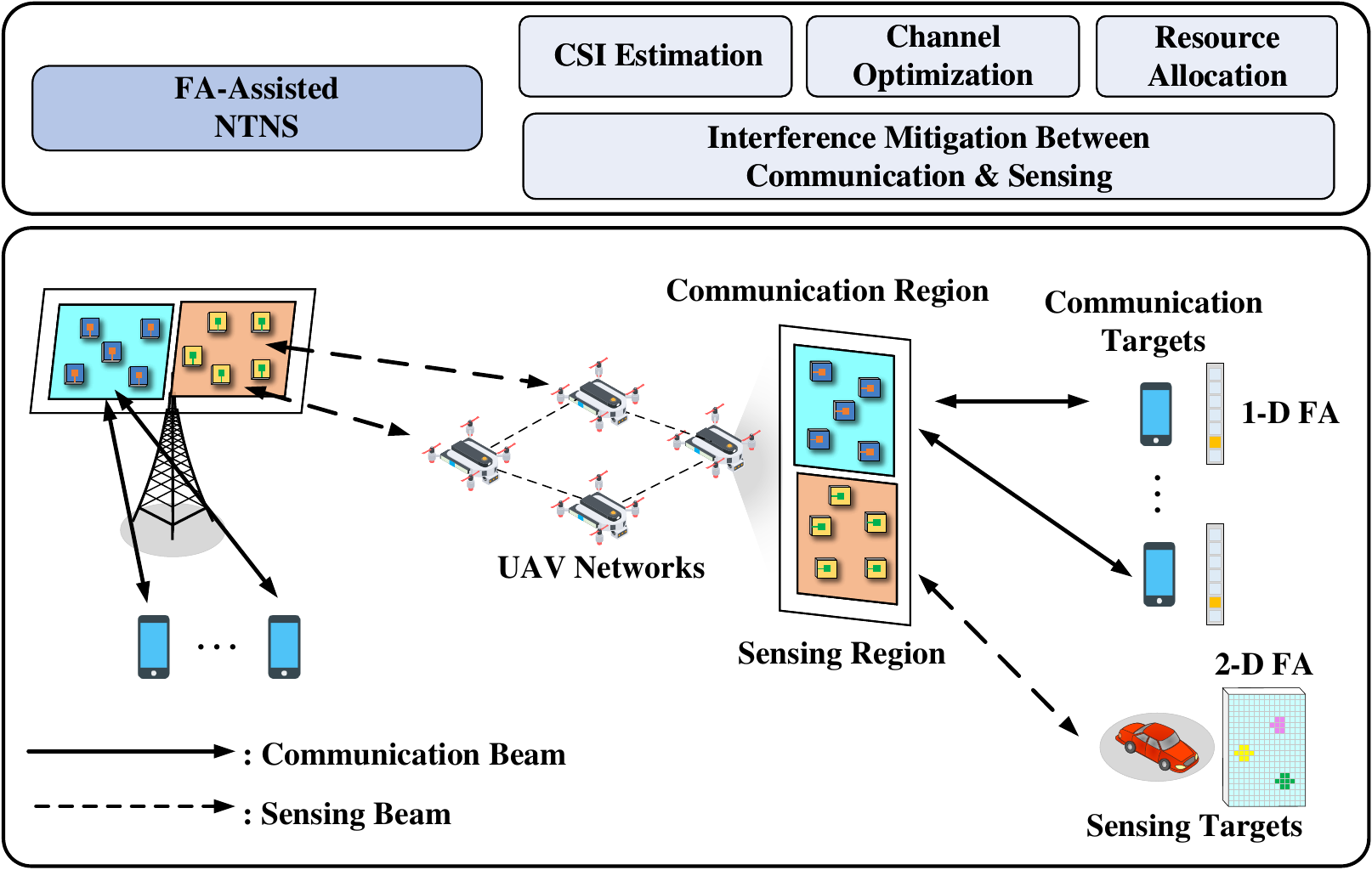}
    \caption{An FA-assisted NTNs integrated ISAC system. FA-assisted NTNs can adapt to complex, dynamic environments and reduce the vector correlation between communication and sensing beams in different directions through spatial flexibility, thereby achieving flexible coexistence.}
    \label{fig:fa-ISAC}
    \vspace{-2ex}
\end{figure}

When communication and sensing signals are designed separately, they are usually orthogonalized first and then jointly optimized through MIMO beamforming and multiple-access resource allocation. Considering the superiority of FA-assisted NTNs in spatial multiplexing gains, diversity gains, beamforming, and flexible deployment, they are promising to be a new paradigm for ISAC systems as shown in Fig. \ref{fig:fa-ISAC}. FA dynamically adjusts configurations (e.g., position and angle) to minimize vector correlations between communication and sensing beams from different directions \cite{141, 141-n3}. As a flexible coexistence approach, FAs utilize spatial resources to expand the array aperture, thereby achieving higher wireless sensing resolution and enabling simultaneous radar sensing and communication. 

When a unified waveform is used for ISAC, performance is often limited by the weak coupling between the communication and sensing channel subspaces, which typically experience different propagation conditions (e.g., LoS for sensing but multipath for communication) \cite{141-n, 142-n}. This mismatch restricts the overall ISAC gains. In contrast, FA-assisted UAVs/HAPs adjust their positions to create more favorable channel conditions for both communication users and sensing targets, thereby increasing channel coupling and strengthening channel gains \cite{19-n, 142}. Particularly, FAs align spatial steering vectors of communication and sensing channels, allowing both functions to operate within a unified signal waveform. The simulations demonstrated that it is useful to reduce interference from sensing requirements.

A key challenge in the FA-ISAC system is the joint optimization of port selection, communication beam, and sensing beam to achieve the trade-off between communication and sensing. For example, Zhou et al. in \cite{143} designed an FA-ISAC system centered on communication rate, where both communication and sensing signals are transmitted simultaneously through a shared beam. To maximize the communication rate, FA positions and transmit beamforming at the base station and communication users are alternately optimized under the sensing power constraints. In \cite{144}, Hao et al. designed a sensing-centric FA-ISAC system, jointly optimized the FA position and the transmit beamforming to maximize the radar SNR. To solve this non-convex problem, the authors proposed an SCA-based AO algorithm to obtain a suboptimal solution. Furthermore, Zou et al. in \cite{145} demonstrated that FA can be effective in the trade-off between communication and sensing in ISAC systems. Under identical sensing and communication constraints, the transmit power of FA-ISAC is substantially lower than baseline methods using FPAs. This indicates that FA offers a superior balance between sensing and communication, enhancing the performance bounds of ISAC systems. The FA-ISAC system can expand the coverage range with the help of UAV networks, because the flexible deployment of UAVs can reduce large-scale fading and provide LoS channels to improve network throughput. In \cite{146}, Kuang et al. proposed a UAV-FA-ISAC system, in which the communication users have dedicated communication beams, and similarly, the sensing targets have dedicated sensing beams. Under the sensing gain constraint, the authors jointly optimized the FA positions, communication beam, and sensing beam to maximize the sum rate. Most current optimization algorithms in prior work rely on SCA and PSO methods, exhibiting slow convergence when computing FA positions and beamforming. 
In \cite{19-n}, Wang et al. introduced a passive 6DMA-enabled UAV-ISAC system that jointly optimizes UAV placement, 6DMA rotation/reflection coefficients, and transmit beamforming to maximize single-target sensing SNR under communication SNR constraints. They adopt an alternating-optimization framework to enhance sensing–communication channel correlation and channel gain, and then derive a closed-form optimal transmit beamformer.
In \cite{147}, Zhang et al. proposed a fast-converging joint optimization scheme for the FA-ISAC system. Based on the majorization-minimization principle, a novel proximal distance algorithm was derived to obtain the closed-form waveform design in each iterative step, alongside an extrapolated projected gradient algorithm to accelerate the convergence of FA position optimization. The proposed algorithm achieves FA position configuration over $60\%$ faster than the conventional SCA and PSO approaches.

Certain studies have already explored the application of AI methods for optimization. In \cite{148}, Wang et al. investigated an advanced actor and critic architecture, jointly optimizing port selection and MIMO precoding matrix to maximize the sum-rate under sensing constraints. Specifically, the authors employed the DRL method, which defined the CSI of all current ports as the state variable and the sum-rate as the reward function, with port selection and precoding matrix design defined as actions. Notably, the critic network leverages a deep neural network to approximate the expected reward of the policy. The actor network integrates a pointer network and a neural precoding network to devise an optimal strategy, where the pointer network handles port selection and the neural precoding network performs MIMO precoding. Unlike supervised learning, the actor-critic-based DRL method trains the neural network through interactive feedback from the environment, rather than using labeled data. During the training process, the authors incorporate ensemble learning to enhance the model's convergence and training stability.

$\textbf{Summary and Insights: }$ 
In the ISAC systems, distinct waveform designs are required for communication users and sensing targets. At the transmitter, FAs introduce new dimensions to establish channel conditions that benefit both sensing and communication tasks. The deployment of UAVs can provide LoS channels to sensing targets, while communication users can adapt to different scattering environments through the FAMA technology.

Furthermore, the ISAC systems can utilize FAs to align the spatial steering vectors of sensing and communication channels so that a unified signal waveform can be used for both tasks. This approach simplifies the system design and reduces interference caused by sensing signals. However, the adoption of the unified waveform signal exposes the system to potential eavesdropping and attacks. The collaboration between UAV networking and FA configurations is promising for improving the PLS of ISAC systems.
\section{Security in FA-Assisted NTNs}
\label{pls}
With the advancement of coverage and network heterogeneity in next-generation wireless communications according to NTNs, there is an increasing concern about the security and privacy of these systems. In the following, we discuss the investigation of PLS and covert communication in FA-assisted NTNs.
\vspace{-2ex}
\subsection{PLS}
To avoid the high computing complexity and key management overhead of conventional encryption/decryption methods, PLS has recently garnered significant attention. PLS primarily ensures high-confidentiality communication between the legitimate transmitter (Alice) and the receiver (Bob) while preventing the eavesdropper (Eve) from intercepting sensitive information \cite{149}. However, Alice and Bob with traditional FPAs, fail to exploit spatial diversity, thereby degrading their secrecy performance. Additionally, the NTN platform is more vulnerable to malicious eavesdropping in strong LoS environments. The utilization of FAs can reshape channel envelopes across spatial and temporal dimensions, enhancing security and performance gains for PLS \cite{150}. Moreover, FAs can perform dynamic beam coverage to enhance directional communication with Bobs and send jamming signals to block eavesdroppers \cite{151}.

Specifically, FAs introduce extra channel gains between Alice and Bob while disrupting links between Eves and them. Considering the entire CSI of a single Eve, Li et al. in \cite{152} jointly optimized FA positions, beamforming, and UAV deployment to reduce channel correlations between legitimate users and Eavesdroppers. Despite the non-convex optimization problems, AO framework of projected gradient ascent, simulated annealing and SCA is used to determine the suboptimal solutions. Moreover, when the perfect CSI of Eves is unavailable, Hu et al. in \cite{153} jointly optimize FA positions and beamforming based on statistical CSI to reduce outage probability. Alternatively, in \cite{154,155}, inferring potential Eve locations from imperfect CSI can adjust FA positions and beamforming toward Eve's weakest channel paths.
In \cite{156}, Wu et al. directly captured the worst-case of potential Eve location area, jointly optimized FA positions, beamforming vectors, artificial noise, and UAV position to maximize the minimum secrecy rate. Notably, the artificial noise is introduced to disrupt unauthorized Eves, ensuring the confidentiality of legitimate user communications. However, it will also introduce interference to Bobs. To address this issue, Xu et al. in \cite{157} proposed an alternative jamming method where Alice transmits encoded codewords instead of Gaussian noise. Therefore, Bob can mitigate the interference by decoding the jamming signal, while Eve cannot. The FA-assisted UAV can also serve as a friendly jammer to ensure secure communication between Bob and Alice. In \cite{158}, Kim et al. exploited the spatial DoFs of the FA array to optimize the directional beamforming for Eve jamming signal mitigation. Then, they combined with the UAV energy consumption model to maximize the secrecy energy efficiency.

However, mechanical latency can aggravate the link reliability for mobile UAVs in complex environments. In \cite{159}, Yu et al. proposed a transformer-LSTM-based FA positioning forecasting framework to solve secrecy rate maximization. On the other hand, Eves are often difficult to detect, while legitimate users’ locations are easily exposed. Introducing FAs into eavesdropping systems can be maliciously exploited, heightening the risks of information leakage. In \cite{160}, Maghrebi et al. demonstrated that FA-enhanced interference systems can severely disrupt legitimate links. Obviously, can Alice’s or Bob’s FAs still improve the system secrecy if Eves' wiretapping capability is also enhanced with FAs and UAVs? There are still many open questions that remain unanswered.

$\textbf{Summary and Insights: }$ 
In satellite communications, the broadcast feature and wide coverage of signals make transmissions easily intercepted, motivating future FA research towards satellite PLS. Conventional PLS schemes often suppress interception by reducing transmit power or injecting artificial noise, but this degrades the target link, particularly for high-loss satellite-terrestrial channels. FAs can enable accurate beam pointing to enhance secure beamforming with less performance sacrifice. Moreover, FAMA and CUMA can preserve the desired signal rate even in the presence of artificial noise and high-loss satellite–ground links.

\subsection{Covert Communication}
In addition to PLS technology, the security of FA-assisted NTNs can also be improved through covert communication. The principle of covert communication is to prevent legitimate users’ signals from being detected by wardens (Willies). Existing works mainly focus on the covert communication with noise uncertainty of Willie, leveraging the position flexibility of FAs to achieve both ultra-reliability and high covertness. In \cite{161}, Wang et al. first developed an FA-assisted covert framework for a single user with one Willie, and analytically derived the sum of the probabilities of the detection errors and the communication outage probability. This work can be further extended to a multi-Willie case where each warden performs independent detection \cite{162}. To satisfy the detection error probability of each Willie, a block coordinate descent algorithm was proposed to jointly optimize FA positions and transmit beamforming. For multiuser covert communication systems under noise uncertainty, Mao et al. in \cite{163} fully utilized spatial DoFs of FAs to enhance target signal strength while reducing mutual interference between users and minimizing power leakage to the Willie. Considering that the link reconfigurability of the RIS can also enhance covert communication performance, Xie et al. in \cite{164} proposed a RIS-FA covert communication framework. This framework maximizes the covert rate over all time slots by jointly optimizing the FA position, transmit beamforming, and RIS phase. To cope with the complex system, a hierarchical paradigm with a DRL-based outer loop and an AO-based inner loop is introduced to obtain the optimal solutions of multiple coupled variables.

$\textbf{Summary and Insights: }$
In NTNs, UAVs acting as Alice and Bob are more susceptible to exposure within Eves' or Willies' coverage due to their higher probability of LoS paths. Meanwhile, Eves and Willies can also leverage FAs to enhance wiretapping capability. Consequently, in NTNs where all parties are equipped with FAs, it is essential to thoroughly investigate the intricate interaction mechanisms and dynamic evolutionary patterns involved for optimal secrecy performance. Additionally, ML methods can help address potential challenges in dynamic environments in NTNs. For instance, in GANs, Alice and Eves function as the generator and discriminator, respectively, to optimize Alice's FA positions and beamforming vectors. Furthermore, incorporating Bob into the GAN framework can enhance the overall system performance.
\section{Future Directions and Opportunities of FA-Assisted NTNs
}\label{challenge}
In this section, we further discuss potential directions and research opportunities, including AI applications, flexible antenna technologies, high-frequency communication, near-field communication, and ISCC systems to unlock the full potential of FA-assisted NTNs.
\vspace{-2ex}
\subsection{AI Applications}
With the demand for analyzing and processing massive data surging in 6G, traditional ML algorithms designed for specific tasks suffer from limited robustness and migration capabilities. Recent years have witnessed advances in generative AI, however, integrating large AI models into FA-assisted NTNs for more intelligent service remains to be explored \cite{165, 166, 166-n3}. Firstly, the vast parameter scale of large models enables precise capture of complex multi-dimensional correlations. FA-assisted NTNs utilize transformer-based models for CSI estimation and prediction, outperforming traditional LSTM methods. Moreover, Large Language Models (LLMs) like DeepSeek \cite{167, 168}, combine Mixture-of-Experts (MoE), Multi-head Latent Attention (MLA), and Group Relative Policy Optimization (GRPO) to reduce computing costs and enhance multitask performance. Model compression and quantization enable lightweight deployment of general large models across the UAVs and satellites. Thus, LLMs can utilize advanced natural language-vision processing and learning capabilities to analyze massive data for task interactions in real-time scenarios. By understanding and evaluating multi-modal data in complex environments, general large models enhance inter-device responsiveness and adaptability to complex environments. It is helpful to optimize resource management, route planning, traffic control, and user experience.

LLMs integrated with reasoning, memory, tool, and action modules form a foundational AI Agent that enables dynamic selection of appropriate action strategies based on real-time surroundings. Furthermore, they can evaluate the outcomes and combine with real-time environmental understanding to iteratively optimize strategies. AI Agents enable systems to adapt to complex and dynamic environments \cite{169}. For instance, in cooperative communication within FA-assisted UAVs and  terrestrial transportation system, AI agents leverage current channel and traffic conditions to perform blockage prediction, trajectory planning, beamforming, and FA position optimization. Meanwhile, multi-agent interactions perform multi-perspective analysis and complex problem decomposition \cite{170}, to promote coordinated executions such as task scheduling, resource allocation, and communication strategy selection. 

AI-driven semantic communication depends on LLMs and DRL to transmit semantically compressed data and decode, enhancing network performance while conserving bandwidth \cite{171}. The main challenges of semantic communication encompass unified semantic representation, accurate semantic extraction and recovery, and channel modeling for semantic communication. In this context, exploring FA-assisted NTNs to enhance the semantic communication capabilities of 6G networks is a promising research direction. Studies have manifested that FAs dynamically adjust antenna configurations based on word importance to improve transmission efficiency and semantic accuracy.

{\revise
\subsection{Flexible Antenna Technologies}
With the continuous evolution of flexible antenna technologies, a series of promising architectures have emerged beyond FA, MA, and 6DMA discussed in this paper, such as RAs and PAs. These technologies introduce reconfigurability from various dimensions, providing new opportunities to enhance NTN coverage, interference mitigation, and ISAC systems.

RAs improve spatial adaptability by exploiting the rotational DoFs of each antenna element. Limited by the 2-D ground coverage design with downtilt configurations optimized for terrestrial users, traditional base-station FPAs struggle to provide effective 3-D coverage and reliable tracking for dynamic aerial users with varying altitudes and flight trajectories. In contrast, RAs can dynamically adjust antenna orientation/boresight direction to fill coverage blind spots, enhance beam focusing, and improve target-tracking accuracy, showing significant potential in low altitude economy and ISAC scenarios \cite{171-nn}. Furthermore, RAs can be integrated into UAVs to jointly optimize the flight trajectory and antenna rotations for more complex task execution \cite{172-nn, 173-nn}. However, the high mobility and rapid attitude variations of UAVs impose stringent requirements on real-time tracking, low-latency antenna adjustment, and control overhead. In this regard, the development of electronically driven FAs with low latency and low hardware cost can offer useful insights for practical RA implementation.

PAs represent another waveguide-based flexible antenna paradigm, which achieves near-field advantages by flexibly deploying multiple candidate PA locations along low-loss dielectric waveguides. In NTNs, these waveguides can be deployed along building facades, roof edges, or road infrastructure. With intelligent configuration of PAs, dynamic control of signal radiation location and propagation characteristics can be achieved. This mechanism is suitable for terrestrial-side assistance in NTNs \cite{174-nn, 175-nn}. For example, in scenarios with high-speed UAV movement and severe urban obstruction, PAs can provide more stable LoS connections, lower path loss, and better coverage continuity for A2G links by extending the waveguide and flexibly selecting radiation points. 

Recent studies have proposed integrating flexible antennas with hybrid MIMO beamforming to create a tri-hybrid MIMO architecture. This approach combines digital and analog beamforming with antenna reconfigurability, optimizing signal transmission and enhancing spatial multiplexing \cite{176-nn, 177-nn}. Compared to conventional massive MIMO, it reduces RF chains and energy consumption while maintaining high performance. Future work should establish physically consistent radiation models for diverse flexible antenna architectures across the digital, analog, and circuit layers. Furthermore, for the different NTN environments, hardware implementation, energy consumption, and algorithm complexity are also needed for consideration.}

\subsection{High-Frequency Communication}
While millimeter-Wave (mmWave) communication has proved significant in unlocking 5G capacity, it falls short of meeting the 6G performance indicators. Terahertz (THz) communication is a promising solution for faster data rates, extreme connectivity, and minimal latency transmission. It can effectively solve the bottleneck of spectrum scarcity and capacity in the network, significantly improving communication efficiency in FA-assisted NTNs.

Large-scale THz communication applications include service/feeder links, satellite networking, and UAV networks. High-capacity satellite networks deliver wide coverage, ultra-reliable, high throughput communication for UAVs, supplemented by services like navigation and remote sensing. However, THz communication in large-scale scenarios suffers from high propagation loss, susceptibility to LoS blockage, and rank-deficiency of the channel matrix. These challenges necessitate the use of massive antenna arrays and RF chains to mitigate long-distance transmission losses, where the costs of hardware and energy consumption are prohibitive. Moreover, high-frequency communication is easily interrupted, and narrow beam directional gain complicates precise localization for both communication and sensing functions. Given the additional spatial DoFs and flexible configurations from FA-assisted NTNs \cite{172}, their integration with THz communication is essential. FAs optimize THz channel gains and beamforming, minimizing transmission losses while achieving precise beam alignment. Moreover, FAMA effectively mitigates inter-cluster interference, enhancing multi-access capabilities for THz communication.

In small-scale scenarios within 100 m, THz applications can supply higher data rates and lower energy consumption to support ultra-reliable and low-latency communication. It is critical for traffic monitoring, real-time route planning, and accident detection to respond rapidly to dynamic events, as even minor delays can have significant repercussions. Additionally, THz communication can optimize computing resource utilization, efficiency, processing capacity, and memory performance. With the continuous generation and analysis of massive data in 6G networks, high data rates are crucial to prevent system overload.
\vspace{-2ex}
\subsection{Near-Field Communication}
As NTNs evolve toward larger FA array apertures and higher carrier frequencies with service scenarios diversifying, near-field effects in short-range links become inevitable. For instance, UAV communications in vehicular networking, smart factories, and UAV swarms often operate over short distances, making near-field communications more likely. Moreover, recently, near-field UAV-enabled ISAC can enhance spatial resolution and expand the design space for PLS \cite{172-n, 173-n}.

Near-field communications occur when the distance between transmitter and receiver falls below the Rayleigh distance, which is proportional to the square of the array aperture and operating frequency. In near-field communications, characteristics such as spherical waves and spatially non-stationary scattering overturn the traditional assumptions of angular domain sparse and plane wave. In this case, electromagnetic waves transition from planar to spherical waves, leading to electromagnetic and beamforming effects in the wireless near-field region, which are non-negligible \cite{173}. Near-field communications exhibit beam effects, including beam focusing and beam splitting. The beam focusing effect generates precise focused beams at specific directions by precoding transmitted signals and weighting the spherical waves of each antenna array element. At designated angles and distances, this effect concentrates energy at targeted locations, mitigating inter-user interference. For instance, FA-assisted UAVs, serving as aerial base stations, adjust antenna distances to reduce the spatial correlation of vectors at different angles, thereby distinguishing between various user clusters \cite{174}. Moreover, beam focusing effect enhances resolution in the distance domain, enabling position-based multiple-access techniques and PLS techniques. On the other hand, in near-field communications, spherical waves at different frequencies focus on distinct physical locations in the near field. This phenomenon, known as the beam splitting effect, causes severe array gain loss because beams in different frequencies fail to align targeted users at specific positions. However, it also brings benefits: by generating multiple beams spatially with the same pilot, FA configurations can be designed to control the angular coverage of beams across frequencies. This advantage enables rapid CSI acquisition and efficient multi-stream transmission.
\vspace{-2ex}
\subsection{Integrated Sensing, Communication, and Computation (ISCC)}
The deep integration of communication, sensing, and AI systems has become a pivotal trend in 6G, driving the exploration of ISCC systems and their enabling technologies \cite{175}. ISCC systems integrate physical-digital spatial sensing, ubiquitous intelligent communication, and computing capabilities. Specifically, ISCC sensing comprises external and intrinsic network sensing. The former focuses on electromagnetic environments (e.g., CSI, spectrum resources, and interference) and physical spaces (e.g., target localization, and motion states), while the latter concentrates on network service awareness (e.g., throughput, latency, and computing demands) and digital space awareness (e.g., device status, resource management, and AI model states). ISCC computing utilizes real-time processing nodes or edge terminals, which handle dynamic data through mobile edge computing and AirComp. ISCC communication serves as a critical enabler for sensing information sharing and distributed computing, establishing the foundation for ubiquitous connectivity and wide-area collaboration. In FA-assisted NTNs, ISCC enhances the sharing of sensing information, coordination of distributed computing, and ubiquitous connectivity.

\subsubsection{Communication and Computation-Enhanced Sensing}
ISCC leverages the flexible deployment and channel spatial adaptability of FA-assisted NTNs to transmit and aggregate sensing information efficiently, enabling multi-node collaborative sensing and expanding the scope and depth of sensing. The distributed computing capabilities of ISCC enable rapid processing of original sensing data at the edge, as well as feature extraction and fusion. By employing advanced AI models, ISCC can capture semantic information from original data to facilitate the shift from environmental sensing to environmental understanding. This significantly improves system’ intent comprehension and performs more accurate decision-making.
 
\subsubsection{Sensing and Computation-Enhanced Communication}
ISCC leverages the advanced sensing capabilities of FA arrays to collect comprehensive data from the user environment, providing prior knowledge for communication systems. FA-assisted NTNs utilize distributed computing to achieve rapid and accurate channel measurements, estimations, and beam alignments. Large AI models reconstruct unknown wireless environments by analyzing multidimensional sensing data and historical information to optimize transmission strategies and enhance communication performance.

\subsubsection{Sensing and Communication-Enhanced Computation} 
Sensing capabilities of ISCC enrich the data sources and prior information for distributed computing in FA-assisted NTNs, enhancing the QoS and robustness of AI models of NTNs through training and fine-tuning. The computing power sensing function can also be realized for flexible resource allocation. Meanwhile, ubiquitous connectivity accelerates collaborative computing resource distribution, improving model training efficiency. Fast model training and inference facilitate high-reliability, low-latency intelligent decision-making in FA-assisted NTNs.
\section{Conclusion}\label{conclusion}
This paper has provided a comprehensive survey on the FA-assisted NTNs in 6G. We first reviewed the fundamentals and applications of NTNs and FAs to outline the key benefits and adaptability of FAs for NTNs. Considering the high latency, dynamics, severe fading, and interference in NTNs, we then delved into the general channel modeling and CSI estimation methods for FA-assisted NTNs from the perspectives of model-based and AI-based approaches.
Next, we conducted a detailed literature review of the SoTA joint optimizations of FA-assisted NTNs, highlighting the crucial roles of FAs in overcoming traditional NTN limitations. Moreover, FA-assisted NTNs combined with other B5G technologies can enhance coverage, connectivity, and reliability. Furthermore, FA-assisted NTNs held great potential in future intelligent function-integration networks to arrange communication, sensing and computation. For the security of FA-assisted NTNs, the PLS and covert communication were investigated in detail.
Finally, we deliberated on promising research directions and opportunities for FA-integrated NTNs, offering insights to guide future investigations. Extensive explorations of academia and industry revealed that FA-assisted NTNs can unlock the enormous potential of 6G, presenting new horizons and innovative applications.

\end{document}